\documentclass[aps,prd,onecolumn,showpacs,nofootinbib,amssymb]{revtex4}
\usepackage{graphicx}
\usepackage{amsmath}
\usepackage{amssymb}
\usepackage{amsfonts}
\usepackage{bm}
\usepackage{color}

\def\be{\begin{equation}}
\def\ee{\end{equation}}
\def\bea{\begin{eqnarray}}
\def\eea{\end{eqnarray}}

\begin{document}

\title{Landau equation for self-gravitating classical and quantum particles: \\
Application to
dark matter}
\author{Pierre-Henri Chavanis}
%\email{chavanis@irsamc.ups-tlse.fr}
\affiliation{Laboratoire de Physique Th\'eorique, Universit\'e de Toulouse,
CNRS, UPS, France}

\begin{abstract}

We develop the kinetic theory of classical and quantum  particles
(fermions and bosons) in gravitational interaction. The kinetic theory of quantum particles may have
applications in the context of dark matter. For simplicity, we consider an
infinite and spatially homogeneous system (or make a local approximation) and neglect
collective effects. This leads to the quantum Landau equation derived heuristically in
[Chavanis, Physica A {\bf 332}, 89 (2004)]. We establish its main
properties: conservation laws, $H$-theorem, equilibrium state, relaxation time, 
quantum diffusion and friction coefficients, quantum Rosenbluth potentials,
self-consistent evolution, (thermal) bath approximation,  quantum Fokker-Planck
equation, quantum King model... For bosonic particles, the Landau equation can
describe the
process of Bose-Einstein condensation. We
discuss the relation of our study with
the works of [Levkov {\it et al.} Phys. Rev. Lett. {\bf 121}, 151301 (2018);
Bar-Or {\it et al.} Astrophys. J. {\bf 871}, 28 (2019)] on fuzzy dark matter
halos and the formation of Bose stars and solitons.

\end{abstract}

\pacs{95.30.Sf, 95.35.+d, 95.36.+x, 98.62.Gq, 98.80.-k}

\maketitle

\section{Introduction}
\label{sec_introduction}

The kinetic theory of self-gravitating systems was pioneered by Chandrasekhar
\cite{chandra}
following previous works by Jeans \cite{jeansk1,jeansk2,jeansbook},
Schwarzschild \cite{schwk}, Ambarzumian \cite{amba} and Spitzer
\cite{spitzer1940}. He evaluated the relaxation
time of a stellar system due to gravitational encounters
and found that it scales as $t_{R}\sim (N/\ln
N)t_D$, where $N$ is
the number of stars in the system and $t_D$ is the dynamical time. For
stellar systems which contain a large number of stars, like spiral or elliptical galaxies,
the relaxation time is larger than the age of the universe by many orders of magnitude
($N\sim 10^{12}$, $t_D\sim 10^8\, {\rm yrs}$, ${\rm age}\sim 10^{10}\, {\rm yrs}\sim 100\, t_D$, $t_R\sim 10^{20}\, {\rm yrs}$ for our Galaxy).  As a
result, galaxies are essentially collisionless.\footnote{Close gravitational encounters may be
important in galactic nuclei.} They are described by the Vlasov-Poisson
equations \cite{jeans,vlasov}. Their apparent regularity is due to a
process of violent collisionless relaxation, operating on a few dynamical times, first
identified by King \cite{kingvrnewref} and H\'enon
\cite{henonvr} 
 and formalized by Lynden-Bell \cite{lb}. For stellar systems
which contain
a moderate number of stars, like globular clusters or galactic nuclei,
the relaxation time is comparable to their age ($N\sim 10^{5}$, $t_D\sim 10^5\, {\rm yrs}$, ${\rm age}\sim 10^{10}\, {\rm yrs}\sim 10^5\, t_D$, $t_{R}\sim 10^{10}\, {\rm yrs}$ for globular clusters).
Therefore, globular clusters are ``collisional'' and undergo a slow (secular) evolution due
to gravitational encounters. In
the simplest description, which is based on a local approximation (equivalent to
assuming that the system is infinite and homogeneous) and neglects
collective effects, their evolution is described by the gravitational Landau
\cite{landau} equation. The Landau equation can be obtained from the Boltzmann
\cite{boltzmann} equation in a limit of weak deflections. An equivalent equation
was obtained independently by Chandrasekhar
\cite{chandra1,chandra2,chandra3} (and generalized by Rosenbluth {\it et al.}
\cite{rosen}) from the Fokker-Planck equation of Brownian theory
\cite{chandrastoch,nice}.\footnote{The Chandrasekhar equation appears in a  less
neat (less symmetric)  form than the Landau equation. The connection between these two
equations is discussed in \cite{aakin} and in Sec. \ref{sec_c} of
the present
paper.} The kinetic theory of self-gravitating systems was initially used to
study the evaporation of globular clusters due to star encounters
\cite{amba,spitzer1940,chandra,chandra2,chandra3,spitzer,michie,king,lcstar}. Recently, this kinetic
theory has been generalized in order to describe fully spatially
inhomogeneous systems and take collective effects into account. This leads to
the inhomogeneous Lenard-Balescu equation \cite{heyvaerts,angleaction2} written
with angle-action variables which is essentially exact at the order $1/N$. The
Landau (or Lenard-Balescu) equation conserves the mass and the
energy, increases the Boltzmann entropy ($H$-theorem), and relaxes towards the
Boltzmann distribution function (DF). In the thermal bath approximation, where the field particles are
at statistical equilibrium, it takes the form of a Kramers
equation with an anisotropic velocity-dependent diffusion tensor. From this
kinetic equation one can derive a truncated model called the  King
model \cite{king,kingmodel}.

In addition to baryonic matter (stars), the universe contains a large amount of dark matter (DM).
The nature of DM is still unknown and remains one of the greatest
mysteries of modern cosmology. DM may be made of quantum particles, either
fermions (like sterile neutrinos) or bosons (like
axions).\footnote{See the introduction of \cite{clm1,clm2,gr1} (resp.
\cite{prd1,tunnel}) for a short history and an
exhaustive list of references on fermionic (resp. bosonic) DM. See also
the reviews  \cite{srm,rds,chavanisbook,marshrevue,leerevue,niemeyer,ferreira}
on bosonic DM.}

The collisional evolution of a gas of self-gravitating fermions is described by
the fermionic Landau equation  \cite{kingen} which involves a term of the form $f(1-f)$ taking into
account the Pauli
exclusion principle. This equation conserves the mass and the
energy, increases the Fermi-Dirac entropy ($H$-theorem), and relaxes towards the
Fermi-Dirac DF. In the thermal bath approximation, where the field particles are
at statistical equilibrium, it takes the form of a fermionic Kramers
equation with an anisotropic velocity-dependent diffusion tensor. From this
kinetic equation one can derive a truncated model called the fermionic King
model \cite{chavmnras,kingen}. However, the
relaxation time towards the Fermi-Dirac DF of
quantum statistical mechanics due to gravitational encounters is extremely long because the
number of particles  in a DM halo is astoundingly  large
(typically $N\sim 10^{75}$ for fermions
of mass $m\sim 100\, {\rm eV/c^2}$). In addition, we will see  that, in the
case of fermions,
quantum  effects (Pauli's blocking) have the tendency to
increase the
relaxation time even more, especially at low temperatures. Therefore,
fermionic DM halos are essentially collisionless. They are described in good approximation
by the classical Vlasov-Poisson equations.\footnote{More fundamentally, one should use the Wigner-Poisson equations for a multistate system. However, in the case of fermions, the quantum potential arising from the Heisenberg uncertainty principle can usually be
neglected (Thomas-Fermi approximation) leading to the classical Vlasov
equation.}
In principle, this should preclude their relaxation towards the Fermi-Dirac
distribution on a relevant timescale. However, Lynden-Bell \cite{lb} has shown
that
collisionless self-gravitating systems described by the Vlasov-Poisson
equations experience a process of violent relaxation on
the coarse-grained scale if they are initially unsteady or dynamically
unstable. Interestingly, the properties of the Vlasov
equation imply that the coarse-grained DF $\overline{f}({\bf r},{\bf
v},t)$ satisfies an exclusion principle
similar to the Pauli exclusion principle in quantum mechanics. Degeneracy in the
sense of Lynden-Bell may be negligible for classical particles \cite{lb} while
it may be relevant for self-gravitating fermions \cite{clm1,clm2}. As a result,
self-gravitating fermions should violently relax towards
an out-of-equilibrium quasistationary state  which is similar to the Fermi-Dirac
DF in quantum
mechanics. This process takes place on a few dynamical times $t_D$. In addition,
the kinetic equation for the coarse-grained DF is similar to the fermionic
Vlasov-Landau equation \cite{kp,sl,chavmnras,kingen}.\footnote{The
fact that this kinetic equation relaxes towards the Lynden-Bell  DF on a few
dynamical times is interpreted by Kadomtsev and Pogutse \cite{kp} in terms of
``collisions'' between  macroparticles with a large effective mass.}
Accordingly, in fermionic DM halos, the Fermi-Dirac distribution and the
fermionic Landau equation may be justified by
the theory of violent relaxation of Lynden-Bell, not by the usual statistical
mechanics and kinetic theory of quantum particles. Since violent
collisionless relaxation occurs on a much shorter timescale than collisional relaxation, the Lynden-Bell theory may solve the timescale problem of fermionic DM halos,
i.e., it may explain why fermionic DM halos are described by the Fermi-Dirac DF although they are
collisionless \cite{clm1,clm2}. Fermionic DM halos described by the Lynden-Bell
(Fermi-Dirac-like) DF generically have a ``core-halo'' structure
\cite{csmnras}. They are made of a quantum core (fermion ball) which
corresponds to the ground state of the self-gravitating Fermi gas at $T=0$
surrounded by an isothermal atmosphere (in the sense of Lynden-Bell) with an
effective temperature $T_{\rm eff}$. The quantum core (fermion ball) is
stabilized against gravitational collapse by the pressure arising from the
exclusion principle (Lynden-Bell and Pauli) so that the density does not diverge
at $r=0$.  This may solve the cusp-core problem of the classical CDM
model.\footnote{Numerical simulations of
classical collisionless CDM lead to a universal density profile called
the NFW profile \cite{nfw}. This profile presents an $r^{-1}$ central cusp which
is not consistent with the observations that rather favor flat cores
\cite{observations}.} On the other hand, an isothermal halo where the density
decreases as $r^{-2}$ at large distances yields flat rotation curves  in
agreement with the observations.

Let us now turn to the case of self-gravitating bosons. At sufficiently high
temperatures (above the condensation temperature $T_c$),\footnote{In the case of
fuzzy dark matter (FDM), with bosons of mass $m\sim 10^{-22}\, {\rm eV/c^2}$, we
are not in this situation (see Appendix
\ref{sec_fdm}). We are rather in the situation, considered below, where $T\ll
T_c$.}
the collisional evolution of uncondensed self-gravitating bosons is described
by the bosonic Landau equation  \cite{kingen} which involves a term of the form $f(1+f)$ taking into account
Bose
enhancement. This equation conserves the mass and the energy,
increases the Bose-Einstein entropy ($H$-theorem), and relaxes towards the
Bose-Einstein DF. In the thermal bath approximation,  where the field particles are
at
statistical equilibrium, it takes the form of a bosonic Kramers equation with an anisotropic
velocity-dependent diffusion tensor. From this kinetic equation one can derive a
truncated model called the bosonic King model \cite{kingen}. As we shall
see, the relaxation time of uncondensed bosons above $T_c$ is very large
and it only slightly
decreases due to Bose enhancement when $T\rightarrow T_c$. Therefore,
self-gravitating bosons above $T_c$ are essentially collisionless. They are
described in good approximation by the classical Vlasov-Poisson
equations.\footnote{More fundamentally, one should use the Wigner-Poisson
equations but, when $T>T_c$, quantum effects are negligible, or weak, during the
collisionless regime.} In this regime, they can experience a process of 
violent relaxation leading to an out-of-equilibrium quasistationary state on a
few dynamical times. Degeneracy in
the sense of Lynden-Bell can be neglected for bosons so the quasistationary state is similar to the Boltzmann distribution. In addition, the kinetic equation for the coarse-grained DF is similar to the classical Vlasov-Landau equation.

We can also use the bosonic Landau equation to study the process of
Bose-Einstein condensation below $T_c$. Let us assume that, at $t=0$, the
bosons are uncondensed but that $T<T_c$. In that case, we expect that, after
some evolution, the DF develops a Dirac peak at ${\bf v}={\bf 0}$ so that
$f({\bf v},t)=f_{\rm gas}({\bf v},t)+M_c(t)\delta({\bf v})$, where $f_{\rm
gas}({\bf v},t)$ is the DF of the uncondensed bosons (gas) and $M_c(t)$ is the mass of
the condensed bosons.  The mass $M_c(t)$ of the Bose-Einstein condensate (BEC)
increases with time until a
statistical equilibrium state is reached. This equilibrium state contains a fraction $(T/T_c)^{3/2}$ of
uncondensed bosons with vanishing chemical potential, the rest being in the form of condensed bosons. For nongravitational
systems, the temporal formation of a BEC (Dirac peak)  when $T<T_c$ has been analytically studied by Sopik {\it et al.} \cite{sopik} in the canonical ensemble (fixed $T$) from the bosonic Kramers equation with
a constant isotropic diffusion coefficient.\footnote{Interestingly, there is a
striking similarity between the Bose-Einstein condensation in the canonical
ensemble  and the gravitational collapse of a gas of classical self-gravitating
Brownian particles \cite{crs}.} They considered the
exact Bose-Einstein model and a simplified model obtained by making the approximation
$f(1+f)\simeq f^2$ to simplify the equations. They obtained a
self-similar solution in the precondensation regime. During this regime, the central DF increases and becomes infinite in a finite time $t_{\rm coll}$ when  $T<T_c$ and in an infinite time when $T=T_c$. At that moment, the chemical potential vanishes. The condensate (Dirac peak) appears
in a postcollapse regime and its mass grows initially as $M_0(t)\propto t-t_{\rm
coll}$ before saturating exponentially rapidly. The statistical equilibrium
state is reached  for $t\rightarrow +\infty$. More recently, Levkov
{\it et al.} \cite{levkov2} considered the mechanism of Bose-Einstein
condensation
and the formation of Bose stars in virialized DM halos and miniclusters by
gravitational interactions (see also \cite{egg,sgbne,cdlmn,kmp}). They started
from an incoherent initial
configuration of bosons described by a random classical field $\psi({\bf r},t)$
satisfying the Schr\"odinger-Poisson equations and assumed that the initial
velocity DF is Maxwellian with $T_0\ll T_c$.  They derived a bosonic Landau
equation in the microcanonical ensemble (fixed $E$) for the DF $f({\bf
v},t)$ in which Bose stimulation is accounted for by
a quadratic term  $f^2$ (their kinetic equation coincides with the bosonic
Landau equation derived in \cite{kingen} when  we make the approximation
$f(1+f)\simeq f^2$). They estimated the kinetic relaxation (or condensation)
time $t_{\rm gr}$ due to gravitational interactions 
and compared it
with direct numerical simulations. They showed that collisional relaxation is
enhanced by the collective interaction of large fluctuations of the boson gas at
large distances. Starting from a spatially homogeneous distribution, they
observed the phenomenon of Bose-Einstein condensation and the formation of
an isolated Bose star. Once the Bose star nucleates, a small $\delta$ peak with
mass $M_s$ appears in the DF. With time the peak grows in
height with an initial rate $M_s(t)\propto (t-t_{\rm gr})^{1/2}$ as
it acquires particles from the surroundings (the scaling changes from $t^{1/2}$ to $t^{1/8}$ at later times \cite{egg,cdlmn}).\footnote{In the precondensation
phase, the central DF grows more and more and diverges at $t_{\rm gr}$ at which
time the Dirac appears. The mass of the Dirac peak grows in the postcondensation
regime. This is very similar to the results of Sopik {\it et al.} \cite{sopik}
except that the exponents characterizing the initial growth of $M_0(t)$ and
$M_s(t)$ are different because the models describe different statistical
ensembles (canonical versus microcanonical).} The Dirac peak $M_s(t)\delta$ in
the velocity or energy distribution gives rise to a Bose star in configuration
space (recall that the wave function $\psi({\bf r},t)$ and the density
$\rho=|\psi|^2$ are related to the Fourier transform of the velocity
distribution $f({\bf v},t)$) while the smooth distribution $f_{\rm gas}$
describes  a sort of homogeneous halo surrounding the Bose star and feeding it.

In the study of Levkov {\it et al.} \cite{levkov2}, it is assumed that the
bosons are initially uncondensed, that they are spatially homogeneously
distributed,  and that their distribution is dynamically stable (they also
consider the formation of Bose stars in virialized miniclusters resulting
from Jeans instability). The bosons initially have an
out-of-equilibrium  Maxwellian DF with
a temperature (or velocity dispersion) $T_0\ll T_c$ so that Bose-Einstein
condensation takes place. The study of Levkov {\it et al.} \cite{levkov2}
therefore describes the slow (secular) evolution of a stable homogeneous system
of bosons, and the process of condensation, caused by gravitational
interactions. Even if the ``collisional'' relaxation  is considerably
accelerated by Bose stimulation, the Bose star still forms on a relatively long
timescale, much longer than the dynamical time $t_{\rm gr}\gg t_D$. Other
authors \cite{ch2,ch3,schwabe,mocz,moczSV,veltmaat,moczprl,moczmnras,veltmaat2}
considered a different situation where the
bosons are in an initial distribution that is spatially inhomogeneous, unsteady,
or dynamically unstable. In that case, a violent collisionless relaxation
leading to a BEC
occurs on a short dynamical timescale before a slower collisional relaxation,
corresponding to the one described by  Levkov {\it et al.} \cite{levkov2}, takes
place. This is the situation that we now review.

Let us consider the dynamical evolution of an inhomogeneous gas of bosons in gravitational interaction in relation to the formation of DM halos. Self-gravitating bosons are basically described by the Schr\"odinger-Poisson equations which govern the evolution of the wavefunction
$\psi({\bf r},t)$.\footnote{If we consider self-interacting bosons with a
scattering length $a_s$, the
Schr\"odinger equation
has to be replaced by the Gross-Pitaevskii equation \cite{prd1}. Basically, one
has to interpret the wavefunction in the Schr\"odinger and Gross-Pitaevskii
equations as an operator (Heisenberg representation). However, when the
occupation
number is large ${\cal N}=n_b\lambda_{\rm dB}^3\gg 1$, which is the case for FDM
(see Appendix \ref{sec_fdm}), the coherent state dynamics of the bosons can be
approximated by a classical field.} These
equations are equivalent to the
Wigner-Poisson equations which govern the evolution of the Wigner DF $f_W({\bf
r},{\bf v},t)$.\footnote{The
Wigner equation is the counterpart of the Klimontovich equation for classical
self-gravitating particles but it takes into account properties specific to
bosons (Heisenberg uncertainty principle and Bose stimulation). In the
collisional (secular) regime, the smoothed-out Wigner DF satisfies the
bosonic Landau equation
\cite{levkov2}.} Let us assume that we start from an initial condition that is
an unsteady or an unstable state of the Schr\"odinger-Poisson or Wigner-Poisson
equations. In a first regime, taking place on a free fall  time, the
Schr\"odinger-Poisson equations  experience a process of gravitational cooling
(like in the case of boson stars \cite{seidel94,gul0,gul}) which is similar to
the violent relaxation of collisionless stellar systems systems
\cite{lb}.\footnote{In this regime, we can neglect collisions and make a mean
field approximation. In that case, the Wigner equation for the smoothed-out
Wigner DF is the
counterpart of the Vlasov  equation for classical collisionless self-gravitating
systems but it takes into account properties specific to
bosons (Heisenberg uncertainty principle) \cite{moczSV}.} The
self-gravitating boson gas  undergoes  gravitational collapse, displays damped
oscillations, and finally settles down on a quasistationary (virialized) state
by radiating part of the scalar field. This is a purely mean field process which
takes place on a few dynamical times.\footnote{In a cosmological setting, if one
starts from an infinite homogeneous distribution of bosons, the gas is expected
to undergo Jeans instability and form clumps (miniclusters). This corresponds to
the linear
regime of structure formation. These clumps grow and, in the nonlinear regime of
structure formation, undergo  free fall, gravitational cooling and violent
relaxation leading to the DM halos that we observe today. During their cosmic
evolution the halos merge  and grow in size to form larger halos. This
corresponds to the hierarchical scenario of bottom-up cosmic structure
formation.} The process of gravitational cooling and
violent relaxation leads to DM halos
with a ``core-halo'' structure. The quantum core (soliton) corresponds to the 
ground state of the Schr\"odinger-Poisson equations. It results from the balance
between the gravitational attraction and the quantum pressure arising from the
Heisenberg uncertainty principle or from the repulsive self-interaction of the
bosons.\footnote{A
repulsive self-interaction ($a_s>0$) stabilizes the quantum
core. By contrast, an attractive  self-interaction destabilizes the quantum core
above a maximum mass $M_{\rm max}=1.012\, \hbar/\sqrt{Gm_b|a_s|}$ first
identified in \cite{prd1}.}  Quantum mechanics is important at ``small'' scales
of the order of the de Broglie length ($\lambda_{\rm dB}=h/(m_b \sigma_b)\sim
1\, {\rm kpc}$) determining the size of the soliton. It stabilizes the system
against gravitational collapse and  may solve the cusp-core problem of classical
CDM. The quantum core (ground state) is surrounded by a halo of scalar radiation
resulting from the quantum interferences of excited states. This extended halo
is made of uncondensed bosons. It is the counterpart of the halo observed in
numerical simulations of classical collisionless CDM. The smooth
density profile of this halo (on a scale larger than de Broglie length) is
similar to the NFW profile \cite{nfw}.
It is also similar to the isothermal profile predicted by the statistical theory
of Lynden-Bell \cite{lb}. It is important to note that this isothermal halo is
an out-of-equilibrium structure with an effective temperature $T_{\rm eff}\ll
T_c$ (it is necessarily thermally out-of-equilibrium otherwise it would be Bose
condensed). An
isothermal halo where the density decreases as $r^{-2}$ at large distances
yields flat rotation curves  in agreement with the observations.\footnote{The
isothermal nature of the halo (Maxwell-Boltzmann distribution) seems to be
confirmed by the numerical results of \cite{veltmaat,moczSV,sgbne}. The
halo cannot
be exactly
isothermal otherwise it would have an infinite
mass \cite{bt}. In reality, the density in the
halo decreases at large distances as $r^{-3}$, similarly to the NFW
\cite{nfw} and Burkert
\cite{observations} profiles, instead of $r^{-2}$ corresponding to the
isothermal
sphere \cite{bt}. This extra-confinement may be due to incomplete
relaxation, tidal
effects, and stochastic
perturbations as discussed in \cite{modeldm}. The $r^{-2}$ isothermal profile
is established in an intermediate region between the core and the external halo.
} Therefore, in the FDM model, the quantum core (soliton) solves the cusp
problem of the CDM model and the approximately
isothermal halo accounts for the flat rotation
curves of the galaxies. In a companion paper \cite{heuristic}, we
have proposed a heuristic
parametrization of the process of gravitational cooling by using  an approach
similar to the kinetic theory \cite{kp,sl,csr,chavmnras,kingen} developed in
connection to the Lynden-Bell theory of violent relaxation
\cite{lb}.\footnote{In that case, the relaxation of the coarse-grained  Wigner
DF $\overline{f}_W({\bf r},{\bf v},t)$ is due to the fluctuations of the
violently changing gravitational potential while the system is unsteady.} In the
case of
bosons, degeneracy in the sense of Lynden-Bell can be
neglected. As a result, the kinetic equation for the  coarse-grained Wigner DF
$\overline{f}_W({\bf r},{\bf v},t)$ has the
form of a Wigner-Landau equation with a Wigner
advection term (taking into account the Heisenberg uncertainty principle)
responsible for the quantum core (soliton)
and
an effective classical Landau ``collision'' term (taking into account the strong
fluctuations of the gravitational potential)
responsible for the
isothermal halo.  This core-halo
structure  has been clearly
evidenced in numerical simulations of the Schr\"odinger-Poisson equations
\cite{ch2,ch3,schwabe,mocz,moczSV,veltmaat,moczprl,moczmnras,veltmaat2}. The
core
mass-halo mass relation
$M_c(M_v)$ has been obtained numerically in Ref. \cite{ch3}  and explained in
Refs. \cite{modeldm,mcmh} from an effective maximum entropy principle in the
sense of Lynden-Bell (see also \cite{ch3,veltmaat,mocz,egg,bbbs} for other
justifications). For
noninteracting bosons, the core mass scales as $M_c\propto M_v^{1/3}$.

On a longer timescale, the bosons in the halo undergo  a collisional evolution.
As shown by Schive {\it et al.} \cite{ch2,ch3}, quantum wave interferences
produce time-dependent small-scale
density granules of the size $\lambda_{\rm dB}=h/(m_b \sigma_b)\sim 1\, {\rm
kpc}$ of the solitonic core. These granules
have been clearly evidenced in numerical simulations of the
Schr\"odinger-Poisson equations
\cite{ch2,ch3,schwabe,mocz,moczSV,veltmaat,moczprl,moczmnras,veltmaat2}. They
have been
interpreted by Hui {\it et al.} \cite{hui} as quasiparticles with an effective
mass $m_{\rm eff}\sim\rho_b \lambda_{\rm dB}^3\sim 10^7\,
M_{\odot}\gg m_b$ that
depends on the local halo density $\rho_b$ and velocity
distribution.\footnote{These quasiparticles are fundamentally different from the
macroparticles introduced by  Kadomtsev and Pogutse \cite{kp} in the context
of the violent relaxation of collisionless stellar systems.} These authors
argued that
the fluctuating gravitational force in a FDM halo of
mean density $\rho_b$ is similar to that of a classical $N$-body system composed
of such quasiparticles.  In CDM halos, the mass $m_b$ of the particles is so
small that the relaxation time is much larger than the age of the universe.
Therefore, CDM halos are collisionless. In FDM halos, the effective mass $m_{\rm
eff}$ of the quasiparticles is large enough that relaxation can be important.
Relaxation between quasiparticles can lead to the formation and/or growth of a
central BEC (soliton). The granules can also cause the diffusion (heating) of
light test particles such as stars (with $m\ll m_{\rm eff}$) on a secular time.
This can heat and expand the central regions of a stellar system embedded in the
halo. On the other hand, heavy test particles such as black holes or globular
clusters (with $m\gg m_{\rm eff}$) experience dynamical friction (cooling). The
variation of the effective mass across the system can stall the inspiral of the
massive object toward the center of the galaxy at the distance where the
effective mass of the quasiparticles $m_{\rm eff}(r)$ becomes equal to the mass
$m$ of the test particles.

The evolution of test particles in a halo of FDM due to
``collisions'' with quasiparticles has been analyzed by Bar-Or {\it et al.}
\cite{bft}.
Starting from the Schr\"odinger-Poisson equations, they
developed a kinetic theory and quantitatively showed that FDM halos indeed behave as classical
self-gravitating systems (similar to globular clusters) made of quasiparticles
with an effective mass $m_{\rm eff}$. Because of Bose enhancement, the
quasiparticles are much heavier than the bosons  ($m_{\rm eff}\gg m_b$). This
strongly accelerates the collisional relaxation with respect to the one that
would be caused by particles of mass $m_b$. They identified a heating time
(coinciding with the condensation time of \cite{levkov2}) strongly reduced by
the effective mass and a cooling time independent of the effective
mass. A similar description has been developed by Marsh and  Niemeyer \cite{mn}
and El-Zant {\it et al.} \cite{zant}. We will see that their results can be
recovered from the bosonic Landau equation of Ref. \cite{kingen} by considering
classical test particles in collision with bosonic field particles in an
out-of-equilibrium quasistationary state (resulting from the violent relaxation
process) for which we can  make the approximation $f(1+f)\simeq f^2$ and assume
that the DF of the field particles $f$ is Maxwellien  with a temperature $T\ll
T_c$.

The granules (quasiparticles) also provoke the secular evolution
of the halo itself. This evolution can be described by the self-consistent
bosonic Landau equation \cite{kingen}. On a secular timescale, the bosons of the
halo condense
(because $T\ll T_c$) and ``feed'' the central soliton. As a result, the mass of
the soliton increases while the halo is slowly depleted.  This is the spatially
``inhomogeneous'' version of the ``homogeneous'' situation studied by Levkov
{\it et al.} \cite{levkov2}. Note that this mechanism of condensation, which
operates on a secular timescale, is physically different from the process of
gravitational cooling and violent relaxation mentioned above, which operates on
a dynamical timescale. In the ``homogeneous''  approach of Levkov {\it et al.}
(where gravitational cooling and violent relaxation are absent because the
initial state is virialized), the formation of the soliton is exclusively due to
the condensation caused by gravitational interactions (collisions). In the
``inhomogeneous'' approach
of \cite{ch2,ch3,schwabe,mocz,moczSV,veltmaat,moczprl,moczmnras,veltmaat2}
(where the
initial state is dynamically unstable), the soliton and the halo form during the
phase of free fall through a process of violent collisionless relaxation.
However, on a longer (secular) timescale, the soliton continues to grow like in
\cite{levkov2},  being fed by the halo. If the bosons have an attractive
self-interaction (see footnote 15), the soliton may grow overcritical
($M_s>M_{\rm max}$ \cite{prd1}) and collapse \cite{collapse} or explode into
relativistic axions (bosenova) and emit radio-photons via parametric resonance
\cite{levkov1}. The distinction
between the phase of violent relaxation (formation of the soliton) and  the phase
of slow collisional relaxation (growth of the soliton) is illustrated in \cite{egg}.
The collapse of the soliton and the formation of multi boson stars in 
miniclusters by a process of fragmentation when the bosons have an attractive
self-interaction and $M_s>M_{\rm max}$ is shown in \cite{cdlmn}.

For completeness, we would like to mention another possible scenario of
evolution \cite{modeldm}. If the bosons have a strongly repulsive
self-interaction, the collisional
evolution of the halo may be due to self-interaction  instead of gravitational
encounters (these comments also apply to fermions). Such an evolution can be
described by the ordinary Boltzmann kinetic
equation.\footnote{If
the bosons are self-interacting, they may experience a relaxation due to direct
``collisions'' with a cross section $\sigma=\pi a_s^2$ rather than a relaxation
due to gravitational scattering with a Rutherford cross section $\sigma_{\rm
gr}\sim \pi G^2m^2\ln\Lambda/\sigma^4$. As discussed in footnote 52 of
\cite{modeldm}, a relevant situation corresponds to bosons with a mass
$m=1.10\times 10^{-3}\, {\rm eV/c^2}$ and scattering length $a_s=4.41\times
10^{-6}\, {\rm fm}$. For QCD axions and ultralight axions, Levkov {\it et
al.} \cite{levkov2} show that the collisional relaxation
due to the self-coupling is slower than the collisional relaxation due to
two-body gravitational encounters. However, for bosons with
$m=1.10\times 10^{-3}\, {\rm eV/c^2}$ and $a_s=4.41\times 10^{-6}\, {\rm fm}$,
we find the opposite.
Self-interaction wins over gravity by $\tau_{\rm gr}/\tau_{\rm self}\sim
\sigma_{\rm self}/\sigma_{\rm gr}\sim \sigma^4 a_s^2/(G^2m^2\ln\Lambda)\sim
10^{77}$. In that case, the self-interaction relaxation time $t_{\rm
self}\sim 1/(n\sigma a_s^2)\sim 10^{11}\, {\rm yrs}$ is comparable to the age
of the universe (it is shorter in the core of the galaxy where the density is
higher). In the numerical applications we have taken  $\sigma\sim 100\, {\rm
km/s}$ and $\rho_b\sim 4\times
10^{-3}\, M_{\odot}/{\rm pc}^3$ (see Appendix
\ref{sec_fdm}).} This
corresponds to self-interacting dark matter (SIDM) \cite{sidm}. Close collisions
can
establish an isothermal distribution, especially in the center of the halo. We
are then led to the  scenario of evolution discussed by
\cite{balberg,modeldm} according to which the DM halo is isothermal and,
because of evaporation,  slowly evolves along the series of equilibria towards
states of higher and higher central density. At some point (when the central
density reaches a critical value corresponding to the minimum energy in the
series of equilibria), it becomes thermodynamically unstable and undergoes a
gravothermal catastrophe on a secular timescale. Core collapse can be followed
by a dynamical instability of general relativistic origin leading to the
formation of a supermassive black hole. This scenario only works for
sufficiently large DM halos such that the gravothermal catastrophe can take
place (see \cite{modeldm} for details). This may explain why supermassive
black holes are observed only at the center of large galaxies.

The goal of the present paper is to establish the general
kinetic equations that describe the collisional evolution of self-gravitating
systems of quantum particles (fermion and bosons). We remain at a general level
and treat all possible situations, even those that do not correspond to
situations of most astrophysical interest (for example, we consider the case of bosons at
$T>T_c$ in spite of the fact that the situation relevant to FDM is $T\ll T_c$). The paper is
organized as follows. In Sec. \ref{sec_mqle}, we recall the heuristic derivation
of the quantum Landau equation presented in \cite{kingen} and extend it to a
multispecies system of particles. In Sec. \ref{sec_c}, we review the kinetic
theory of classical self-gravitating particles. In Sec. \ref{sec_q}, we extend
these results to the case of quantum particles. In Sec. \ref{sec_m}, we consider
the case of classical particles in collision with quantum particles. In Sec.
\ref{sec_gbm}, we show at a general level that the diffusion caused by quantum
particles may be interpreted in terms of quasiparticles with an effective mass
$m_{\rm eff}$ and recover, in the case of FDM, the results of \cite{bft}. In
Sec. \ref{sec_ssl}, we consider the evolution of the system as a whole governed
by the self-consistent Landau equation. We discuss the possibility, in the case of
bosons, to describe the process of Bose-Einstein condensation. In Appendix
\ref{sec_fdm}, we discuss the validity of the approximations made in the context
of FDM. In Appendix \ref{sec_cl}, we establish the main properties of the
multi-species quantum Landau equation. In Appendix \ref{sec_logo}, we discuss
the simplification of the bosonic Landau equation in the limit of large
occupation numbers. In Appendix \ref{sec_mapp}, we study the rate of change of
the energy of classical particles due to collisions with other classical
particles. In Appendix \ref{sec_mapp}, we extend  these results to the case of
classical particles experiencing collision with quasiparticles in FDM halos. In
Appendix \ref{sec_ssss}, we derive a self-similar solution of the
Spitzer-Schwarzschild equation describing the diffusion of classical
self-gravitating systems when $m\ll m_b$.

\section{Multispecies classical and quantum Landau equation}
\label{sec_mqle}

\subsection{Classical Landau equation}
\label{sec_cle}

When considering the ``collisional'' evolution of an isolated
system of classical particles, it is natural to start from the Boltzmann
equation \cite{boltzmann}
\begin{eqnarray}
\label{cle1}
\frac{\partial f}{\partial t}=\int d{\bf v}_1d\Omega\, w({\bf v},{\bf v}_1;{\bf
v}',{\bf v}'_{1})\lbrace f({\bf v}',t) f({\bf v}'_1,t)- f({\bf v},t) f({\bf
v}_1,t)\rbrace,
\end{eqnarray}
where $d\Omega$ is the element of solid angle and $w({\bf v},{\bf v}_1;{\bf
v}',{\bf v}'_{1})$ is the density probability of a collision transforming the
velocities ${\bf v}$, ${\bf v}_1$ into ${\bf v}'$, ${\bf v}'_1$ or the converse
(see, e.g., \cite{balescubook} for details).
This transition rate depends only on the nature of the two-body particle
interaction. According to the principle of detailed balance, the transition
rate satisfies the symmetry property $w({\bf v},{\bf v}_1;{\bf
v}',{\bf v}'_{1})=w({\bf v}',{\bf v}'_1;{\bf
v},{\bf v}_{1})$ expressing the fact that the probability densities of the
direct and inverse collisions are equal. The above description assumes that the
system is a dilute (low density) gas so that it can be described by a
one-particle DF. We also assume that the system is spatially homogeneous so that
the DF depends only on the velocity and
time, i.e., $f=f({\bf v},t)$.

If we now
consider weakly coupled gases of particles in interaction, like charges in a
plasma or stars in a stellar system, we can make an additional approximation. In
the case of Coulombian or Newtonian potentials, the interactions are very weak
compared to the mean kinetic energy of the particles and this implies that the
deviation resulting from a collision is small (on the average) compared to the
initial velocity of the particles. In that case, we can assume that each
encounter provokes a weak deflection of the particle trajectory.  This is the
approach followed by Landau \cite{landau} in the case of Coulombian plasmas. His
results also apply to gravitational plasmas and we shall write them directly in
that context. Considering the weak deflection limit of the Boltzmann
equation, Landau obtained a kinetic equation of the form \cite{landau}
\begin{eqnarray}
\label{cle2}
\frac{\partial f}{\partial t}=2\pi G^2m\ln\Lambda\frac{\partial}{\partial
v_i}\int d{\bf v}' K_{ij}\left ( f' \frac{\partial
f}{\partial v_j}-  f \frac{\partial f'}{\partial v'_j}\right )
\end{eqnarray}
with
\begin{eqnarray}
\label{cle3}
K_{ij}=\frac{u^2\delta_{ij}-u_iu_j}{u^3},
\end{eqnarray}
where ${\bf u}={\bf v}'-{\bf v}$ is the relative velocity between the particles
engaged in a collision. We have noted $f$ for $f({\bf v},t)$ and  $f'$ for
$f({\bf v}',t)$. The Landau equation can be viewed as a Fokker-Planck
equation involving a diffusion term and a friction term. There is a well-known
difficulty with the Landau approach in
the sense that the Landau equation yields a logarithmically diverging quantity
$\ln\Lambda=\int db/b$, where $b$ is the impact parameter ($\ln\Lambda$ is
called the Coulomb logarithm).\footnote{In a Coulombian plasma, the divergence
at large scales can be avoided by taking into account
collective effects (Debye shielding). This leads to the Lenard-Balescu equation
\cite{lenard,balescu}. In a gravitational plasma (stellar system), the
divergence at large scales can be avoided by taking into account spatial
inhomogeneity. This leads to the inhomogeneous Landau equation \cite{aakin}. One
can also take into account collective effects (anti-shielding) leading to the
inhomogeneous Lenard-Balescu equation \cite{heyvaerts,angleaction2}. These
equations can be obtained from systematic methods, the BBGKY hierarchy (based on
the Liouville equation) or the quasilinear theory (based on the Klimontovich
equation), by considering an expansion of the exact equations of motion in power
of $1/N$ (see
the introduction of \cite{epjp2,aakin} for a short history on kinetic
theories). In this
paper, for simplicity, we will neglect collective effects and assume that the
system is spatially homogeneous.}
Since the divergence is weak (logarithmic), the procedure is marginally
valid. The divergence can be regularized  by introducing cut-offs.
The small
scale cut-off is the Landau length $\lambda_L\sim Gm/\sigma^2\sim Gm^2/k_B T$
(where $\sigma\sim (k_B T/m)^{1/2}$ denotes the typical root mean square velocity of
the particles and $T$ the temperature) corresponding to the distance
of closest
approach at which a collision produces a deflection at $90^o$. This cut-off eliminates
very close encounters which produce large deflections. The large scale
cut-off is the Jeans length
$\lambda_J\sim\sigma t_D\sim \sigma/\sqrt{G\rho}\sim (k_B T/G\rho m)^{1/2}$
(where $t_D\sim 1/\sqrt{G\rho}$ is the dynamical time and $\rho=n m$ is the mass density)
which is the
presumable analogue
of the Debye length in plasma physics.\footnote{According to the  virial
theorem, we have $\sigma^2\sim {GM}/{R}$, where $M$ is the total mass of the
stellar system and $R$ is its radius. Therefore, $\lambda_J\sim R$ showing that
the Jeans length is of the order of the system's size.} This yields
\begin{eqnarray}
\label{cle4}
\ln\Lambda\sim \ln\left (\frac{\lambda_J}{\lambda_L}\right )\sim \ln\left
(n\lambda_J^3\right )\sim \ln N,
\end{eqnarray}
where $N$ is the number of particles in the system.

The Landau equation can be generalized to a multi-species system as
\begin{eqnarray}
\label{cle5}
\frac{\partial f_a}{\partial t}=2\pi G^2
\sum_b \ln\Lambda_{ab} \frac{\partial}{\partial
v_i}\int d{\bf v}' K_{ij}\left ( m_b f'_b \frac{\partial
f_a}{\partial v_j}- m_a f_a \frac{\partial f'_b}{\partial
v'_j}\right ),
\end{eqnarray}
where $f_a({\bf v},t)$ is the DF of particles of species ``a'' with mass $m_a$,
$f_b({\bf v},t)$ is the DF of particles of species ``b'' with mass
$m_b$ and $\ln\Lambda_{ab}$ is the Coulomb logarithm constructed with the Landau
length $\lambda_L=G(m+m_b)/V_{ab}^2$ (where $V_{ab}$ is a typical 
relative velocity).

The Landau equation conserves the mass $M_a$ of each species and the total
energy $E$ of the system. It also satisfies an $H$-theorem for the Boltzmann entropy (see
Appendix \ref{sec_cl}). As a result, it relaxes towards the Boltzmann
distribution
\begin{eqnarray}
\label{cle6}
f_a^{\rm eq}({\bf v})=\rho_a \left (\frac{\beta m_a}{2\pi}\right
)^{3/2}e^{-\beta m_a
\frac{v^2}{2}},
\end{eqnarray}
which maximizes the Boltzmann entropy $S$ at fixed masses $M_a$ and energy $E$ (see
Appendix \ref{sec_cl}).
Here $\rho_a=\int f_a\, d{\bf v}$ is the mass density of species a and
$\beta=1/k_B T$ is the inverse temperature. We note that the temperature $T$
is the same for all the species of particles (this is the Lagrange multiplier
associated with the conservation of the total energy). This leads to the theorem of
equipartition of energy:
\begin{eqnarray}
\label{cle7}
\frac{1}{2}m_a\langle v^2\rangle_a=\frac{3}{2}k_B T.
\end{eqnarray}
Equipartition of energy usually implies that heavy particles sink at the center
of the system while light particles wander around.

{\it Remark:} We can obtain an estimate of the relaxation time as
follows.\footnote{We consider a single species system for simplicity. More
accurate expressions of the relaxation time for single and multispecies systems
are given in Sec. \ref{sec_dke}.} Writing $v\sim \sigma$ and $f\sim
\rho/\sigma^3$ and considering the scaling of the different terms
appearing in the classical Landau equation (\ref{cle2}) we obtain
\begin{eqnarray}
\label{cle8}
t_R\sim \frac{\sigma^3}{G^2 \rho m \ln\Lambda}.
\end{eqnarray}
If we extend this formula to inhomogeneous systems by making a local
approximation (see, e.g., \cite{aakin}), we find that the relaxation time
is inversely proportional to the local density $\rho({\bf r})$ [assuming that $\sigma\sim (k_B T/m)^{1/2}$ is uniform throughout the system]. Therefore, the
relaxation time is shorter in regions of high density (core) and longer in
regions of low densities (halo). Introducing the
dynamical time
\begin{eqnarray}
\label{cle9}
t_D\sim \frac{\lambda_J}{\sigma}\sim \frac{1}{(G\rho)^{1/2}},
\end{eqnarray}
we can rewrite Eq. (\ref{cle8}) under the form \cite{chandra,bt}
\begin{eqnarray}
\label{cle10}
t_R\sim \frac{n\lambda_J^3}{\ln \Lambda}t_D\sim \frac{N}{\ln N}t_D.
\end{eqnarray}
This shows that the relaxation time measured in terms of the dynamical
time scales like $N$ (up to a logarithmic correction). This is connected to the
fact that the Landau equation can be obtained as the first deviation from the
Vlasov equation (or collisionless Boltzmann equation) in an expansion of the
equations of the BBGKY hierarchy in powers of $1/N\ll 1$ (see, e.g.,
\cite{aakin}).

\subsection{Generalized Landau equation}
\label{sec_gle}

In Ref. \cite{kingen}, we have introduced a generalized Landau
equation
associated with a formalism of generalized thermodynamics. We started from the
generalized Boltzmann equation \cite{kaniadakis}
\begin{eqnarray}
\label{gle1}
\frac{\partial f}{\partial t}=\int d{\bf v}_1d\Omega\, w({\bf v},{\bf v}_1;{\bf
v}',{\bf v}'_{1})\lbrace a(f')b(f)a(f'_1)b(f_1)-a(f)b(f')a(f_1)b(f'_1)\rbrace,
\end{eqnarray}
where $f=f({\bf v},t)$, $f_1=f({\bf v}_1,t)$, $f'=f({\bf v}',t)$, and
$f'_1=f({\bf v}'_1,t)$. This equation can be obtained from a kinetical interaction
principle (KIP) which
allows the probabilities of transition to depend on the
occupation numbers (concentration) of the starting and arrival sites. The
factor $a(f)$ is an arbitrary function of the particle population of the
starting site and the factor $b(f')$ is an arbitrary function of the arrival
site particle population. Usually, the probability of transition is
proportional to the density of the starting site and independent of the
density of the arrival site so that $a(f)=f$ and $b(f')=1$. This leads to the
ordinary Boltzmann equation (\ref{cle1}). However, we can consider a more
general
dependence on the occupancy in the starting and arrival sites. This can account
for microscopic constraints leading to exclusion
(close-packing effects, steric hindrance, Pauli exclusion principle for
fermions, Lynden-Bell exclusion principle for collisionless
self-gravitating systems...) or
inclusion (Bose enhancement) effects that can inhibite or stimulate the
particle transition ${\bf v}\rightarrow {\bf v}'$. These constraints
may be of classical or quantum origin. The factors
$a(f)$ and
$b(f')$ take these constraints into account. For example, if we take $a(f)=f$
and $b(f')=1\mp f'/\eta_0$ we recover the fermionic and bosonic Boltzmann
equations 
\begin{eqnarray}
\label{gle1q}
\frac{\partial f}{\partial t}=\int d{\bf v}_1d\Omega\, w({\bf v},{\bf v}_1;{\bf
v}',{\bf v}'_{1})\lbrace f' (1\mp f/\eta_0) f'_1 (1\mp f_1/\eta_0)-f (1\mp f'/\eta_0) f_1 (1\mp f'_1/\eta_0)\rbrace
\end{eqnarray}
introduced by Nordheim \cite{nordheim} and  Uehling and Uhlenbeck \cite{uu}. For
more general functions, this
generalization encompasses the case of quantum particles (fermion and bosons)
with exclusion or inclusion principles.

Considering the gravitational interaction and making a weak deflection
approximation, we obtained in Ref. \cite{kingen} the generalized Landau equation
\begin{eqnarray}
\label{gle2}
\frac{\partial f}{\partial t}=A\frac{\partial}{\partial
v_i}\int d{\bf v}_1 K_{ij}\left\lbrace a_1b_1(ba'-b'a)\frac{\partial
f}{\partial v_j}-ab(b_1a'_1-b'_1a_1)\frac{\partial f_1}{\partial
v_{1j}}\right\rbrace,
\end{eqnarray}
where $a=a(f)$, $b=b(f)$,  $a_1=a(f_1)$, $b_1=b(f_1)$ and $'$ denotes the
derivative with respect to $f$, e.g., $a'=a'(f)$. On the other hand, $A=2\pi G^2m\ln\Lambda$. If
we define the
functions $g$ and $h$ by
\begin{eqnarray}
\label{gle3}
g(f)=a(f)b(f)\qquad {\rm and} \qquad h(f)=b(f)a'(f)-b'(f)a(f),
\end{eqnarray}
the generalized Landau equation (\ref{gle2}) can be rewritten as \cite{kingen}
\begin{eqnarray}
\label{gle4}
\frac{\partial f}{\partial t}=A\frac{\partial}{\partial
v_i}\int d{\bf v}' K_{ij}\left\lbrace g(f')h(f)\frac{\partial
f}{\partial v_j}- g(f)h(f') \frac{\partial f'}{\partial
v'_{j}}\right\rbrace,
\end{eqnarray}
where $f=f({\bf v},t)$ and $f'=f({\bf v}',t)$.
It is associated with a generalized entropy of the form
\begin{eqnarray}
\label{gle5}
S=-\int C(f)\, d{\bf v},
\end{eqnarray}
where $C(f)$ is a convex function determined by $h$ and $g$ according to
\begin{eqnarray}
\label{gle5b}
C''(f)=\frac{h(f)}{g(f)}.
\end{eqnarray}
One can show \cite{kingen} that the  generalized Landau equation
(\ref{gle4}) conserves the mass and
the energy
and satisfies an $H$-theorem for the generalized entropy (\ref{gle5}). As a
result, it
relaxes towards the distribution
\begin{eqnarray}
\label{gle6}
f_{\rm eq}({\bf v})=(C')^{-1}\left (\alpha-\beta m\frac{v^2}{2}\right ),
\end{eqnarray}
which maximizes the generalized entropy at fixed mass and energy. It
is
obtained by writing the variational problem as $\delta S-\beta\delta
E+(\alpha/m)\delta M=0$ where $\beta$ (inverse temperature) and $\alpha$
(chemical potential) are Lagrange multipliers (see Ref. \cite{kingen} for a more
detailed discussion).

\subsection{Quantum Landau equation}
\label{sec_qle}

We can use the formalism of \cite{kingen} to obtain a kinetic equation for
quantum particles in gravitational interaction in a semiclassical
approximation. To that aim, we can proceed as follows.

It is natural to assume that the
transition probability is proportional to the density of the starting site so
that $a(f)=f$. In that case, the generalized Landau equation (\ref{gle2})
reduces to
\begin{eqnarray}
\label{qle1}
\frac{\partial f}{\partial t}=A\frac{\partial}{\partial
v_i}\int d{\bf v}_1 K_{ij}\left\lbrace f_1b_1(b-b'f)\frac{\partial
f}{\partial v_j}-fb(b_1-b'_1f_1)\frac{\partial f_1}{\partial
v_{1j}}\right\rbrace.
\end{eqnarray}
We note that the coefficients of diffusion and friction are not
independent since they both depend on $b(f)$.
Choosing $b(f)=1$, i.e., a probability of transition which does not
depend on the population of the arrival site, leads to the ordinary Landau
equation (\ref{cle2}). If we now assume that the transition probability is
blocked (inhibited) if the concentration of the arrival site is $\eta_0$, it is
natural to take $b(f)=1-f/\eta_0$. This corresponds to the case of fermions in
quantum mechanics where
\begin{eqnarray}
\label{etaz}
\eta_0=g \frac{m^4}{h^3}
\end{eqnarray}
is the maximum value of
the DF fixed by the Pauli exclusion principle ($h=2\pi\hbar$ is the Planck constant and $g=2s+1$ is the spin multiplicity of the quantum states). A similar exclusion
principle also arises in the theory of violent relaxation developed by Lynden-Bell
\cite{lb} for collisionless self-gravitating systems described by the
Vlasov-Poisson equations (in that case, $\eta_0$ is related to the initial
value of the DF).
Inversely, if we assume that the transition probability is stimulated if the
arrival site is occupied, it is natural to take $b(f)=1+f/\eta_0$. This
corresponds to the case of bosons in quantum mechanics which experience Bose
enhancement. In that case, $\eta_0$ is given by Eq. (\ref{etaz}) with $g=1$ for spin-zero bosons. Therefore, we shall take
\begin{eqnarray}
\label{qle2}
b(f)=1\mp\frac{f}{\eta_0},
\end{eqnarray}
where the upper sign corresponds to fermions and the lower sign corresponds to
bosons. In that case, we find that
\begin{eqnarray}
\label{qle3}
b-b'f=1,
\end{eqnarray}
implying that the diffusion coefficient in Eq. (\ref{qle1}) does not explicitly
depend on
$f({\bf v},t)$.\footnote{It is, however, a functional of $f$ since it is
expressed as an integral of $f({\bf v}_1,t)$ over ${\bf v}_1$.}
Conversely, if we impose that the diffusion coefficient does not explicitly
depend on $f({\bf v},t)$,
we obtain the differential equation (\ref{qle3}) which can be integrated into
\begin{eqnarray}
\label{qle4}
b(f)=1-\kappa\frac{f}{\eta_0}.
\end{eqnarray}
Interestingly, this condition
selects
the
case of fermions ($\kappa=1$) and bosons ($\kappa=-1$), and also the case of
intermediate quantum statistics ($\kappa$ arbitrary). Regrouping the previous
results, we obtain the quantum Landau equation \cite{kingen}
\begin{eqnarray}
\label{qle5}
\frac{\partial f}{\partial t}=2\pi G^2m\ln\Lambda\frac{\partial}{\partial
v_i}\int d{\bf v}' K_{ij}\left\lbrace f' \left (1-\kappa
\frac{f'}{\eta_0}\right )\frac{\partial f}{\partial v_j}-  f \left
(1-\kappa
\frac{f}{\eta_0}\right )\frac{\partial f'}{\partial v'_j}\right\rbrace.
\end{eqnarray}
For quantum particles, the Coulomb logarithm $\ln\Lambda$ is constructed with 
 the de Broglie length $\lambda_{\rm
dB}=h/(m\sigma)$ instead of the Landau length $\lambda_L$.\footnote{The
quantum Coulomb logarithm has been discussed in detail by Bar-Or {\it et al.}
\cite{bft} in the context of FDM so we shall not re-discuss it here; we leave it
under the generic form $\ln\Lambda$ both in the classical and quantum cases.} It
can be generalized to several
species as 
\begin{eqnarray}
\label{qle6}
\frac{\partial f_a}{\partial t}=2\pi G^2
\sum_b \ln\Lambda_{ab} \frac{\partial}{\partial
v_i}\int d{\bf v}' K_{ij}\left\lbrace m_b f'_b \left (1-\kappa_b
\frac{f'_b}{\eta_b}\right )\frac{\partial f_a}{\partial v_j}- m_a f_a \left
(1-\kappa_a
\frac{f_a}{\eta_a}\right )\frac{\partial f'_b}{\partial v'_j}\right\rbrace,
\end{eqnarray}
where
\begin{eqnarray}
\eta_a=g\frac{m_a^4}{h^3}\quad {\rm and}\quad \eta_b=g\frac{m_b^4}{h^3}.
\end{eqnarray}

The quantum Landau equation conserves the total mass $M_a$ of each species and
the total
energy $E$ of the system. It also satisfies an $H$-theorem for the Fermi-Dirac or Bose-Einstein
entropy (see
Appendix \ref{sec_cl}). As a result, it relaxes towards the Fermi-Dirac or
Bose-Einstein
distribution\footnote{In the case of bosons, the Bose-Einstein distribution is valid only for $T>T_c$ (see Appendix \ref{sec_bec}).}
\begin{eqnarray}
\label{qle7}
f_a^{\rm eq}({\bf v})=\frac{\eta_a}{\lambda_a e^{\beta m_a
\frac{v^2}{2}}+\kappa_a},
\end{eqnarray}
which maximizes the Fermi-Dirac or Bose-Einstein entropy $S$ at fixed masses
$M_a$
and energy $E$ (see
Appendix \ref{sec_cl}). The temperature $T$
is the same for all the species. The inverse fugacity $\lambda_a$ is determined by the mass density
$\rho_a=\int f_a\, d{\bf v}$ as detailed in Appendix \ref{sec_aeos}.

{\it Remark:} We can obtain an estimate of the relaxation time
as follows.\footnote{We consider a single species system for simplicity. More
accurate expressions of the relaxation time for single and multispecies systems (including the case of FDM) are given in Sec. \ref{sec_gbm}.} Writing $v\sim \sigma$ and $f\sim
\rho/\sigma^3$ and considering the scaling of the different terms
appearing in the quantum Landau equation (\ref{qle5}), we obtain
\begin{eqnarray}
\label{qle8}
t_R\sim \frac{\sigma^3}{G^2 \rho m (1-\kappa \chi) \ln\Lambda}
\end{eqnarray}
with
\begin{eqnarray}
\label{qle9}
\chi\equiv \frac{f}{\eta_0}\sim \frac{\rho h^3}{\sigma^3 m^4}.
\end{eqnarray}
Expression (\ref{qle8}) differs from the classical relaxation time from Eq.
(\ref{cle8}) by the fact that the mass $m$ of the particles is replaced by an
effective mass $m_{\rm eff}\equiv m (1-\kappa \chi)$, corresponding to the mass of quasiparticles. In the case of fermions
($\kappa=1$), we have $m_{\rm eff}<m$ so that quantum mechanics (Pauli's
blocking) has the effect of increasing the relaxation time ($t_{\rm R}^{\rm
Fermi}>t_{\rm R}^{\rm class}$). In the
case of bosons ($\kappa=-1$), we have $m_{\rm eff}>m$ so that  quantum mechanics
(Bose enhancement)  has the effect of reducing the relaxation time ($t_{\rm
R}^{\rm Bose}<t_{\rm R}^{\rm class}$). The
classical limit is recovered when $\chi\ll 1$ and $m_{\rm eff}\sim m$. On the other hand, for
bosons ($\kappa=-1$), there are situations where $\chi\gg 1$ (see Appendix
\ref{sec_fdm}). In that case, we can make the approximation $f(1+f/\eta_0)\simeq
f^2/\eta_0$ so the bosonic Landau
equation becomes (see Appendix \ref{sec_logo})
\begin{eqnarray}
\label{qle10}
\frac{\partial f}{\partial t}=2\pi  \frac{G^2h^3}{gm^3}\ln\Lambda\frac{\partial}{\partial
v_i}\int d{\bf v}' K_{ij}\left ( f'^2\frac{\partial f}{\partial v_j}-  f^2\frac{\partial f'}{\partial v'_j}\right ).
\end{eqnarray}
The stationary solution of this equation is the Rayleigh-Jeans DF (\ref{tlogo2})
which
is associated with the log-entropy (\ref{tlogo1}) \cite{logo}. The
corresponding relaxation time is
\begin{eqnarray}
\label{qle11}
t_R\sim \frac{\sigma^3}{G^2 \rho m \chi \ln\Lambda}\sim \frac{\sigma^3}{G^2 \rho m_{\rm eff} \ln\Lambda}\sim \frac{\sigma^6m^3}{G^2
\rho^2 h^3 \ln\Lambda}.
\end{eqnarray}
The  effective mass of the particles is given by
\begin{eqnarray}
\label{qle12}
m_{\rm eff}\sim m\chi\sim \frac{\rho h^3}{\sigma^3 m^3}\sim \rho \lambda_{\rm
dB}^3,
\end{eqnarray}
where  $\lambda_{\rm dB}=h/(m\sigma)$ is the de
Broglie length. The effective mass $m_{\rm eff}$ corresponds to the mass
contained within the de Broglie sphere. The effective number of particles is
$N_{\rm eff}=M/m_{\rm eff}$ and the 
relaxation time can be written as
\begin{eqnarray}
\label{qle12b}
t_R\sim \frac{\rho R^3}{m_{\rm eff}\ln\Lambda}t_D\sim \left
(\frac{R}{\lambda_{\rm dB}}\right
)^3\frac{1}{\ln\Lambda}t_D\sim\frac{N_{\rm eff}}{\ln N_{\rm eff}}t_D,
\end{eqnarray}
where $t_D\sim R/\sigma\sim (G\rho)^{-1/2}$ is the dynamical time. It
can be shown that
$t_R$ represents the
condensation time for the self-consistent evolution of a system of bosons
in gravitational interaction below the critical temperature $T_c$ \cite{levkov2}
or the heating time for a
system of classical particles in ``collisions'' with quasiparticles
of mass $m_{\rm eff}$ arising from wave interferences in FDM halos \cite{bft}.
Furthermore, the kinetic theory is valid for $R\gg\lambda_{\rm dB}$,
i.e., $N_{\rm eff}\gg 1$.
Numerical applications to justify the previous approximations and give
characteristic values of the relaxation time for FDM are made in Appendix
\ref{sec_fdm}.

\section{Classical kinetic theory}
\label{sec_c}

In this section, we review and complete the kinetic theory of classical
particles in gravitational interaction such as stars in globular clusters. As
discussed previously, we use an idealization in which the system is assumed to
be spatially homogeneous and we ignore collective effects.

\subsection{Classical Landau equation}
\label{sec_clee}

We consider a two-species system consisting of test particles of mass $m$ and DF
$f({\bf v},t)$
experiencing ``collisions'' with field (background) particles of mass $m_b$ and
DF $f_b({\bf v},t)$. The DF of the field particles is assumed
to be
given. It may be independent of time (fixed) or may evolve with time
according
to another equation that is not specified here (it could be the Landau equation
of species $b$ in the self-consistent treatment of collisions described above or, possibly, another equation of evolution).
If we neglect the
``collisions'' between the test particles (when $b\neq a$) and only consider the ``collisions''
between the test particles and the field particles, the
classical Landau equation writes
\begin{eqnarray}
\label{c1}
\frac{\partial f}{\partial t}=2\pi G^2\ln\Lambda\frac{\partial}{\partial
v_i}\int d{\bf v}' K_{ij}\left ( m_b f'_b \frac{\partial
f}{\partial v_j}- m f \frac{\partial f'_b}{\partial v'_j}\right ).
\end{eqnarray}
If we assume that $f_b({\bf v})$ is fixed (independent of time), then Eq.
(\ref{c1}) is just a partial differential equation of the Fokker-Planck type
(see below).
This corresponds to the ``bath'' approach in which the test particles interact
with field particles that have a prescribed DF. In
particular, the ``thermal bath'' approach corresponds to the situation
where the field particles are at statistical equilibrium with the
Maxwell-Boltzmann
DF (this assumes that they are not disturbed by collisions with the test
particles).  On the other hand, by taking $m_b=m$
and $f_b=f$, Eq. (\ref{c1}) describes a single species system of
particles of mass $m$ and DF $f({\bf v},t)$ evolving in time in a
self-consistent manner. In that case, we recover the classical Landau equation
(\ref{cle2}) which is an integrodifferential equation.

{\it Remark:} As emphasized in previous works (see, e.g.,
\cite{epjp}), the ``bath''
approximation transforms an integrodifferential  equation (Landau)  where
the diffusion and friction coefficients are functionals of $f$ into a
differential equation (Fokker-Planck) where the diffusion and friction
coefficients are independent of $f$.

\subsection{Classical Fokker-Planck equation}
\label{sec_cfp}

The classical Landau equation  (\ref{c1}) can be written in the form of a
Fokker-Planck equation
\begin{eqnarray}
\label{c2}
\frac{\partial f}{\partial t}=\frac{\partial}{\partial
v_i}\left ( D_{ij}\frac{\partial f}{\partial v_j}- f  F_i^{\rm pol}
\right )
\end{eqnarray}
involving a diffusion tensor\footnote{The diffusion arises from the fluctuations of the
field particles  This is why it is proportional to the mass $m_b$ of the field
particles.}
\begin{eqnarray}
\label{c3}
D_{ij}=2\pi G^2 m_b \ln\Lambda \int d{\bf v}'
K_{ij} f'_b
\end{eqnarray}
and a friction by polarisation\footnote{This terminology was
introduced in
\cite{hb4} in order to distinguish the friction force ${\bf F}_{\rm pol}$ that
appears naturally in the Landau equation from the ``true'' friction ${\bf
F}_{\rm friction}$ that appears in the Fokker-Planck equation (see below). As
explained
in \cite{hb4}, the friction by polarisation arises from the retroaction of the
field particles to the perturbation caused by the test particles. This is why it
is proportional to the mass $m$ 
of the test particles.}
\begin{eqnarray}
\label{c4}
F_i^{\rm pol}=2\pi G^2 m \ln\Lambda\int d{\bf v}'
K_{ij}\frac{\partial f'_b}{\partial v'_j}.
\end{eqnarray}
The usual form of the Fokker-Planck equation  \cite{risken} is
\begin{eqnarray}
\label{c4b}
\frac{\partial f}{\partial t}=\frac{\partial^2}{\partial
v_i\partial v_j}( D_{ij} f)-\frac{\partial}{\partial v_i}(f
F_i^{\rm
friction}),
\end{eqnarray}
where the first two moments of the velocity increment $\Delta {\bf v}$ are
\begin{eqnarray}
\label{c5}
D_{ij}=\frac{\langle \Delta v_i \Delta v_j\rangle}{2\Delta t}\qquad {\rm
and}\qquad
F_i^{\rm friction}=\frac{\langle \Delta v_i \rangle}{\Delta t}.
\end{eqnarray}
Equation (\ref{c5}) defines the true (total) friction force ${\bf
F}_{\rm friction}$. The relation between the friction by polarisation and the
true friction is
\begin{eqnarray}
\label{c6}
F_i^{\rm friction}=F_i^{\rm pol}+\frac{\partial D_{ij}}{\partial v_j}.
\end{eqnarray}
For classical particles,  we have
\begin{eqnarray}
\label{c7}
\frac{\partial D_{ij}}{\partial v_j}&=&2\pi G^2 m_b \ln\Lambda \int d{\bf v}'
\frac{\partial K_{ij}}{\partial v_j} f'_b\nonumber\\
&=&-2\pi G^2 m_b \ln\Lambda \int d{\bf v}'
\frac{\partial K_{ij}}{\partial v'_j} f'_b
\nonumber\\
&=&2\pi G^2 m_b \ln\Lambda \int d{\bf v}'
K_{ij} \frac{\partial f'_b}{\partial v'_j}\nonumber\\
&=&\frac{m_b}{m}F_i^{\rm pol}.
\end{eqnarray}
Therefore, we obtain
\begin{eqnarray}
\label{c8}
{\bf F}_{\rm friction}=\frac{m+m_b}{m}{\bf F}_{\rm pol}.
\end{eqnarray}
When $m_b=m$ we find that ${\bf F}_{\rm friction}=2{\bf F}_{\rm pol}$. In that
case, the true friction and the friction by polarisation differ by a factor $2$.
This factor $2$, or more generally the ratio $(m+m_b)/m$, appeared at several
occasions in the literature and was not always clearly
understood.

\subsection{Rosenbluth potentials}
\label{sec_cr}

Recalling Eq. (\ref{cle3}), the diffusion tensor (\ref{c3}) is given explicitly
by
\begin{eqnarray}
\label{c12}
D_{ij}=2\pi G^2 m_b \ln\Lambda \int d{\bf v}' \frac{u^2\delta_{ij}-u_iu_j}{u^3}
f'_b.
\end{eqnarray}
On the other hand, integrating Eq. (\ref{c4}) by parts and using the identity
\begin{eqnarray}
\label{c9}
\frac{\partial K_{ij}}{\partial v'_j}=2\frac{u_i}{u^3}
\end{eqnarray}
we can rewrite the friction by polarisation as 
\begin{eqnarray}
\label{c10}
F_i^{\rm pol}=-4\pi G^2 m \ln\Lambda \int d{\bf v}'\,
\frac{u_i}{u^3} f'_b.
\end{eqnarray}
Then, using Eq. (\ref{c8}), the true friction is
\begin{eqnarray}
\label{c11}
F_i^{\rm friction}=-4\pi G^2 (m+m_b) \ln\Lambda \int d{\bf v}'
\frac{u_i}{u^3} f'_b.
\end{eqnarray}
Now,using the identities
\begin{eqnarray}
\label{c13}
K_{ij}=\frac{u^2\delta_{ij}-u_iu_j}{u^3}=\frac{\partial^2 u}{\partial v_i\partial v_j}\qquad {\rm and}\qquad
\frac{\partial K_{ij}}{\partial
v_{j}}=-2\frac{u_i}{u^3}=2\frac{\partial}{\partial v_i}\left (\frac{1}{u}\right
),
\end{eqnarray}
the diffusion tensor and the friction can be rewritten as
\begin{eqnarray}
\label{c14}
D_{ij}=2\pi G^2 m_b \ln\Lambda \frac{\partial^2\chi}{\partial v_i\partial
v_j}({\bf v}),
\end{eqnarray}
\begin{eqnarray}
\label{c15}
{\bf F}_{\rm pol}=4\pi G^2 m \ln\Lambda  \frac{\partial \lambda}{\partial {\bf
v}}({\bf v}),
\end{eqnarray}
\begin{eqnarray}
\label{c15y}
\frac{\partial D_{ij}}{\partial v_j}=4\pi G^2 m_b \ln\Lambda  \frac{\partial
\lambda}{\partial {\bf
v}}({\bf v}),
\end{eqnarray}
\begin{eqnarray}
\label{c15b}
{\bf F}_{\rm friction}=4\pi G^2 (m+m_b) \ln\Lambda  \frac{\partial
\lambda}{\partial {\bf
v}}({\bf v})
\end{eqnarray}
with
\begin{eqnarray}
\label{gpp1}
\chi({\bf v})=\int   f'_b  |{\bf v}-{\bf v}'|\, d{\bf v}'\qquad {\rm and}
\qquad \lambda({\bf v})=\int   \frac{f'_b}{|{\bf v}-{\bf v}'|}\, d{\bf v}'.
\end{eqnarray}
The functions $\chi({\bf v})$ and $\lambda({\bf v})$ are the so-called
Rosenbluth potentials \cite{rosen}. They are the solutions of the
differential
equations
\begin{eqnarray}
\Delta_{\bf v}\chi=2\lambda,\qquad \Delta_{\bf v}\lambda=-4\pi f_b,
\end{eqnarray}
where we have used $\Delta |{\bf v}-{\bf v}'|=2/|{\bf v}-{\bf
v}'|$ and  $\Delta
(1/|{\bf v}-{\bf v}'|)=-4\pi \delta({\bf v}-{\bf v}')$.

{\it Remark:} The expressions of the diffusion tensor [Eq. (\ref{c12})] and true
friction [Eq. (\ref{c11})] can be directly obtained from the two-body
encounters theory developed by Chandrasekhar \cite{chandra,chandra1,nice} and
Rosenbluth {\it et
al.} \cite{rosen}. In these approaches, contrary to the Landau approach, the
Coulomb logarithm does not
diverge at short distances since strong collisions occurring at small impact
parameters are taken into account exactly through the Rutherford cross
section of gravitational scattering. These coefficients of diffusion and
friction can then be substituted into the general form of the
 Fokker-Planck equation  [Eq. (\ref{c4b})] to obtain a self-consistent kinetic
equation.
This equation is equivalent to the Landau equation but it appears in a less
symmetric form which is not as elegant as the Landau equation (see
\cite{aakin} for a comparison between the Chandrasekhar and Landau
approaches).

\subsection{Isotropic bath}
\label{sec_cib}

When $f_b({\bf v})$ is isotropic, i.e., $f_b=f_b(v)$, the Rosenbluth potentials
can be simplified
and the coefficients of diffusion and friction can be  calculated explicitly
(see, e.g., \cite{rosen,bt}). The
diffusion tensor takes the form
\begin{eqnarray}
\label{c17}
D_{ij}=\left (D_{\|}-\frac{1}{2}D_{\perp} \right
)\frac{v_iv_j}{v^2}+\frac{1}{2}D_{\perp}\delta_{ij},
\end{eqnarray}
where $D_{\|}$ and $D_{\perp}$ are the diffusion coefficients in the
directions parallel and perpendicular to the velocity ${\bf v}$ of the test
particle. They are given by
\begin{eqnarray}
\label{c18}
D_{\|}=\frac{16\pi^2}{3}G^2m_b\ln\Lambda\frac{1}{v}\left\lbrack \int_0^v
\frac{v_1^4}{v^2}f_b(v_1) \,
dv_1+v\int_v^{+\infty} v_1 f_b(v_1) \, dv_1 \right\rbrack,
\end{eqnarray}
\begin{eqnarray}
\label{c19}
D_{\perp}=\frac{16\pi^2}{3}G^2m_b\ln\Lambda\frac{1}{v}\left\lbrack \int_0^v
\left (3v_1^2-\frac{v_1^4}{v^2}\right )f_b(v_1) \,
dv_1+2v\int_v^{+\infty} v_1 f_b(v_1)\, dv_1 \right\rbrack.
\end{eqnarray}
On the other hand, the friction force is given by
\begin{eqnarray}
\label{c21}
{\bf F}_{\rm pol}=-16\pi^2 G^2 m \ln\Lambda  \frac{{\bf v}}{v^3}\int_0^v v_1^2
f_b(v_1)\, dv_1,
\end{eqnarray}
\begin{eqnarray}
\label{c22}
\frac{\partial D_{ij}}{\partial v_j}=-16\pi^2 G^2 m_b \ln\Lambda  \frac{{\bf
v}}{v^3}\int_0^v v_1^2
f_b(v_1)\, dv_1,
\end{eqnarray}
\begin{eqnarray}
\label{c23}
{\bf F}_{\rm friction}=-16\pi^2 G^2 (m+m_b)  \ln\Lambda \frac{{\bf
v}}{v^3} \int_0^v v_1^2
f_b(v_1)\, dv_1.
\end{eqnarray}
Equation (\ref{c23}) is the celebrated Chandrasekhar formula of
dynamical friction \cite{chandra1}. We note that it
depends only on the DF of the
field particles with a velocity $v_1$ smaller than $v$.
This is because the Rosenbluth potential $\lambda({\bf r})$ given by Eq. (\ref{gpp1}) is
similar to the gravitational potential produced by a distribution of matter
with density $\rho({\bf r})$ (if we identify ${\bf v}$ with ${\bf r}$ and
$f_b({\bf v})$ with $\rho({\bf r})$). Consequently, the total friction force
given by Eq. (\ref{c15b}) is similar to the gravitational force. If the DF of
the field particles is isotropic (corresponding to a spherically symmetric
distribution of matter in the gravitational analogy), we immediately obtain Eq. (\ref{c23}) which is the counterpart of Newton's law ${\bf a}=-GM(r){\bf r}/r^3$ with $M(r)=\int_0^r \rho(r')4\pi r^2\, dr$.

\subsection{Thermal bath: Einstein relation}
\label{sec_ctb}

We now assume that the field particles are in a statistical equilibrium state
described by the Maxwellian DF:
\begin{eqnarray}
\label{c24}
f_b({\bf v})=\rho_b \left (\frac{\beta m_b}{2\pi}\right )^{3/2}e^{-\beta m_b
\frac{v^2}{2}}.
\end{eqnarray}
This corresponds to the so-called ``thermal bath''. This DF may
be written as
\begin{eqnarray}
\label{c24b}
f_b({\bf v})=\frac{\rho_b}{(2\pi\sigma_b^2)^{3/2}}e^{-
\frac{v^2}{2\sigma_b^2}},
\end{eqnarray}
where we have introduced the velocity dispersion of
the field particles in one
direction (see Appendix \ref{sec_class})
\begin{eqnarray}
\sigma_b^2=\frac{k_B T}{m_b}=\frac{1}{\beta m_b}=\frac{\langle
v_b^2\rangle}{3}.
\end{eqnarray}
Substituting the identity
\begin{eqnarray}
\label{c25}
\frac{\partial f_b}{\partial {\bf v}}=-f_b\beta m_b {\bf v}
\end{eqnarray}
into Eq. (\ref{c4}), we obtain
\begin{eqnarray}
\label{c26}
F_i^{\rm pol}&=&-2\pi G^2 \beta m m_b  \ln\Lambda\int d{\bf v}'
K_{ij} f'_b v'_j\nonumber\\
&=&-2\pi G^2 \beta m m_b  \ln\Lambda\int d{\bf v}'
K_{ij} f'_b (u_j+v_j)\nonumber\\
&=&-2\pi G^2 \beta m m_b v_j  \ln\Lambda\int d{\bf v}'
K_{ij} f'_b,
\end{eqnarray}
where we have used $K_{ij}u_j=0$ according to Eq. (\ref{cle3}) to get the third
line. Comparing this
expression with Eq.
(\ref{c3}) we
obtain
\begin{eqnarray}
\label{c27}
F_i^{\rm pol}=- \beta m D_{ij} v_j.
\end{eqnarray}
Using  the identity $D_{ij}v_j=D_{\|}
v_i$ [see Eq. (\ref{c17})], we can also write
\begin{eqnarray}
\label{c28}
{\bf F}_{\rm pol}=- \beta m D_{\|} {\bf v}.
\end{eqnarray}
We see that the friction by polarisation is proportional and opposite to the
velocity of the test particle. Furthermore, the friction coefficient is given
by the Einstein relation
\begin{eqnarray}
\label{c29}
\xi= \beta m D_{\|},
\end{eqnarray}
which expresses the fluctuation-dissipation theorem. We note that the Einstein
relation is valid for the friction by polarisation,
not for the true friction.\footnote{This is because $D_{ij}$
depends on the
velocity. We do not have this subtlety for ordinary
Brownian motion where
the diffusion coefficient is constant. In his seminal paper on dynamical
friction, Chandrasekhar \cite{chandra1} 
calculates the true friction and
obtains the following relation between the friction force and the diffusion coefficient
\begin{eqnarray}
\label{erc}
{\bf F}_{\rm friction}=- \beta (m+m_b) D_{\|} {\bf v}\quad {\rm or}\quad \frac{\langle \Delta v_i\rangle}{\Delta t}=-\beta (m+m_b)\frac{\langle \Delta v_i \Delta v_j\rangle}{2\Delta t}v_j
\end{eqnarray}
that he calls the Einstein relation.  The r.h.s. differs from Eq. (\ref{c28}) in the ratio $(m+m_b)/m$
(see the comment at the end of Sec. \ref{sec_cfp}). However, Eq.  (\ref{erc}) is valid only for classical
particles while Eq. (\ref{c28}) remains valid for quantum particles (see below)
and, more generally, for an arbitrary form of entropy \cite{kingen}. In
addition,
Eq. (\ref{erc}) is valid only when collective effects are neglected while Eq.
(\ref{c28}) remains valid when they are taken into account (see, e.g.,
\cite{epjp}). Therefore, Eq. (\ref{c28}) involving the friction by
polarisation appears to be more fundamental than Eq. (\ref{erc}) involving the
true friction. }
Substituting Eq. (\ref{c27}) into Eq. (\ref{c2}), we obtain the
Kramers-Chandrasekhar equation\footnote{The ordinary Kramers
\cite{kramers} equation, first derived by Klein \cite{klein}, is a Fokker-Planck
equation describing the evolution of the DF $f({\bf r},{\bf v},t)$ of a
spatially inhomogeneous system of Brownian particles submitted to an external
force ${\bf F}$. It involves an
advection term in phase space and a constant isotropic
diffusion tensor $D_{ij}=D\delta_{ij}$. It writes
\begin{eqnarray}
\label{kori}
\frac{\partial f}{\partial t}+{\bf v}\cdot
\frac{\partial f}{\partial {\bf r}}+\frac{{\bf F}}{m}\cdot\frac{\partial
f}{\partial {\bf v}} =\frac{\partial}{\partial
{\bf v}}\cdot \left \lbrack D \left (\frac{\partial f}{\partial {\bf v}}+\beta
m f {\bf v}\right ) \right \rbrack.
\end{eqnarray}
A drift-diffusion equation in velocity space similar to Eq. (\ref{kori}) but
without the advection term  was
first derived by Lord Rayleigh  \cite{lr} to describe the dynamics of massive
particles bombarded by numerous small projectiles. This paper can
be seen as a precursor of the theory of
Brownian motion that is usually considered to
start with the seminal work of Einstein \cite{einstein} (see \cite{bectsallis}
for additional
comments about the connection between the paper of Lord Rayleigh 
\cite{lr} and Brownian theory). The Fokker-Planck equation (\ref{kori}) with
a velocity-dependent diffusion coefficient was studied by Chandrasekhar
\cite{chandra1,chandra2,chandra3} in relation to the
evaporation of globular
clusters.}
\begin{eqnarray}
\label{c30}
\frac{\partial f}{\partial t}=\frac{\partial}{\partial
v_i}\left \lbrack D_{ij} \left (\frac{\partial f}{\partial v_j}+\beta m f
v_j\right ) \right \rbrack.
\end{eqnarray}
This equation relaxes towards  the Maxwellian DF:
\begin{eqnarray}
\label{c31}
f_{\rm eq}({\bf v})=\rho \left (\frac{\beta m}{2\pi}\right )^{3/2}e^{-\beta m
\frac{v^2}{2}},
\end{eqnarray}
involving the mass $m$ of the test particles.
This DF may be written as
\begin{eqnarray}
\label{c31b}
f_{\rm eq}({\bf v})=\frac{\rho}{(2\pi\sigma_{\rm eq}^2)^{3/2}}e^{-
\frac{v^2}{2\sigma_{\rm eq}^2}},
\end{eqnarray}
where we have introduced the equilibrium velocity dispersion of the test
particles in one direction
\begin{eqnarray}
\sigma_{\rm eq}^2=\frac{k_B T}{m}=\frac{1}{\beta m}=\frac{\langle
v^2\rangle_{\rm eq}}{3}.
\end{eqnarray}
At statistical equilibrium, the test particles and the field
particles have the same temperature $T$ (equipartition of
energy) but not the same typical velocities since, according to Eq.
(\ref{cle7}), we have
\begin{eqnarray}
\frac{1}{2}m\langle v^2\rangle_{\rm eq}=\frac{1}{2}m_b\langle
v_b^2\rangle=\frac{3}{2}k_B T\qquad\Rightarrow \qquad \sigma_{\rm eq}=\left
(\frac{m_b}{m}\right )^{1/2}\sigma_b.
\end{eqnarray}
If $m>m_b$  (resp. $m<m_b$) the typical velocity of the test particles is smaller (resp. larger) than the typical velocity of the field particles. On the other hand, the DF of the test particles at statistical
equilibrium is related to the DF of the field particles (thermal bath) by
\begin{eqnarray}
\label{c31c}
f_{\rm eq}({\bf
v})\propto f_b({\bf v})^{m/m_b}.
\end{eqnarray}

\subsection{Diffusion and friction terms in the thermal bath
approximation}
\label{sec_cdtb}

For a thermal bath [see Eq. (\ref{c24})], the diffusion coefficients of the test
particles obtained from Eqs. (\ref{c18}) and (\ref{c19}) are given by
\begin{eqnarray}
\label{c32}
D_{\|}=4\pi G^2 m_b\ln\Lambda \rho_b G(x)\frac{1}{v},
\end{eqnarray}
\begin{eqnarray}
\label{c33}
D_{\perp}=4\pi G^2 m_b\ln\Lambda \rho_b \left\lbrack {\rm
erf}(x)-G(x)\right\rbrack \frac{1}{v},
\end{eqnarray}
where $x$ and $G(x)$ are defined below. On the other hand, the friction force  [obtained from
Eqs.
(\ref{c21})-(\ref{c23})] is given by
\begin{eqnarray}
\label{c36}
{\bf F}_{\rm pol}&=&-   4\pi G^2 \beta m  m_b\ln\Lambda \rho_b
G(x)\frac{{\bf v}}{v}\nonumber\\
&=&- \frac{4\pi G^2 m \ln\Lambda \rho_b}{\sigma_b^2}
G(x)\frac{{\bf v}}{v},
\end{eqnarray}
\begin{eqnarray}
\label{c37}
\frac{\partial D_{ij}}{\partial v_j}&=&-   4\pi G^2 \beta  m_b^2\ln\Lambda
\rho_b
G(x)\frac{{\bf v}}{v}\nonumber\\
&=&- \frac{4\pi G^2 m_b \ln\Lambda \rho_b}{\sigma_b^2}
G(x)\frac{{\bf v}}{v},
\end{eqnarray}
\begin{eqnarray}
\label{c38}
{\bf F}_{\rm friction}&=&-   4\pi G^2 \beta (m+m_b)m_b\ln\Lambda \rho_b
G(x)\frac{{\bf v}}{v}\nonumber\\
&=&- \frac{4\pi G^2 (m+m_b) \ln\Lambda \rho_b}{\sigma_b^2}
G(x)\frac{{\bf v}}{v}.
\end{eqnarray}
We can check from these expressions that the Einstein relation
(\ref{c29}) is satisfied. In the foregoing equations, we have defined
\begin{eqnarray}
\label{c34}
{\bf x}=\sqrt{\frac{\beta m_b}{2}}{\bf v}=\frac{\bf
v}{\sqrt{2}\, \sigma_b}
\end{eqnarray}
and
\begin{eqnarray}
\label{c35}
G(x)=\frac{2}{\sqrt{\pi}}\frac{1}{x^2}\int_0^x t^2 e^{-t^2}\,
dt=\frac{1}{2x^2}\left\lbrack {\rm
erf}(x)-\frac{2x}{\sqrt{\pi}}e^{-x^2}\right\rbrack,
\end{eqnarray}
where
\begin{eqnarray}
{\rm erf}(x)=\frac{2}{\sqrt{\pi}}\int_{0}^{x}e^{-t^{2}}dt
\label{diff10}
\end{eqnarray}
is the error function. We have $G(x)\sim 2x/(3\sqrt{\pi})$ for $x\rightarrow 0$
and $G(x)\sim 1/(2 x^2)$ for $x\rightarrow +\infty$. We can then write the
diffusion tensor as
\begin{eqnarray}
\label{dke2}
D_{ij}({\bf v})&=&\left (\frac{2}{\pi}\right )^{1/2}G^2 m_b \ln\Lambda \rho_b
\sqrt{\beta m_b} G_{ij}({\bf x})\nonumber\\
&=&\left (\frac{2}{\pi}\right )^{1/2}\frac{G^2 m_b \ln\Lambda
\rho_b}{\sigma_b}G_{ij}({\bf x}),
\end{eqnarray}
where
\begin{eqnarray}
G_{ij}=\left (G_{\|}-\frac{1}{2}G_{\perp}\right )\frac{x_ix_j}{
x^{2}}+\frac{1}{2}G_{\perp}\delta_{ij}
\label{ps38}
\end{eqnarray}
with
\begin{eqnarray}
G_{\|}=\frac{2\pi^{3/2}}{x}G(x),\qquad
G_{\perp}=\frac{2\pi^{3/2}}{x}\lbrack {\rm erf}(x)-G(x)\rbrack.
\label{ps39}
\end{eqnarray}
We have the asymptotic behaviours
\begin{eqnarray}
G_{\|}(0)=\frac{4\pi}{3}, \qquad G_{\perp}(0)=\frac{8\pi}{3},
\label{diff11}
\end{eqnarray}
\begin{eqnarray}
G_{\|}(x)\sim_{+\infty} \frac{\pi^{3/2}}{x^3}, \qquad
G_{\perp}(x)\sim_{+\infty}\frac{2\pi^{3/2}}{x}.
\label{diff12}
\end{eqnarray}
The diffusion coefficients $D_{\|}(v)$ and $D_{\perp}(v)$ are plotted in Fig. 1 of \cite{aakin}. We note in particular that  $D_{\|}(v)$ decreases as $v^{-3}$ for $v\rightarrow +\infty$.

\subsection{Dimensionless Kramers-Chandrasekhar equation}
\label{sec_dke}

It is convenient to write the Kramers-Chandrasekhar equation (\ref{c30}) under a
dimensionless form \cite{ss,spitzerastro,aakin}. Making the change of variables
from Eq. (\ref{c34}) and using Eq. (\ref{dke2}), we obtain
\begin{eqnarray}
\label{dke3}
\frac{\partial f}{\partial t}=\frac{1}{t_H}\frac{\partial}{\partial
x_i}\left \lbrack G_{ij}({\bf x}) \left (\frac{\partial f}{\partial
x_j}+2\frac{m}{m_b} f x_j\right ) \right \rbrack,
\end{eqnarray}
where we have introduced the relaxation time (heating time)
\cite{ss,spitzerastro,aakin}\footnote{We can
estimate the heating time  by writing
$\langle (\Delta v)^2\rangle \sim  D t_{H} \sim \sigma_b^2$. This corresponds to
the typical time needed by the test particles to acquire the velocity of the
field particles through diffusion. According to Eq. (\ref{dke2}), the diffusion coefficient
scales as $D\sim {G^2 \rho_b m_b \ln\Lambda }/{\sigma_b}$. This gives $t_{H}\sim
{\sigma_b^3}/({G^2 \rho_b m_b \ln\Lambda })$ in agreement with Eq.
(\ref{dke4}).} 
\begin{eqnarray}
\label{dke4}
t_H&=&\frac{\sqrt{2\pi}}{G^2 \rho_b m_b (\beta m_b)^{3/2} \ln\Lambda }\nonumber\\
&=&\frac{\sqrt{2\pi}\sigma_b^3}{G^2 \rho_b m_b \ln\Lambda }.
\end{eqnarray}
The heating time depends only on the mass of the field particles. When the field stars have a larger
mass than the test
stars ($m_b>m$) the heating time $t_H$ is reduced as compared to the case
where all the stars have the same mass ($m_b=m$). On the other hand, when $m\ll m_b$, the friction term is negligible and we just
have a diffusive heating (see Sec. \ref{sec_sbd}).

We can also write
\begin{eqnarray}
\label{dke5}
\frac{\partial f}{\partial t}=\frac{1}{t_C}\frac{\partial}{\partial
x_i}\left \lbrack G_{ij}({\bf x}) \left (\frac{m_b}{m}\frac{\partial f}{\partial
x_j}+2 f x_j\right ) \right \rbrack,
\end{eqnarray}
where we have introduced the relaxation time (cooling time)
\cite{ss,spitzerastro,aakin}\footnote{We can
estimate the cooling  time  by writing $\langle (\Delta
v)^2\rangle \sim  D t_{C}
\sim \sigma_{\rm eq}^2$. This corresponds to
the typical time needed by the test particles to acquire their equilibrium
velocity. Using  $D\sim {G^2 \rho_b m_b \ln\Lambda }/{\sigma_b}$ and $m\sigma_{\rm eq}^2=m_b\sigma_b^2$, we obtain
$t_{C}\sim {\sigma_b^3}/({G^2 \rho_b m \ln\Lambda })$ in agreement with Eq.
(\ref{dke6}).
Alternatively, we can  estimate the cooling time by writing $t_{C}\sim
\xi^{-1}$. This is the friction time. Using the Einstein relation
$\xi\sim D \beta m$ we get $t_{C}\sim {\sigma_b^3}/({G^2 \rho_b m \ln\Lambda })$
as before.}
\begin{eqnarray}
\label{dke6}
t_C&=&\frac{\sqrt{2\pi}}{G^2 \rho_b m (\beta m_b)^{3/2} \ln\Lambda }\nonumber\\
&=&\frac{\sqrt{2\pi}\sigma_b^3}{G^2 \rho_b m \ln\Lambda }
\end{eqnarray}
The cooling time depends only on the mass of the test particles. It has the same expression as when all the stars have the same mass ($m_b=m$).
When $m\gg m_b$, the diffusion term is negligible and we just
have a frictional cooling  (see Sec. \ref{sec_ss}).

We note that
\begin{eqnarray}
\label{dke7}
\frac{t_C}{t_H}=\frac{m_b}{m}.
\end{eqnarray}
The dimensionless Kramers-Chandrasekhar equation can therefore
be written as
\begin{eqnarray}
\label{dke5a}
\frac{\partial f}{\partial t}=\frac{\partial}{\partial
x_i}\left \lbrack G_{ij}({\bf x}) \left (\frac{1}{t_H}\frac{\partial f}{\partial
x_j}+\frac{2}{t_C} f x_j\right ) \right \rbrack.
\end{eqnarray}

When $f({\bf x},t)$ is isotropic, i.e., $f=f(x,t)$, using
$G_{ij}x_j=G_{\|}(x)x_i$, the foregoing equations take the form
\begin{eqnarray}
\label{dke8}
\frac{\partial f}{\partial t}=\frac{1}{t_H}\frac{1}{x^2}\frac{\partial}{\partial
x}\left \lbrack x^2 G_{\|}(x) \left (\frac{\partial f}{\partial
x}+2\frac{m}{m_b} f x\right ) \right \rbrack,
\end{eqnarray}
\begin{eqnarray}
\label{dke9}
\frac{\partial f}{\partial t}=\frac{1}{t_C}\frac{1}{x^2}\frac{\partial}{\partial
x}\left \lbrack x^2 G_{\|}(x) \left (\frac{m_b}{m}\frac{\partial f}{\partial
x}+2 f x\right ) \right \rbrack,
\end{eqnarray}
\begin{eqnarray}
\label{dke9b}
\frac{\partial f}{\partial t}=\frac{1}{x^2}\frac{\partial}{\partial
x}\left \lbrack x^2 G_{\|}(x) \left (\frac{1}{t_H}\frac{\partial
f}{\partial
x}+\frac{2}{t_C} f x\right ) \right \rbrack.
\end{eqnarray}

\subsection{Diffusion equation ($m\ll m_b$)}
\label{sec_sbd}

When the test particles are much lighter than the
field particles ($m\ll m_b$)
we can neglect the force by polarisation:\footnote{This amounts
to making $m\rightarrow 0$ in Eq. (\ref{c4}). When $m\ll m_b$ the test
particles do not significantly perturb the field particles (see footnote 28). As
a result, there
is no retroaction from the field particles, hence no friction by polarisation.
Note, however, that the true friction is nonzero [see Eq. (\ref{c40})].}
\begin{eqnarray}
\label{c39b}
{\bf F}_{\rm pol}={\bf 0}.
\end{eqnarray}
In that case, the
stars experience just a process of diffusion (heating). This
is the situation considered by Spitzer and Schwarzschild \cite{ss}.
The Fokker-Planck equation (\ref{c2}) reduces to a pure diffusion
equation
\begin{eqnarray}
\label{c39}
\frac{\partial f}{\partial t}=\frac{\partial}{\partial
v_i}\left ( D_{ij}\frac{\partial f}{\partial v_j}\right ),
\end{eqnarray}
where the diffusion coefficient is given by Eq. (\ref{c3}). The mass
$m$ of the test particles does not appear in this equation. According to
Eq. (\ref{c6}), the total friction force is given by
\begin{eqnarray}
\label{c40}
F_i^{\rm friction}=\frac{\partial D_{ij}}{\partial v_j}.
\end{eqnarray}
Using Eq. (\ref{c5}), this relation can be rewritten as
\begin{eqnarray}
\label{c41}
\frac{\langle \Delta v_i \rangle}{\Delta t}=\frac{\partial}{\partial
v_j}\left (\frac{\langle \Delta v_i \Delta v_j\rangle}{2\Delta t}\right ).
\end{eqnarray}
This relation is sometimes referred to as the fluctuation-dissipation theorem but
this terminology is misleading because the friction by polarisation (which
is at the origin of the Einstein relation (\ref{c28})) is zero in the present
case (see footnote 29).

If we assume that the field stars have the Maxwellian distribution
(\ref{c24}) and that the DF of the test stars is isotropic, we find, using the
normalized
variables introduced in Sec. \ref{sec_dke}, that the
diffusion equation
(\ref{c39}) takes the form
\begin{eqnarray}
\label{ss1}
\frac{\partial f}{\partial
t}=\frac{1}{t_H}\frac{1}{x^2}\frac{\partial}{\partial
x}\left ( x^2 G_{\|}(x)\frac{\partial f}{\partial x}\right ).
\end{eqnarray}
This is the Spitzer-Schwarzschild diffusion equation \cite{ss}. We obtain in
Appendix \ref{sec_ssss} an analytical self-similar solution of this equation and
compare this solution with the numerical results of Spitzer and Schwarzschild
\cite{ss}.

\subsection{Deterministic equation ($m\gg m_b$)}
\label{sec_ss}

When the test particles are much heavier than the field particles ($m\gg m_b$)
we can neglect the diffusion:\footnote{This amounts
to making $m_b\rightarrow 0$ in Eq. (\ref{c3}).  When $m_b\ll m$ the field
particles do not produce significant fluctuations to induce a diffusion of the
test particles (see footnote 27).}
\begin{eqnarray}
\label{c42}
D_{ij}=0.
\end{eqnarray}
In that case, the test particles
experience just a
process of friction (cooling).  The Fokker-Planck equation
(\ref{c2}) reduces to the purely frictional equation
\begin{eqnarray}
\label{c43}
\frac{\partial f}{\partial t}=\frac{\partial}{\partial
v_i}\left (-f F_i^{\rm pol} \right ),
\end{eqnarray}
where the friction force is given by Eq. (\ref{c4}).  The mass
$m_b$ of the field particles does not appear in this equation. According to Eq.
(\ref{c6}), we have
\begin{eqnarray}
\label{c44}
{\bf F}_{\rm friction}={\bf F}_{\rm pol}.
\end{eqnarray}
Eq. (\ref{c43}) is equivalent to the deterministic equation of motion
\begin{eqnarray}
\label{c45}
\frac{d{\bf v}}{dt}={\bf F}_{\rm pol},
\end{eqnarray}
where ${\bf v}$ is the velocity of the test particle. For an isotropic bath,
${\bf F}_{\rm pol}$ is given by Eq. (\ref{c21}) and for a thermal bath it takes
the form of Eq. (\ref{c28}).

{\it Remark:} Dynamical friction is responsible for the decay of globular
clusters and black hole orbits in a galaxy (this corresponds to the so-called
sinking satellite problem). As they orbit through a galaxy they are subject to
dynamical friction and spiral toward 
the galaxy center \cite{bt,tos}.

\section{Quantum kinetic theory}
\label{sec_q}

In this section, we develop the kinetic theory of quantum
particles in gravitational interaction such as fermions (massive neutrinos) or
bosons (axions) in DM halos. Again, we use an idealization in which the
system is assumed to be spatially homogeneous and we ignore collective effects.

\subsection{Quantum Landau equation}
\label{sec_ql}

Under the same assumptions as those made in Sec. \ref{sec_clee}, but considering
now quantum particles instead of classical particles, the quantum Landau
equation writes
\begin{eqnarray}
\label{q1}
\frac{\partial f}{\partial t}=2\pi G^2\ln\Lambda\frac{\partial}{\partial
v_i}\int d{\bf v}' K_{ij}\left\lbrace m_b f'_b \left (1-\kappa_b
\frac{f'_b}{\eta_b}\right )\frac{\partial f}{\partial v_j}- m f \left
(1-\kappa
\frac{f}{\eta_0}\right )\frac{\partial f'_b}{\partial v'_j}\right\rbrace
\end{eqnarray}
with $\kappa=+1$ for fermions and $\kappa=-1$ for bosons (and $\kappa=0$
for classical particles). As in Sec. \ref{sec_qle}, we have defined
\begin{eqnarray}
\label{q2}
\eta_0=g\frac{m^4}{h^3},\qquad \eta_b=g\frac{m_b^4}{h^3}.
\end{eqnarray}

\subsection{Quantum Fokker-Planck equation}
\label{sec_qfp}

The quantum Landau equation (\ref{q1}) can be written in the form of a quantum
Fokker-Planck equation
\begin{eqnarray}
\label{q3}
\frac{\partial f}{\partial t}=\frac{\partial}{\partial
v_i}\left \lbrack D_{ij}\frac{\partial f}{\partial v_j}- f \left
(1-\kappa
\frac{f}{\eta_0}\right ) F_i^{\rm pol} \right \rbrack
\end{eqnarray}
with a diffusion tensor
\begin{eqnarray}
\label{q4}
D_{ij}=2\pi G^2 m_b \ln\Lambda \int d{\bf v}'
K_{ij} f'_b \left (1-\kappa_b
\frac{f'_b}{\eta_b}\right )
\end{eqnarray}
and a friction by polarisation
\begin{eqnarray}
\label{q5}
F_i^{\rm pol}=2\pi G^2 m \ln\Lambda\int d{\bf v}'
K_{ij}\frac{\partial f'_b}{\partial v'_j}.
\end{eqnarray}
We note that the force by polarisation has the same form as in the classical
case. The quantum factor $  f'_b  (1-\kappa_b
{f'_b}/{\eta_b})$ occurs only in the diffusion coefficient.
The usual form of the quantum Fokker-Planck equation \cite{kaniadakis} is
\begin{eqnarray}
\label{q6}
\frac{\partial f}{\partial t}=\frac{\partial}{\partial
v_i}\left \lbrack D_{ij}\frac{\partial f}{\partial v_j}- f \left
(1-\kappa
\frac{f}{\eta_0}\right ) \left (F_i^{\rm friction}-\frac{\partial
D_{ij}}{\partial v_j}\right ) \right \rbrack,
\end{eqnarray}
where the first two moments of the velocity increment $\Delta {\bf v}$ are
given by Eq. (\ref{c5}). Equation (\ref{c5}) defines the true (complete)
friction force ${\bf
F}_{\rm friction}$. The relation between the friction by polarisation and the
true friction is
\begin{eqnarray}
\label{q8}
F_i^{\rm friction}=F_i^{\rm pol}+\frac{\partial D_{ij}}{\partial v_j},
\end{eqnarray}
as in the classical case. We can also write the quantum Fokker-Planck equation
as
\begin{eqnarray}
\label{q6b}
\frac{\partial f}{\partial t}=\frac{\partial}{\partial
v_i}\left \lbrack \frac{\partial}{\partial v_j}(D_{ij}f)-f\frac{\partial
D_{ij}}{\partial v_j} - f \left
(1-\kappa
\frac{f}{\eta_0}\right )F_i^{\rm pol}\right \rbrack.
\end{eqnarray}

\subsection{Quantum Rosenbluth potentials}
\label{sec_qrp}

Using the
identities
from Eq. (\ref{c13}), the diffusion tensor
(\ref{q4}) may be
rewritten as
\begin{eqnarray}
\label{q9}
D_{ij}=2\pi G^2 m_b \ln\Lambda \frac{\partial^2\chi}{\partial v_i\partial
v_j}({\bf v})
\end{eqnarray}
with
\begin{eqnarray}
\label{q10}
\chi({\bf v})=\int   f'_b \left (1-\kappa_b
\frac{f'_b}{\eta_b}\right ) |{\bf v}-{\bf v}'|\, d{\bf v}'.
\end{eqnarray}
Similarly, the friction
by polarisation (\ref{q5}) may be rewritten as (using an integration by parts)
\begin{eqnarray}
\label{q11}
{\bf F}_{\rm pol}=4\pi G^2 m \ln\Lambda  \frac{\partial \lambda}{\partial {\bf
v}}({\bf v})
\end{eqnarray}
with
\begin{eqnarray}
\label{q12}
\lambda({\bf v})=\int   \frac{f'_b}{|{\bf v}-{\bf v}'|}\, d{\bf v}'.
\end{eqnarray}
Using Eq. (\ref{c13}), we also have
\begin{eqnarray}
\label{q13}
\frac{\partial D_{ij}}{\partial v_j}=4\pi G^2 m_b \ln\Lambda  \frac{\partial
\sigma}{\partial
v_i}({\bf v})
\end{eqnarray}
with
\begin{eqnarray}
\label{q14}
\sigma({\bf v})=\int   \frac{f'_b\left (1-\kappa_b
\frac{f'_b}{\eta_b}\right )}{|{\bf v}-{\bf v}'|}\, d{\bf v}'.
\end{eqnarray}
The functions $\chi({\bf v})$, $\lambda({\bf v})$ and $\sigma({\bf v})$ are
the quantum Rosenbluth potentials introduced in \cite{kingen}.\footnote{In this
paper, they are introduced for an  arbitrary form of entropy including the
Fermi-Dirac and Bose-Einstein entropies.} They are the
solutions of the differential equations
\begin{eqnarray}
\Delta_{\bf v}\lambda=-4\pi f_b,\qquad \Delta_{\bf v}\sigma=-4\pi f_b  \left
(1-\kappa_b
\frac{f_b}{\eta_b}\right ), \qquad \Delta_{\bf v}\chi=2\sigma.
\end{eqnarray}

\subsection{Isotropic bath}
\label{sec_qib}

When $f_b({\bf v})$ is isotropic, i.e., $f_b=f_b(v)$, the quantum Rosenbluth
potentials can be
simplified \cite{kingen}. In that
case, the diffusion tensor is of the form of Eq. (\ref{c17}) with
\begin{eqnarray}
\label{q15}
D_{\|}=\frac{16\pi^2}{3}G^2m_b\ln\Lambda\frac{1}{v}\left\lbrack \int_0^v
\frac{v_1^4}{v^2}f_b(v_1) \left (1-\kappa_b \frac{f_b(v_1)}{\eta_b}\right )\,
dv_1+v\int_v^{+\infty} v_1 f_b(v_1) \left (1-\kappa_b
\frac{f_b(v_1)}{\eta_b}\right )\, dv_1 \right\rbrack,
\end{eqnarray}
\begin{equation}
\label{q16}
D_{\perp}=\frac{16\pi^2}{3}G^2m_b\ln\Lambda\frac{1}{v}\left\lbrack \int_0^v
\left (3v_1^2-\frac{v_1^4}{v^2}\right )f_b(v_1) \left (1-\kappa_b
\frac{f_b(v_1)}{\eta_b}\right )\,
dv_1+2v\int_v^{+\infty} v_1 f_b(v_1) \left (1-\kappa_b
\frac{f_b(v_1)}{\eta_b}\right )\, dv_1 \right\rbrack.
\end{equation}
On the other hand, the friction force is given by
\begin{eqnarray}
\label{q17}
{\bf F}_{\rm pol}=-16\pi^2 G^2 m \ln\Lambda  \frac{{\bf v}}{v^3}\int_0^v v_1^2
f_b(v_1)\, dv_1,
\end{eqnarray}
\begin{eqnarray}
\label{q18}
\frac{\partial D_{ij}}{\partial v_j}=-16\pi^2 G^2 m_b \ln\Lambda
\frac{v_i}{v^3}\int_0^v v_1^2
f_b(v_1) \left (1-\kappa_b
\frac{f_b(v_1)}{\eta_b}\right )\, dv_1,
\end{eqnarray}
\begin{eqnarray}
\label{q19}
{\bf F}_{\rm friction}=-16\pi^2 G^2  \ln\Lambda \frac{{\bf
v}}{v^3} \left\lbrack m \int_0^v v_1^2
f_b(v_1)\, dv_1+m_b \int_0^v v_1^2
f_b(v_1) \left (1-\kappa_b
\frac{f_b(v_1)}{\eta_b}\right )\, dv_1\right\rbrack.
\end{eqnarray}

\subsection{Thermal bath: Einstein relation}
\label{sec_qtb}

We now assume that the field particles are in a statistical equilibrium state
described by the Fermi-Dirac or Bose-Einstein DF:
\begin{eqnarray}
\label{q20}
f_{b}({\bf v})=\frac{\eta_b}{\lambda_b e^{\beta m_b \frac{v^2}{2}}+\kappa_b}.
\end{eqnarray}
This corresponds to the so-called ``thermal bath''.\footnote{For bosons, the DF
(\ref{q20}) is valid for $T>T_c$, where $T_c$ is constructed with $m_b$ and
$\rho_b$ (see Appendix \ref{sec_bec}).} 
The inverse
fugacity $\lambda_b$ is related to the
density $\rho_b$ by the relation from Eq. (\ref{eos2}). Using the relation
\begin{eqnarray}
\label{q21}
f_b\left (1-\kappa_b
\frac{f_b}{\eta_b}\right )=\frac{\eta_b\lambda_b e^{\beta m_b v^2/2}}{(\lambda_b
e^{\beta m_b \frac{v^2}{2}}+\kappa_b)^2},
\end{eqnarray}
we find that
\begin{eqnarray}
\label{q22}
\frac{\partial f_b}{\partial {\bf v}}=-f_b\left (1-\kappa_b
\frac{f_b}{\eta_b}\right )\beta m_b {\bf v}.
\end{eqnarray}
Substituting this identity into Eq. (\ref{q5}), we get
\begin{eqnarray}
\label{q23}
F_i^{\rm pol}&=&-2\pi G^2 \beta m m_b  \ln\Lambda\int d{\bf v}'
K_{ij} f'_b\left (1-\kappa_b
\frac{f'_b}{\eta_b}\right ) v'_j\nonumber\\
&=&-2\pi G^2 \beta m m_b  \ln\Lambda\int d{\bf v}'
K_{ij} f'_b\left (1-\kappa_b
\frac{f'_b}{\eta_b}\right ) (u_j+v_j)\nonumber\\
&=&-2\pi G^2 \beta m m_b v_j  \ln\Lambda\int d{\bf v}'
K_{ij} f'_b\left (1-\kappa_b
\frac{f'_b}{\eta_b}\right ),
\end{eqnarray}
where we have used $K_{ij}u_j=0$ according to Eq. (\ref{cle3}) to get the third
line. Comparing this
expression with Eq.
(\ref{q4}), we obtain
\begin{eqnarray}
\label{q24}
F_i^{\rm pol}=- \beta m D_{ij} v_j.
\end{eqnarray}
Using  the identity $D_{ij}v_j=D_{\|}
v_i$ [see Eq. (\ref{c17})], we can also write
\begin{eqnarray}
\label{q25}
{\bf F}_{\rm pol}=- \beta m D_{\|} {\bf v}.
\end{eqnarray}
We see that the friction by polarisation is proportional and opposite to the
velocity of the test particle. Furthermore, the friction coefficient is given
by the Einstein relation
\begin{eqnarray}
\label{q26}
\xi= \beta m D_{\|},
\end{eqnarray}
as in the classical case [see Eq. (\ref{c29})].\footnote{It is shown in
\cite{kingen} that relation (\ref{q25}) is valid for an arbitrary form of
entropy.} As noted previously, the
Einstein relation is valid for the friction by
polarisation,
not for the true friction which cannot be easily related to the diffusion
coefficient in the case of quantum particles. Substituting Eq.
(\ref{q24}) into Eq. (\ref{q3}), we
obtain the quantum
Kramers-Chandrasekhar equation \cite{kingen}\footnote{This is a particular case
of nonlinear Fokker-Planck equations \cite{nfp}. It satisfies an H-theorem in
the canonical ensemble for a generalized free energy of the form $F=E-TS$ where
$S$ is the generalized entropy (\ref{gle5}) associated with the Landau
equation from which it is issued (here the Fermi-Dirac ou Bose-Einstein
entropy).}
\begin{eqnarray}
\label{q27}
\frac{\partial f}{\partial t}=\frac{\partial}{\partial
v_i}\left \lbrack D_{ij} \left (\frac{\partial f}{\partial v_j}+\beta m f \left
(1-\kappa
\frac{f}{\eta_0}\right ) v_j\right ) \right \rbrack.
\end{eqnarray}
This equation relaxes towards the Fermi-Dirac or Bose-Einstein DF:
\begin{eqnarray}
\label{q28}
f_{\rm eq}({\bf v})=\frac{\eta_0}{\lambda e^{\beta m \frac{v^2}{2}}+\kappa}.
\end{eqnarray}
At statistical equilibrium, the test particles and the field
particles have the same temperature.

{\it Remark:} In the case of bosons ($\kappa=-1$), the
Kramers-Chandrasekhar equation (\ref{q27}) relaxes towards the Bose-Einstein DF from Eq. (\ref{q28}) only
if $T>T_c$, where $T_c$ is constructed with $m$ and $\rho$ (see Appendix \ref{sec_bec}). When  $T<T_c$, this equation displays a process of Bose-Einstein condensation leading to the formation  of a Dirac peak at ${\bf v}={\bf 0}$. In that case, the DF can be written as
$f({\bf v},t)=f_{\rm gas}({\bf v},t)+M_c(t)\delta({\bf v})$ where $f_{\rm
gas}({\bf v},t)$ is the DF of the uncondensed bosons (gas) and $M_c(t)=M-M_{\rm
gas}(t)=M-\int_0^{+\infty} f_{\rm gas}(v,t)4\pi v^2\, dv$ is the mass of the
condensed bosons. This mass  increases with time until a statistical equilibrium
state, comprising both uncondensed and condensed bosons, 
is reached [see Eq. (\ref{bec4})]. Since the bosonic Kramers-Chandrasekhar
equation assumes the existence of a thermal bath, it corresponds to a canonical
description of Bose-Einstein
condensation. The formation of a BEC has been studied by Sopik {\it et al.}
\cite{sopik} by solving the bosonic Kramers equation in the case
$D_{ij}=D\delta_{ij}$ with $D$ constant.

\subsection{Diffusion and friction terms in the thermal bath
approximation}
\label{sec_qdtb}

For a thermal bath [see Eq. (\ref{q20})], the diffusion coefficients of the test
particles obtained from Eqs. (\ref{q15}) and (\ref{q16}) are given by
\begin{eqnarray}
\label{q29}
D_{\|}=\frac{16\pi^2}{3}G^2m_b\ln\Lambda\frac{1}{v^3}( L_4+v^3 M_1),
\end{eqnarray}
\begin{eqnarray}
\label{q30}
D_{\perp}=\frac{16\pi^2}{3}G^2m_b\ln\Lambda\frac{1}{v}\left
(3L_2-\frac{1}{v^2}L_4+2v M_1 \right )
\end{eqnarray}
with
\begin{eqnarray}
\label{q31}
M_1=\int_v^{+\infty} v_1 f_b(v_1) \left (1-\kappa_b
\frac{f_b(v_1)}{\eta_b}\right )\, dv_1=\frac{\eta_b}{\beta
m_b}\frac{1}{\lambda_b e^{\beta m_b \frac{v^2}{2}}+\kappa_b},
\end{eqnarray}
\begin{eqnarray}
\label{q32}
L_2=\int_0^{v} v_1^2 f_b(v_1) \left (1-\kappa_b
\frac{f_b(v_1)}{\eta_b}\right )\, dv_1=-\frac{\eta_b}{\beta
m_b}\frac{v}{\lambda_b e^{\beta m_b
\frac{v^2}{2}}+\kappa_b}+\frac{\eta_b}{\beta
m_b}\int_0^v \frac{1}{\lambda_b e^{\beta m_b \frac{v_1^2}{2}}+\kappa_b}\,
dv_1,
\end{eqnarray}
\begin{eqnarray}
\label{q33}
L_4=\int_0^{v} v_1^4 f_b(v_1) \left (1-\kappa_b
\frac{f_b(v_1)}{\eta_b}\right )\, dv_1=-\frac{\eta_b}{\beta
m_b}\frac{v^3}{\lambda_b e^{\beta m_b
\frac{v^2}{2}}+\kappa_b}+\frac{3\eta_b}{\beta
m_b}\int_0^v \frac{v_1^2}{\lambda_b e^{\beta m_b \frac{v_1^2}{2}}+\kappa_b}\,
dv_1.
\end{eqnarray}
To obtain the second equalities in Eqs. (\ref{q31})-(\ref{q33}), we have used
Eq. (\ref{q21}) and performed straightforward integrations by parts. Combining
these expressions, we get
\begin{eqnarray}
\label{q34}
D_{\|}=16\pi^2 G^2 \ln\Lambda\frac{1}{v^3}\frac{\eta_b}{\beta}\int_0^v \frac{v_1^2}{\lambda_b e^{\beta m_b \frac{v_1^2}{2}}+\kappa_b}\,
dv_1,
\end{eqnarray}
\begin{eqnarray}
\label{q35}
D_{\perp}=16\pi^2 G^2 \ln\Lambda\frac{1}{v}\frac{\eta_b}{\beta}\left
(\int_0^v \frac{1}{\lambda_b e^{\beta m_b \frac{v_1^2}{2}}+\kappa_b}\,
dv_1-\frac{1}{v^2}\int_0^v \frac{v_1^2}{\lambda_b e^{\beta
m_b \frac{v_1^2}{2}}+\kappa_b}\,
dv_1 \right ).
\end{eqnarray}
Performing the change of variables $x_1=\beta m_b v_1^2/2$ and using Eq. (\ref{eos2}), we finally obtain
\begin{eqnarray}
\label{q36}
D_{\|}=2\pi G^2m_b\rho_b\ln\Lambda
\frac{1}{x^2}I_{1/2}(\lambda_b,x)\frac{1}{v},
\end{eqnarray}
\begin{eqnarray}
\label{q37}
D_{\perp}=2\pi G^2m_b\rho_b\ln\Lambda \left\lbrack
\frac{I_{-1/2}(\lambda_b)}{I_{1/2}(\lambda_b)}
I_{-1/2}(\lambda_b,x)-\frac{I_{1/2}(\lambda_b,x)}{x^2}\right\rbrack \frac{1}{v},
\end{eqnarray}
where we have defined
\begin{eqnarray}
\label{q38}
I_n(t,x)=\frac{1}{I_n(t)}\int_0^{x^2} \frac{y^n}{te^y+\kappa}\, dy
\end{eqnarray}
and
\begin{eqnarray}
\label{q39}
x=\sqrt{\frac{\beta m_b}{2}}v.
\end{eqnarray}
Here, $I_n(t,x)$ denote the incomplete Fermi and Bose
integrals \cite{kingen}. The complete Fermi and Bose integrals $I_n(t)$ are
defined in Appendix \ref{sec_aeos}. Using Eqs. (\ref{q17})-(\ref{q19}) and
(\ref{q25}), we similarly obtain
\begin{eqnarray}
{\bf F}_{\rm pol}=-2\pi G^2 \beta m m_b\rho_b\ln\Lambda
\frac{1}{x^2}I_{1/2}(\lambda_b,x)\frac{{\bf v}}{v},
\end{eqnarray}
\begin{eqnarray}
\frac{\partial D_{ij}}{\partial v_j}=2\pi G^2 \beta m_b^2 \rho_b\ln\Lambda
\frac{1}{x^2}\frac{1}{I_{1/2}(\lambda_b)}\frac{{\bf v}}{v}\left\lbrace \frac{x}{\lambda_b e^{x^2}+\kappa_b}-\frac{1}{2}I_{-1/2}(\lambda_b,x)I_{-1/2}(\lambda_b)\right\rbrace,
\end{eqnarray}
\begin{eqnarray}
{\bf F}_{\rm friction}={\bf F}_{\rm pol}+\frac{\partial D_{ij}}{\partial v_j}.
\end{eqnarray}

In the classical limit where $\lambda_b\rightarrow +\infty$  (see Appendix
\ref{sec_class}), using the identities \cite{kingen}
\begin{eqnarray}
\label{q39a}
I_n(t)\sim \frac{1}{t}\Gamma(n+1) \qquad (t\rightarrow +\infty),
\end{eqnarray}
\begin{eqnarray}
\label{q39b}
I_n(+\infty,x)=\frac{2}{\Gamma(n+1)}\int_0^x y^{2n+1}e^{-y^2}\, dy,
\end{eqnarray}
\begin{eqnarray}
\label{q39c}
I_{1/2}(+\infty,x)=2x^2 G(x),\qquad I_{-1/2}(+\infty,x)={\rm erf}(x),
\end{eqnarray}
we recover the results from Sec. \ref{sec_cdtb}.

For completely degenerate fermions described by the DF (see
Appendix \ref{sec_compdeg})
\begin{eqnarray}
\label{q40}
f_b(v_1)=\eta_b H(v_1-v_F),
\end{eqnarray}
where $v_F$ is the Fermi velocity, we see from Eq. (\ref{q4}) that
$D_{ij}\rightarrow 0$. This is because the fermionic field particles are in the
ground state at $T=0$ so there is no fluctuation. Therefore, the diffusion
tensor vanishes. The test particles undergo only a  friction force ${\bf F}_{\rm
friction}={\bf F}_{\rm pol}$ given by Eq. (\ref{q17}). We are in
the same situation as in the sinking satellite problem of
Sec. \ref{sec_ss}. Using Eq. (\ref{q40}), we find that 
\begin{eqnarray}
\label{q41}
{\bf F}_{\rm pol}=-\frac{16}{3}\pi^2G^2m\ln\Lambda \eta_b {\bf v}\qquad (v<v_F),
\end{eqnarray}
\begin{eqnarray}
\label{q42}
{\bf F}_{\rm pol}=-\frac{16}{3}\pi^2G^2m\ln\Lambda \eta_b \frac{v_F^3}{v^3}{\bf
v}\qquad (v>v_F).
\end{eqnarray}

\subsection{Fermionic and bosonic King models}
\label{sec_qkm}

We can use the kinetic theory to derive a truncated DF accounting for an escape
of particles above a limit energy
$\epsilon_m$. Let us consider a possibly inhomogeneous system of
quantum particles in gravitational interaction. Making a local approximation,
the evolution of the DF is governed by the quantum Vlasov-Landau
equation\footnote{The classical Vlasov-Landau equation ($\kappa=0$) relying on a local
approximation was used  by Ipser
\cite{ipser} to study spatially inhomogeneous stellar systems. This approximation was criticized by Kandrup \cite{kandrup} who derived a more complicated kinetic equation. We refer to the introduction of
\cite{aakin} for a review of different  approaches attempting to take into
account spatial inhomogeneity in the kinetic theory of self-gravitating systems
going beyond the local approximation. These efforts finally
led to the inhomogeneous Lenard-Balescu equation \cite{heyvaerts,angleaction2}
which is the exact kinetic equation of self-gravitating systems at the order
$O(1/N)$.}
\begin{eqnarray}
\label{q1loc}
\frac{\partial f}{\partial t}+{\bf
v}\cdot\frac{\partial f}{\partial {\bf r}}-\nabla\Phi\cdot \frac{\partial
f}{\partial {\bf v}}=2\pi G^2\ln\Lambda\frac{\partial}{\partial
v_i}\int d{\bf v}' K_{ij}\left\lbrace m_b f'_b \left (1-\kappa_b
\frac{f'_b}{\eta_b}\right )\frac{\partial f}{\partial v_j}- m f \left
(1-\kappa
\frac{f}{\eta_0}\right )\frac{\partial f'_b}{\partial v'_j}\right\rbrace,
\end{eqnarray}
where $\Phi({\bf r},t)$ is the gravitational potential and we have noted
$f=f({\bf r},{\bf v},t)$ and  $f_b'=f_b({\bf r},{\bf v}',t)$. Making a thermal
bath approximation, we obtain the quantum Vlasov-Kramers equation\footnote{When
self-consistently coupled to gravity, we obtain the Kramers-Poisson equation. In
the strong friction limit $\xi\rightarrow +\infty$ (which may be relevant for
certain self-gravitating systems), we obtain the Smoluchowski-Poisson equation
that has been extensively
studied in relation to a model of self-gravitating Brownian particles
\cite{crs}.}
\begin{eqnarray}
\label{qkm1}
\frac{\partial f}{\partial t}+{\bf
v}\cdot\frac{\partial f}{\partial {\bf r}}-\nabla\Phi\cdot \frac{\partial
f}{\partial {\bf v}}=\frac{\partial}{\partial
{\bf v}}\cdot \left \lbrack \frac{K}{v^3} \left (\frac{\partial f}{\partial {\bf
v}}+\beta m f \left
(1-\kappa
\frac{f}{\eta_0}\right ) {\bf v}\right ) \right \rbrack.
\end{eqnarray}
To simplify the equation, we have assumed that the velocity DF of the field particles is isotropic
and replaced the diffusion coefficient $D_{\|}(v)$  by its asymptotic expression
$D_{\|}\sim K/v^3$ valid for high velocities $v\rightarrow +\infty$.\footnote{Recall that the diffusion coefficient of quantum particles at large $v$ behaves like the diffusion coefficient of classical particles studied in Sec. \ref{sec_cdtb}.} We now assume that high energy
particles are removed
by a tidal field. In the case of globular clusters or DM halos, this field may be due to
the gravitational attraction of a nearby galaxy. We then look for a stationary
solution of Eq. (\ref{qkm1}) of the form $f=f(\epsilon)$ satisfying the boundary
condition $f(\epsilon_m)=0$. Here, $\epsilon\equiv m[v^2/2+\Phi({\bf r})]$ denotes the energy of a particle and $\epsilon_m$ is the escape energy above which $f=0$. A DF of
the form $f=f(\epsilon)$ cancels the advection term (l.h.s. of Eq. (\ref{qkm1}))
according to the Jeans theorem \cite{bt}. Then, using the identity
$(\partial/\partial {\bf v})({\bf v}/v^3)=0$ for ${\bf v}\neq
{\bf 0}$, we get
\begin{eqnarray}
\label{qkm2}
\frac{d}{d\epsilon} \left \lbrack \frac{df}{d\epsilon}+\beta  f \left
(1-\kappa
\frac{f}{\eta_0}\right )\right\rbrack =0
\end{eqnarray}
or, equivalently,
\begin{eqnarray}
\label{qkm3}
\frac{df}{d\epsilon}+\beta  f \left
(1-\kappa
\frac{f}{\eta_0}\right )=-J,
\end{eqnarray}
where $J$ is a constant of integration representing the diffusion current in
energy space. If we set $J=0$, we recover the Fermi-Dirac and Bose-Einstein DFs of statistical
equilibrium [see Eq. (\ref{qle7})]  with $\epsilon$ instead of $m v^2/2$.
However, for spatially inhomogeneous self-gravitating systems, these DFs have an
infinite mass \cite{bt}. This is why it is important to take into
account the escape of high energy particles. If $J\neq 0$, Eq. (\ref{qkm3})
accounts for an escape of particles at a constant rate $J$. The system is not
truly static  since it looses gradually particles but we can
consider that it passes by a succession of quasistationary states that are
the solutions of Eq. (\ref{qkm3}). This is a Riccatti equation that can be solved analytically by usual means.
Repeating the calculations of \cite{chavmnras}, that are valid both for fermions
and
bosons, we find in good
approximation that\footnote{This expression is valid for
$J/\beta\eta_0\ll 1$.  The general case will be
treated in a specific paper \cite{spec} (see also \cite{chavmnras}). For bosons,
there is a maximum current $J_{\rm
max}=\beta\eta_0/4$ corresponding to
$f=(\eta_0/2)\beta(\epsilon_m-\epsilon)/[2-\beta(\epsilon_m-\epsilon)]$ with
$\beta\epsilon_m<2$.}
\begin{eqnarray}
\label{qkm}
f=\eta_0\frac{e^{-\beta \epsilon}-e^{-\beta \epsilon_m}}{\lambda+\kappa e^{-\beta
\epsilon}}.
\end{eqnarray}
In the classical limit ($\kappa=0$), we recover the King model \cite{king}.
Therefore, the DF from Eq. (\ref{qkm}) may be called the quantum (fermionic or
bosonic) King model. This DF could describe DM
halos made of massive neutrinos or axions limited in extension by tidal
forces. On a secular timescale, we should  take into account the slow change of
the constants $\lambda$ and $\beta$ in Eq. (\ref{qkm}) due to evaporation in a
sort of 
adiabatic approximation.

\section{Classical particles in collision with quantum particles}
\label{sec_m}

In this section, we consider the case of classical test particles
in ``collision'' with quantum field particles. For
example, the classical particles may be stars, globular clusters or black holes moving
in a DM halo made of fermions or bosons (see Sec. \ref{sec_gbmt}). As before,  we use an idealization in which the
system is assumed to be spatially homogeneous and we ignore collective effects.

Assuming that the test particles are classical ($\kappa=0$) and that the field
particles are quantum ($\kappa_{b}=\pm 1$), we obtain the Landau
equation
\begin{eqnarray}
\label{m1}
\frac{\partial f}{\partial t}=2\pi G^2\ln\Lambda\frac{\partial}{\partial
v_i}\int d{\bf v}' K_{ij}\left\lbrace m_b f'_b \left (1-\kappa_b
\frac{f'_b}{\eta_b}\right )\frac{\partial f}{\partial v_j}- m f \frac{\partial
f'_b}{\partial v'_j}\right\rbrace.
\end{eqnarray}
It can be written in the form of a Fokker-Planck
equation
\begin{eqnarray}
\label{m2}
\frac{\partial f}{\partial t}=\frac{\partial}{\partial
v_i}\left ( D_{ij}\frac{\partial f}{\partial v_j}- f  F_i^{\rm pol}
\right )
\end{eqnarray}
with a diffusion tensor
\begin{eqnarray}
\label{m3}
D_{ij}=2\pi G^2 m_b \ln\Lambda \int d{\bf v}'
K_{ij} f'_b \left (1-\kappa_b
\frac{f'_b}{\eta_b}\right )
\end{eqnarray}
and a friction by polarisation
\begin{eqnarray}
\label{m4}
F_i^{\rm pol}=2\pi G^2 m \ln\Lambda\int d{\bf v}'
K_{ij}\frac{\partial f'_b}{\partial v'_j}.
\end{eqnarray}
As before, the diffusion coefficient is affected by quantum mechanics while the friction is not (it has the same expression as for classical particles). The usual form of the Fokker-Planck
equation is
\begin{eqnarray}
\label{m4b}
\frac{\partial f}{\partial t}=\frac{\partial^2}{\partial
v_i\partial v_j}( D_{ij} f)-\frac{\partial}{\partial v_i}(f
F_i^{\rm
friction}),
\end{eqnarray}
with the total friction
\begin{eqnarray}
\label{m5}
F_i^{\rm friction}=F_i^{\rm pol}+\frac{\partial D_{ij}}{\partial v_j}.
\end{eqnarray}
In the thermal bath approximation, where the field particles have the
Fermi-Dirac or Bose-Einstein DF (\ref{q20}), the
friction by polarisation is given by (see Sec. \ref{sec_qtb})
\begin{eqnarray}
\label{m6}
{\bf F}_{\rm pol}=- \beta m D_{\|} {\bf v}
\end{eqnarray}
and we get the Kramers-Chandrasekhar equation
\begin{eqnarray}
\label{m7}
\frac{\partial f}{\partial t}=\frac{\partial}{\partial
v_i}\left \lbrack D_{ij} \left (\frac{\partial f}{\partial v_j}+\beta m f
v_j\right ) \right \rbrack,
\end{eqnarray}
where the diffusion tensor is given by Eq. (\ref{c17}) with Eqs. (\ref{q36}) and (\ref{q37}). This equation relaxes towards the Maxwell-Boltzmann
distribution (\ref{c31}). At statistical equilibrium, the (classical) test
particles and the (quantum) field particles have the same temperature.

\section{Effective mass of quasiparticles}
\label{sec_gbm}

In this section, we consider the case of classical or quantum  particles in ``collision'' with quantum particles and introduce the notion of quasiparticles with effective mass $m_{\rm eff}$.

\subsection{General expression}
\label{sec_gbmu}

We redefine the DF of the field particles through the relation
\begin{eqnarray}
\label{m8}
F_{\rm eff}({\bf v})=\frac{\int f_b\, d{\bf v}}{\int f_b \left (1-\kappa_b
\frac{f_b}{\eta_b}\right )\, d{\bf v}} \, f_b({\bf v}) \left (1-\kappa_b
\frac{f_b({\bf v})}{\eta_b}\right ).
\end{eqnarray}
The effective  DF
$F_{\rm eff}({\bf
v})$ is normalized such that $\int
F_{\rm eff}({\bf v})\,
d{\bf v}=\int f_b({\bf v})\,
d{\bf v}=\rho_b$. The diffusion tensor from Eq. (\ref{q4}) or (\ref{m3}) can then be rewritten as
\begin{eqnarray}
\label{m9}
D_{ij}=2\pi G^2 m_{\rm eff}
\ln\Lambda \int d{\bf v}'
K_{ij} F'_{\rm eff},
\end{eqnarray}
where we have introduced  the effective mass
\begin{eqnarray}
\label{m10}
m_{\rm eff}= m_b
\frac{\int f_b \left (1-\kappa_b
\frac{f_b}{\eta_b}\right )\, d{\bf v}}{\int f_b\, d{\bf v}}.
\end{eqnarray}
In this manner, we see that the diffusion coefficient of the test particles due
to collisions
with quantum particles of mass $m_b$ and DF $f_b$ is the same as the
diffusion
coefficient  (\ref{c3}) due to collisions with
classical particles of effective mass $m_{\rm eff}$ and effective DF $F_{\rm
eff}$. We note that $m_{\rm eff}<m_b$ for fermions and  $m_{\rm eff}>m_b$ for
bosons. We stress that this interpretation in terms of quasiparticles of effective mass $m_{\rm eff}$ is valid only for the diffusion
coefficient, not for the friction by polarisation [see Eq. (\ref{q5}) or (\ref{m4})] which keeps the same expression as for classical particles.

\subsection{Thermal bath}
\label{sec_gbmd}

If the quantum particles are at statistical equilibrium, using Eq. (\ref{q21}),
we get
\begin{eqnarray}
\label{m12}
\int f_b \left (1-\kappa_b
\frac{f_b}{\eta_b}\right )\, d{\bf v}=\int_0^{+\infty} \frac{\eta_b\lambda_b
e^{\beta m_b v^2/2}}{(\lambda_b
e^{\beta m_b \frac{v^2}{2}}+\kappa_b)^2} 4\pi v^2\, dv=\frac{4\pi}{\beta
m_b}\int_0^{+\infty} f_b\, dv=\frac{1}{\beta
m_b}\int \frac{f_b}{v^2}\, d{\bf v},
\end{eqnarray}
where we have used an integration by parts to obtain the second
equality. This
relation can also be obtained by using Eq. (\ref{q22}), implying
\begin{eqnarray}
\label{m14}
{\bf v}\cdot \frac{\partial f_b}{\partial {\bf v}}=-f_b\left (1-\kappa_b
\frac{f_b}{\eta_b}\right )\beta m_b v^2.
\end{eqnarray}
Then,
\begin{eqnarray}
\label{m15}
\int f_b \left (1-\kappa_b
\frac{f_b}{\eta_b}\right )\, d{\bf v}=-\frac{1}{\beta
m_b}\int \frac{1}{v^2}{\bf v}\cdot \frac{\partial f_b}{\partial {\bf v}} \,
d{\bf v}=-\frac{1}{\beta
m_b}\int \frac{f'_b(v)}{v}\, d{\bf v}
=\frac{1}{\beta
m_b}\int \frac{f_b}{v^2}\, d{\bf v},
\end{eqnarray}
where the last equality is obtained by integrating by parts.
As a result, the effective mass of the field particles may be rewritten as
\begin{eqnarray}
\label{m16}
m_{\rm eff}= \frac{1}{\beta}\frac{\int \frac{f_b}{v^2}\, d{\bf v}}{\int
f_b\, d{\bf v}}=\frac{1}{\beta}\left\langle \frac{1}{v^2}\right\rangle.
\end{eqnarray}
More specifically, since
\begin{eqnarray}
\label{m17}
\int f_b \left (1-\kappa_b
\frac{f_b}{\eta_b}\right )\, d{\bf v}=\frac{4\pi\eta_b}{\beta
m_b}\int_0^{+\infty} \frac{dv}{\lambda_b e^{\beta m_b
\frac{v^2}{2}}+\kappa_b}=\frac{4\pi\eta_b}{\sqrt{2}(\beta
m_b)^{3/2}}I_{-1/2}(\lambda_b),
\end{eqnarray}
we obtain
\begin{eqnarray}
\label{m18}
m_{\rm eff}= \frac{1}{2} m_b \frac{I_{-1/2}(\lambda_b)}{I_{1/2}(\lambda_b)}.
\end{eqnarray}

\begin{figure}[!h]
\begin{center}
\includegraphics[clip,scale=0.3]{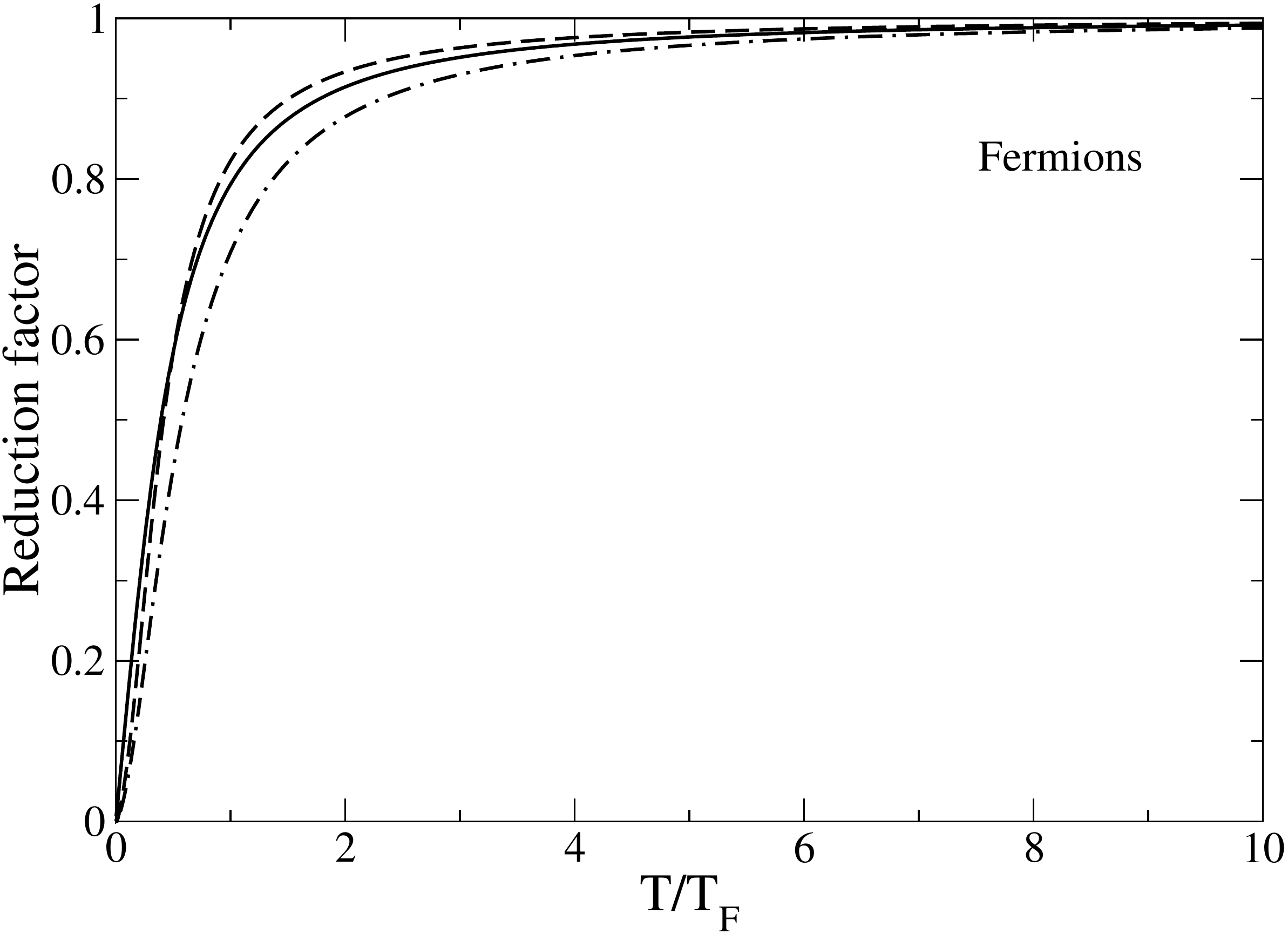}
\caption{Reduction factor for fermions as a function of the temperature: $m_{\rm eff}/m_b$ (solid line);
$D_{\|}^{\rm F}/D_{\|}^{\rm C}$ (dashed line); $D_{\perp}^{\rm F}/D_{\perp}^{\rm
C}$ (dashed-dotted line) for $x=1$. The relaxation time is  increased with respect to the classical case.}
\label{tmFtotal}
\end{center}
\end{figure}

In the case of fermions, using the results of
Appendix \ref{sec_compdeg}, we find that
\begin{eqnarray}
\label{nm19}
\frac{T}{T_{F}}=\left (\frac{2}{3}\right
)^{2/3}\frac{1}{I_{1/2}(\lambda_b)^{2/3}},\qquad \frac{m_{\rm
eff}}{m_b}=\frac{1}{2}\frac{I_{-1/2}(\lambda_b)}{I_{1/2}(\lambda_b)}.
\end{eqnarray}
The first relation of Eq. (\ref{nm19}) links the temperature to the inverse
fugacity ($T_F$ is the Fermi
temperature). The second relation of Eq. (\ref{nm19}) links the effective mass
to the inverse fugacity. Eliminating the inverse fugacity between these two
relations, we obtain the effective mass as a function of the temperature. This
relation is plotted in Fig. \ref{tmFtotal}. Because of
Pauli's blocking, the effective mass $m_{\rm eff}$ of the quasiparticles is
smaller than the mass $m_b$ of 
the fermions. Furthermore, the effective mass decreases as the temperature
decreases. For
$\lambda_b\rightarrow +\infty$ (classical limit), using Eq. (\ref{eos9b}), we
find $T/T_F\sim (16/9\pi)^{1/3}\lambda_b^{2/3}\rightarrow +\infty$ and $m_{\rm
eff}/m_b\simeq 1-1/(2\sqrt{2}\lambda_b)\rightarrow 1$ yielding
\begin{eqnarray}
\label{nm19b}
\frac{m_{\rm eff}}{m_b}\simeq 1-\frac{1}{3}\sqrt{\frac{2}{\pi}}\left (\frac{T_F}{T}\right )^{3/2}...\qquad
(T\rightarrow +\infty).
\end{eqnarray}
For $\lambda_b\rightarrow 0$ (completely
degenerate limit), using Eq. (\ref{fermi8}), we find $T/T_F\sim
1/(-\ln\lambda_b)\rightarrow 0$ and $m_{\rm eff}/m_b\sim
(3/2)/(-\ln\lambda_b)\rightarrow 0$ yielding
\begin{eqnarray}
\label{nm19bb}
\frac{m_{\rm eff}}{m_b}\sim \frac{3}{2}\frac{T}{T_F}\qquad
(T\rightarrow 0).
\end{eqnarray}

\begin{figure}[!h]
\begin{center}
\includegraphics[clip,scale=0.3]{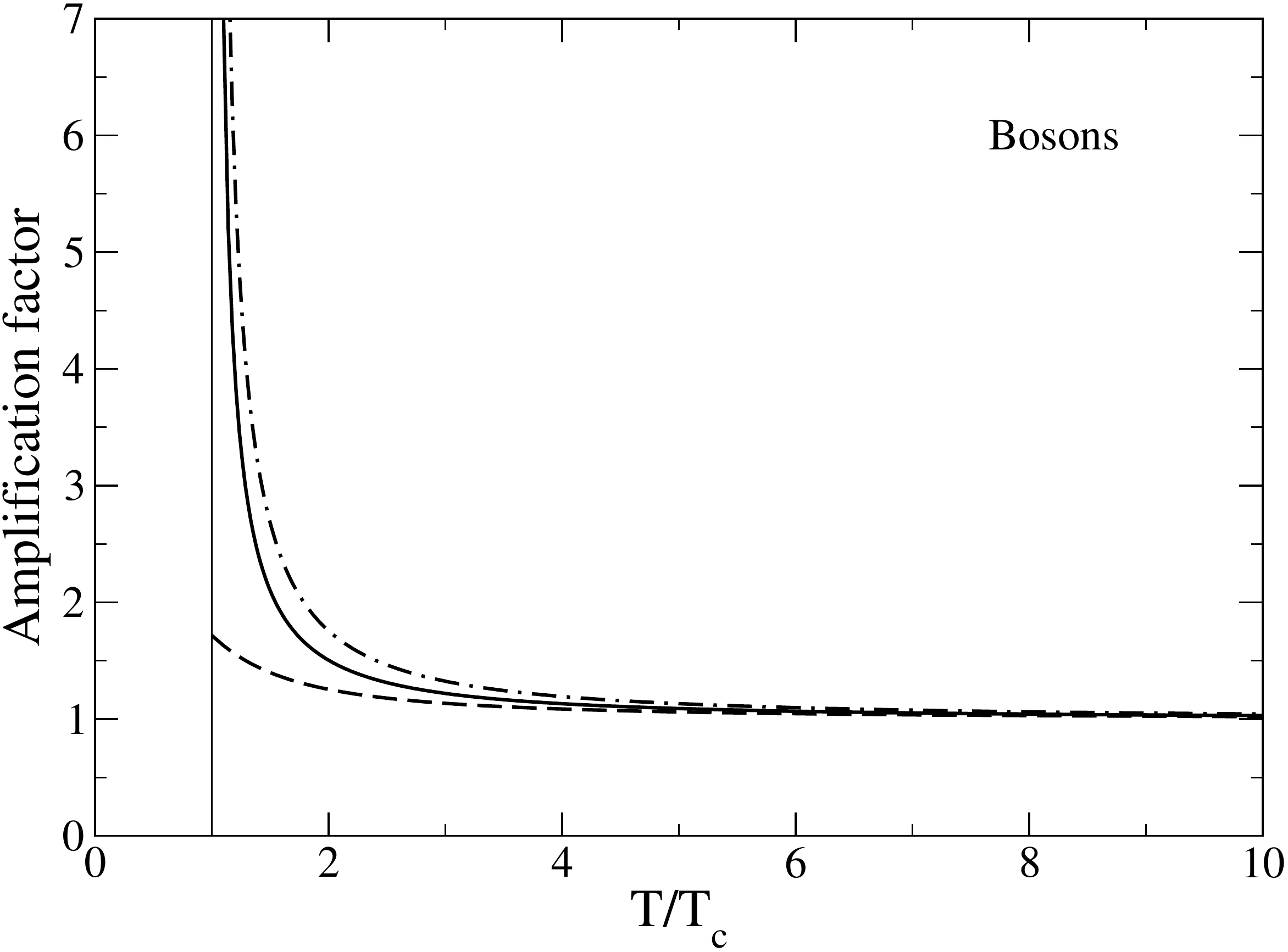}
\caption{Amplification factor for bosons as a function of the temperature:
$m_{\rm eff}/m_b$ (solid line);
$D_{\|}^{\rm B}/D_{\|}^{\rm C}$ (dashed line); $D_{\perp}^{\rm B}/D_{\perp}^{\rm
C}$ (dashed-dotted line) for $x=1$. The relaxation time is reduced with respect to the classical case.}
\label{tmBtotal}
\end{center}
\end{figure}

In the case of bosons, using the results of
Appendix \ref{sec_bec}, we find that
\begin{eqnarray}
\label{nm20}
\frac{T}{T_{c}}=\left
\lbrack \frac{I_{1/2}(1)}{I_{1/2}(\lambda_b)}\right \rbrack^{2/3},\qquad
\frac{m_{\rm
eff}}{m_b}=\frac{1}{2}\frac{I_{-1/2}(\lambda_b)}{I_{1/2}(\lambda_b)}.
\end{eqnarray}
The first relation of Eq. (\ref{nm20}) links the  temperature to the inverse
fugacity ($T_c$ is the condensation temperature). The second
relation of Eq. (\ref{nm20}) links the effective mass to the inverse fugacity.
Eliminating the inverse fugacity between these two relations, we obtain the
effective mass as a function of the temperature. This relation is plotted in
Fig. \ref{tmBtotal}. Because of Bose enhancement, the effective mass $m_{\rm
eff}$ of the quasiparticles is larger than the mass $m_b$ of the bosons.
Furthermore, the effective mass increases as the temperature decreases. For
$\lambda_b\rightarrow +\infty$ (classical limit), using Eq. (\ref{eos9b}), we
find $T/T_c\sim \zeta(3/2)^{2/3}\lambda_b^{2/3}\rightarrow +\infty$ and $m_{\rm
eff}/m_b\simeq 1+1/(2\sqrt{2}\lambda_b)\rightarrow 1$ yielding
\begin{eqnarray}
\label{nm21b}
\frac{m_{\rm eff}}{m_b}\simeq 1+\zeta\left (\frac{3}{2}\right ) \left (\frac{1}{2}\right )^{3/2}\left (\frac{T_c}{T}\right )^{3/2}+...\qquad
(T\rightarrow +\infty).
\end{eqnarray}
For $\lambda_b\rightarrow 1$
(limit of condensation), using Eqs. (\ref{we1}) and
(\ref{we2}), we get $T/T_c\simeq
1+[4\sqrt{\pi}/3\zeta(3/2)]\sqrt{\ln \lambda_b}$ and $m_{\rm eff}/m_b\sim
[\sqrt{\pi}/\zeta(3/2)]\sqrt{\ln \lambda_b}$, yielding
\begin{eqnarray}
\label{nm21}
\frac{m_{\rm eff}}{m_b}\sim \frac{4\pi}{3\zeta\left (\frac{3}{2}\right )^2}\,
\left (\frac{T}{T_c}-1\right )^{-1}\qquad
(T\rightarrow T_c^+).
\end{eqnarray}

The previous calculations show that $m_{\rm eff}\le m_b$ for fermions and
$m_{\rm eff}\ge m_b$ for bosons. Using the heuristic arguments of Sec. \ref{sec_qle} this
implies that the relaxation time due to collision with fermions (resp. bosons) is larger (resp.
smaller) than the relaxation time due to collision with  classical particles of the same mass. We also note that the
effective mass of bosons diverges when $T\rightarrow T_c$. This could suggest
that the relaxation time diverges at $T_c$. Actually,
we must be careful that the diffusion coefficient (\ref{m9}) does not
necessarily diverge at $T_c$ because it involves an integral over $F_{\rm eff}$
that may tend to zero. To determine the relaxation time, it is therefore better
to come back to the explicit expressions (\ref{q36}) and (\ref{q37}) of the
diffusion coefficient. The relaxation time  can then be estimated by
\begin{eqnarray}
\label{qle8b}
t_R\sim \frac{\sigma^2}{D},
\end{eqnarray}
where $\sigma$ is the typical value of the velocity and $D$ is the typical value
of the diffusion coefficient (see
footnotes 31 and 32). The relaxation time is inversely proportional to
the diffusion coefficient. To measure the effect of quantum mechanics, instead
of considering the ratio $m_{\rm eff}/m_b$ between the effective mass and the bare mass, we can consider the ratio between the
quantum diffusion coefficients (\ref{q36}) and (\ref{q37}) and the
classical diffusion coefficients (\ref{c32}) and (\ref{c33}). We find
\begin{eqnarray}
\label{rat1}
\frac{D_{\|}^{Q}}{D_{\|}^{\rm C}}=\frac{I_{1/2}(\lambda_b,x)}{2x^2G(x)},
\end{eqnarray}
\begin{eqnarray}
\label{rat2}
\frac{D_{\perp}^{Q}}{D_{\perp}^{\rm C}}=\frac{1}{2}\frac{1}{{\rm
erf}(x)-G(x)}\left\lbrack
\frac{I_{-1/2}(\lambda_b)}{I_{1/2}(\lambda_b)}I_{-1/2}(\lambda_b,x)-\frac{I_{1/2
}(\lambda_b,x)}{x^2}\right\rbrack.
\end{eqnarray}
In the classical limit $\lambda_b\rightarrow +\infty$ we
recover ${D^{Q}}/{D^{\rm C}}=1$. In the case of fermions (resp. bosons), one can show that
${D^{\rm F}}/{D^{\rm C}}\le 1$ (resp. ${D^{\rm B}}/{D^{\rm C}}\ge 1$) which
confirms that the relaxation time for fermions (resp. bosons) is larger (resp. smaller) than for classical
particles of the same mass. On the other hand, in the case of bosons, we see that
$D_{\|}^{Q}$ does {\it not} diverge at $T_c$ (contrary to $m_{\rm eff}$) while
$D_{\perp}^{Q}$ diverges at $T_c$ (similarly to $m_{\rm eff}$). To get more
quantitative results, we can consider particular 
values of $x$. For
$x\rightarrow 0$, we get
\begin{eqnarray}
\frac{D^{Q}}{D^{\rm C}}\sim \frac{\sqrt{\pi}}{2}\frac{1}{I_{1/2}(\lambda_b)}\frac{1}{\lambda_b+\kappa}.
\end{eqnarray}
For $x\rightarrow +\infty$, we get
\begin{eqnarray}
\label{qle8c}
\frac{D_{\|}^{Q}}{D_{\|}^{\rm C}}\rightarrow 1,
\end{eqnarray}
\begin{eqnarray}
\label{qle8d}
\frac{D_{\perp}^{Q}}{D_{\perp}^{\rm C}}\sim \frac{1}{2}
\frac{I_{-1/2}(\lambda_b)}{I_{1/2}(\lambda_b)}=\frac{m_{\rm eff}}{m_b}.
\end{eqnarray}
For $x=1$, the factors ${D_{\|}^{Q}}/{D_{\|}^{\rm C}}$ and
${D_{\perp}^{Q}}/{D_{\perp}^{\rm C}}$ are plotted as a function of $T/T_Q$ in
Figs. \ref{tmFtotal} and \ref{tmBtotal} and compared with $m_{\rm eff}/m_b$. In
the case of fermions, we see on Fig. \ref{tmFtotal}  that the reduction factors
${D_{\|}^{\rm F}}/{D_{\|}^{\rm C}}$ and ${D_{\perp}^{\rm F}}/{D_{\perp}^{\rm
C}}$ both behave similarly to $m_{\rm eff}/m_b$. This implies that the increase
of the relaxation time due to Pauli's blocking is relatively important for $T\ll
T_F$. In the case of bosons, we see on Fig.  \ref{tmBtotal} that the
amplification factor  ${D_{\perp}^{\rm B}}/{D_{\perp}^{\rm C}}$ behaves
similarly to $m_{\rm eff}/m_b$ (in particular, 
it diverges at $T_c$)  while the amplification factor ${D_{\|}^{\rm
B}}/{D_{\|}^{\rm C}}$ remains approximately constant (it tends to $1.72$ at 
$T_c$). If we consider isotropic systems, this implies that the reduction of the
relaxation time due to Bose enhancement is not very important for $T\ge T_c$. If
we consider anisotropic systems, the divergence of the diffusion coefficient 
${D_{\perp}^{\rm B}}$ at $T_c$ implies that system becomes rapidly isotropic
when $T\rightarrow T_c$ (the corresponding relaxation time tends to
zero).

\subsection{A nice relation}
\label{sec_nice}

Using the formalism of \cite{kingen}, the results of the previous sections
can be generalized to an arbitrary entropy of the form (\ref{gle5}). According
to Eq. (\ref{gle6}), the DF  of the field particles is
\begin{eqnarray}
\label{gle6new}
f_{b}({\bf v})=(C')^{-1}\left (\alpha_b-\beta m_b\frac{v^2}{2}\right ).
\end{eqnarray}
The
effective DF and the effective mass can be written in the general case as
\begin{eqnarray}
\label{effgen1}
F_{\rm eff}({\bf v})=\frac{\int f_b\, d{\bf v}}{\int \frac{d{\bf v}}{C''(f_b)}\, d{\bf v}}\frac{1}{C''(f_b)}
\end{eqnarray}
and
\begin{eqnarray}
\label{effgen2}
m_{\rm eff}=m_b\frac{\int \frac{d{\bf v}}{C''(f_b)}}{\int f_b\, d{\bf v}}
=-\frac{1}{\beta}\frac{\int \frac{{\bf v}}{v^2}\cdot \frac{\partial
f_b}{\partial {\bf v}}\, d{\bf v}}{\int f_b\, d{\bf
v}}=-\frac{1}{\beta}\frac{\int \frac{f'_b(v)}{v}\, d{\bf v}}{\int f_b\, d{\bf
v}}= \frac{1}{\beta}\frac{\int \frac{f_b}{v^2}\, d{\bf v}}{\int
f_b\, d{\bf v}}=\frac{1}{\beta}\left\langle \frac{1}{v^2}\right\rangle.
\end{eqnarray}
To obtain the second and subsequent equalities, we have used the
identity
\begin{eqnarray}
\label{effgen3}
C'(f_b)=\alpha_b-\beta m_b \frac{v^2}{2}\quad\Rightarrow \quad \frac{\partial
f_b}{\partial {\bf v}}=-\frac{\beta m_b}{C''(f_b)}{\bf v},
\end{eqnarray}
which follows from Eq. (\ref{gle6new}) and performed integrations par parts.

On the other hand, introducing the density $\rho_b=\int
f_b\, d{\bf v}$ and the pressure  $P_b=\frac{1}{3}\int f_b v^2\, d{\bf v}$,
where $f_b$ is given by Eq. (\ref{gle6new}), and eliminating $\alpha_b$ between
these two expressions, we obtain the equation of state $P_b(\rho_b)$ of the
field particles. The squared speed of sound $c_s^2=P_b'(\rho_b)$ is then given
by (see also Eq. (50) of \cite{nyquistgrav}) 
\begin{eqnarray}
\label{effgen4}
c_s^2=\frac{\frac{1}{3}\int F'(x)v^2\, d{\bf v}}{\int F'(x)\,
d{\bf v}}=\frac{\frac{1}{3}\int {\bf v}\cdot\frac{\partial f_b}{\partial {\bf
v}}\, d{\bf v}}{\int \frac{\bf v}{v^2}\cdot\frac{\partial f_b}{\partial
{\bf v}}\, d{\bf v}}=-\frac{\rho_b}{\int \frac{f'_b(v)}{v}\, d{\bf
v}}=\frac{\rho_b}{\int \frac{f_b}{v^2}\, d{\bf v}}=\frac{1}{\left\langle
\frac{1}{v^2}\right\rangle},
\end{eqnarray}
where $F(x)=(C')^{-1}(-x)$ with $x=\beta m_b \frac{v^2}{2}-\alpha_b$. To obtain
the second and subsequent equalities, we have used the
identity
\begin{eqnarray}
\label{effgen5}
f_b({\bf v})=F\left (\beta m_b \frac{v^2}{2}-\alpha_b\right
)\quad\Rightarrow \quad \frac{\partial f_b}{\partial {\bf v}}=F'(x)\beta m_b
{\bf v},
\end{eqnarray}
which follows from Eq. (\ref{gle6new}) and performed integrations par parts.
Comparing Eqs. (\ref{effgen3}) and
(\ref{effgen5}), we obtain $F'(x)=-1/C''(f_b)$. Finally, comparing Eqs.
(\ref{effgen2}) and
(\ref{effgen4}), we obtain the nice relation
\begin{eqnarray}
\label{effgen6}
m_{\rm eff}=\frac{1}{\beta c_s^2},
\end{eqnarray}
which is valid when the field particles are at statistical equilibrium.

\subsection{The case of FDM}
\label{sec_gbmt}

In this section we focus on the case of bosons ($\kappa_b=-1$). We consider the situation studied by  Bar-Or {\it et al.} \cite{bft} in the context of FDM (see the introduction and Appendix \ref{sec_fdm}), namely the collisional evolution of test particles (like stars, globular clusters, black holes...) moving in a FDM halo. We assume that the bosons that compose the halo are described by an out-of-equilibrium DF with an effective temperature $T_{\rm eff}\ll T_c$ resulting from a process of violent collisionless relaxation. As
shown by Hui {\it et al.} \cite{hui}, quantum interferences produce a form of scalar radiation made of quasiparticles (granules) of effective mass $m_{\rm eff}\gg m_b$ and approximately Maxwellian DF. Everything happens as if the test particles were in collision with these quasiparticles. To simplify the expression of the effective mass $m_{\rm eff}$ of these quasiparticles, we make two
approximations:

(i) We assume that the amplification due to Bose stimulation is large ($f_b\gg \eta_b$) so that
\begin{eqnarray}
\label{m19}
f_b \left (1+
\frac{f_b}{\eta_b}\right )\simeq \frac{f_b^2}{\eta_b}.
\end{eqnarray}
With this approximation, the effective  DF of the bosons and their effective
mass defined by Eqs. (\ref{m8}) and (\ref{m10}) become
\begin{eqnarray}
\label{m20}
F_{\rm eff}({\bf v})=\frac{\int f_b\, d{\bf v}}{\int f_b^2\,
d{\bf v}} \, f_b({\bf v})^2
\end{eqnarray}
and
\begin{eqnarray}
\label{m21}
m_{\rm eff}= \frac{m_b}{\eta_b}
\frac{\int f_b^2\, d{\bf v}}{\int f_b\, d{\bf v}}.
\end{eqnarray}
We note that the effective mass of the quasiparticles is much higher than the mass of the bosons ($m_{\rm eff}\gg m_b$). Therefore, Bose stimulation significantly accelerates the collisional relaxation.

(ii) We assume that $f_b$ is  Maxwellian
\begin{eqnarray}
\label{m22}
f_b({\bf v})=\frac{\rho_b}{(2\pi \sigma_b^2)^{3/2}}
e^{-\frac{v^2}{2\sigma_b^2}}
\end{eqnarray}
with a velocity dispersion $\sigma_b^2$. Substituting Eq.
(\ref{m22}) into Eq.
(\ref{m20}), we find that the
effective DF of the bosons is a Maxwellian with velocity dispersion
$\sigma_b^2/2$, i.e.,
\begin{eqnarray}
\label{m23}
F_{\rm eff}({\bf v})=\frac{\rho_b}{(\pi \sigma_b^2)^{3/2}}
e^{-\frac{v^2}{\sigma_b^2}}.
\end{eqnarray}
On the other hand, their effective mass from Eq. (\ref{m21}) is
\begin{eqnarray}
\label{m27}
m_{\rm eff}&=&\frac{m_b}{\eta_b}
\frac{\rho_b}{(2\pi \sigma_b^2)^{3/2}}\frac{\int_0^{+\infty}
e^{-v^2/\sigma_b^2}v^2\,
dv}{\int_0^{+\infty} e^{-
v^2/2\sigma_b^2}v^2\, dv}\nonumber\\
&=&\frac{m_b}{\eta_b}\frac{\rho_b}{(4\pi \sigma_b^2)^{3/2}}.
\end{eqnarray}
Recalling Eq. (\ref{q2}), and taking $g=1$, we obtain
\begin{eqnarray}
\label{m29}
m_{\rm eff}=\frac{\pi^{3/2}\hbar^3\rho_b}{ m_b^3\sigma_b^3}
\end{eqnarray}
This can be written as
\begin{eqnarray}
\label{m30}
m_{\rm eff}=\frac{1}{8\pi^{3/2}}\rho_b \lambda_{\rm dB}^3
\end{eqnarray}
where $\lambda_{\rm dB}=h/(m_b\sigma_b)$ is the de Broglie length.

According to Eqs. (\ref{m9}),
(\ref{m23}) and (\ref{m27}), the
diffusion tensor is  the same as the one due to classical particles with
effective mass $m_{\rm eff}$ and Maxwellian DF with a velocity
dispersion
$\sigma_{\rm eff}=\sigma_b/\sqrt{2}$. Using Eqs. (\ref{c32}) and
(\ref{c33}), we therefore obtain 
\begin{eqnarray}
\label{m31}
D_{\|}=4\pi G^2 m_{\rm eff}\ln\Lambda \rho_b G(x_{\rm eff})\frac{1}{v},
\end{eqnarray}
\begin{eqnarray}
\label{m32}
D_{\perp}=4\pi G^2 m_{\rm eff}\ln\Lambda \rho_b \left\lbrack {\rm
erf}(x_{\rm eff})-G(x_{\rm eff})\right\rbrack \frac{1}{v},
\end{eqnarray}
with
\begin{eqnarray}
\label{m33}
{\bf x}_{\rm eff}=\frac{\bf v}{\sigma_b}.
\end{eqnarray}
The diffusion tensor can be written as
\begin{eqnarray}
\label{m42}
D_{ij}=\frac{2}{\sqrt{\pi}}G^2m_{\rm
eff}\ln\Lambda\rho_b\frac{1}{\sigma_b}G_{ij}({\bf x}_{\rm eff}),
\end{eqnarray}
where $G_{ij}$ is defined by Eq. (\ref{ps38}). According to Eq. (\ref{c37}), we have
\begin{eqnarray}
\label{m34}
\frac{\partial D_{ij}}{\partial v_j}
=-     \frac{8\pi G^2 m_{\rm
eff}\ln\Lambda \rho_b}{\sigma^2}
G(x_{\rm eff})\frac{{\bf v}}{v}.
\end{eqnarray}
On the other hand,
according to Eqs. (\ref{m4}) and (\ref{m22}),
the force by polarisation is  the same as the one due to classical particles
with mass $m_{b}$ and Maxwellian DF with a velocity dispersion $\sigma_b$.
Using Eq. (\ref{c36}), we therefore
obtain
\begin{eqnarray}
\label{m35}
{\bf F}_{\rm pol}=-     \frac{4\pi G^2 m \ln\Lambda \rho_b}{\sigma_b^2}
G(x)\frac{{\bf v}}{v},
\end{eqnarray}
with
\begin{eqnarray}
\label{m36}
{\bf x}=\frac{\bf v}{\sqrt{2}\sigma_b}.
\end{eqnarray}
We note that ${\bf x}_{\rm eff}=\sqrt{2}{\bf x}$. According to Eqs. (\ref{m5}), (\ref{m34}) and (\ref{m35}) the total friction
force is
\begin{eqnarray}
\label{m37}
{\bf F}_{\rm friction}=- \frac{4\pi G^2\ln\Lambda \rho_b}{\sigma_b^2}
\left\lbrack m G(x)+2m_{\rm eff}G(x_{\rm eff})\right\rbrack \frac{{\bf v}}{v}.
\end{eqnarray}
We note that it is affected by Bose stimulation through the term ${\partial D_{ij}}/{\partial v_j}$. The friction by polarisation can be written as
\begin{eqnarray}
\label{m38}
{\bf F}_{\rm pol}=- \frac{1}{\sigma_b^2} \frac{m}{m_{\rm
eff}}\frac{G(x)}{G(x_{\rm
eff})} D_{\|} {\bf v}.
\end{eqnarray}
It differs from the Einstein relation (\ref{m6}) obtained by assuming
that the field particles (bosons) are at statistical equilibrium with the
Bose-Einstein DF at $T>T_c$. This is because we are in a completely
different situation where the bosons are in a out-of-equilibrium state with $T_{\rm eff}\ll T_c$. Finally, assuming that the test particles are classical and using $D_{ij}v_j=D_{\|}v_i$ according to Eq. (\ref{c17}), the Fokker-Planck equation
(\ref{m2}) can  be
written as
\begin{eqnarray}
\label{m39}
\frac{\partial f}{\partial t}=\frac{\partial}{\partial v_i}\left\lbrack
D_{ij}\left (\frac{\partial f}{\partial v_j}+\frac{1}{\sigma_b^2}
\frac{m}{m_{\rm
eff}}\frac{G(x)}{G(x_{\rm
eff})} f v_j\right )\right\rbrack.
\end{eqnarray}
It relaxes towards an equilibrium DF determined by the differential equation
\begin{eqnarray}
\label{m40}
\frac{\partial f}{\partial v}+\frac{1}{\sigma_b^2} \frac{m}{m_{\rm
eff}}\frac{G(x)}{G(x_{\rm
eff})} f v=0.
\end{eqnarray}
Its solution is
\begin{eqnarray}
\label{m41}
f_{\rm eq}({\bf v})=A e^{-\frac{1}{\sigma_b^2} \frac{m}{m_{\rm
eff}}\int_0^v \frac{G(x)}{G(x_{\rm
eff})} v \, dv}.
\end{eqnarray}
We note that the equilibrium state of the test (classical) particles is
non-Maxwellian (see Fig. 1 of \cite{bft}) contrary to the
case where the bosons are at statistical
equilibrium [see Eq. (\ref{c31})]. Again, this is because we are
considering a completely different situation where the bosons are out-of-equilibrium.\footnote{Using the asymptotic behaviors of $G(x)$ (see Sec. \ref{sec_cdtb}), we find that $f_{\rm eq}\propto {\rm exp}\lbrack -\frac{1}{2\sigma_b^2} \frac{m}{\sqrt{2}m_{\rm
eff}}v^2\rbrack$ for $v\rightarrow 0$ and $f_{\rm eq}\propto {\rm exp}\lbrack
-\frac{1}{2\sigma_b^2} \frac{2 m}{m_{\rm eff}}v^2\rbrack$ for $v\rightarrow
+\infty$. The first expression is also valid for all $v$ when $m\gg m_{\rm eff}$
and the second when $m\ll m_{\rm eff}$. Therefore, the DF of the test particles
is Maxwellian in these limits. Furthermore, the velocity dispersion of the test
particles varies from $(\sigma_t)_{\rm eq}^2/\sigma_b^2=m_{\rm eff}/(2m)$ when
$m\ll m_{\rm eff}$ to $(\sigma_t)_{\rm eq}^2/\sigma_b^2=\sqrt{2}m_{\rm eff}/m$
when $m\gg m_{\rm eff}$. These results are consistent with Figs. 1 and 2 of
\cite{bft} (note that our expression of $f_{\rm eq}$ is slightly simpler than
their Eq. (90)).   They can also be recovered from Eq.
(\ref{m11c}).}

The Fokker-Planck equation (\ref{m39}) can be written under a dimensionless form
as
\begin{eqnarray}
\label{m42b}
\frac{\partial f}{\partial t}=\frac{1}{t_H}\frac{\partial}{\partial x_i}\left\lbrack
G_{ij}({\bf x}_{\rm eff})\left (\frac{\partial f}{\partial x_j}+2\frac{m}{m_{\rm
eff}}\frac{G(x)}{G(x_{\rm
eff})} f x_j\right )\right\rbrack,
\end{eqnarray}
where we have introduced  the heating time \cite{bft}
\begin{eqnarray}
\label{m43}
t_H=\frac{\sqrt{\pi}\sigma_b^3}{G^2\rho_b m_{\rm
eff}\ln\Lambda}=\frac{m_b^3\sigma_b^6}{\pi G^2\rho_b^2 \hbar^3\ln\Lambda}.
\end{eqnarray}
Alternatively, it can be written as
\begin{eqnarray}
\label{m44}
\frac{\partial f}{\partial t}=\frac{1}{t_C}\frac{\partial}{\partial x_i}\left\lbrack
G_{ij}({\bf x}_{\rm eff})\left (\frac{m_{\rm
eff}}{m}\frac{\partial f}{\partial x_j}+2\frac{G(x)}{G(x_{\rm
eff})} f x_j\right )\right\rbrack,
\end{eqnarray}
where we have introduced the cooling time \cite{bft}
\begin{eqnarray}
\label{m45}
t_C=\frac{\sqrt{\pi}\sigma_b^3}{G^2\rho_b m\ln\Lambda}.
\end{eqnarray}
We can also write
\begin{eqnarray}
\label{m46}
\frac{\partial f}{\partial t}=\frac{\partial}{\partial x_i}\left\lbrack
G_{ij}({\bf x}_{\rm eff})\left (\frac{1}{t_H}\frac{\partial f}{\partial x_j}+\frac{1}{t_C} 2\frac{G(x)}{G(x_{\rm
eff})} f x_j\right )\right\rbrack
\end{eqnarray}
with
\begin{eqnarray}
\label{m47}
\frac{t_C}{t_H}=\frac{m_{\rm eff}}{m}.
\end{eqnarray}
Apart from a numerical factor, the cooling time has the same expression as in
the case of classical field particles of mass $m_b$ and velocity dispersion
$\sigma_b$. By contrast, the heating time is
very different as
it involves the effective mass of the quasiparticles $m_{\rm eff}$ instead of the boson mass
$m_b$. Since $m_{\rm eff}\gg m_b$, the heating time is considerably
reduced.

{\it Remark:} Bar-Or {\it et al.} \cite{bft} obtained these 
results by a very different method based on the Schr\"odinger-Poisson equations.
It is interesting to recover their results
directly from the formalism of Ref. \cite{kingen} based on the bosonic Landau
equation (\ref{qle5}).

\section{Self-consistent Landau equation in the single species case}
\label{sec_ssl}

In this section we consider the self-consistent (integrodifferential) Landau
equation for a single
species system. We  make a change of variables to write it in energy space.

\subsection{Quantum Landau equation in energy space}
\label{sec_qles}

The self-consistent quantum Landau equation can be written as
\begin{eqnarray}
\label{qles1}
\frac{\partial f}{\partial t}=2\pi G^2 m \ln\Lambda\frac{\partial}{\partial
v_i}\int d{\bf v}' K_{ij}\left\lbrace f' \left (1-\kappa
\frac{f'}{\eta_0}\right )\frac{\partial f}{\partial v_j}- f \left
(1-\kappa
\frac{f}{\eta_0}\right )\frac{\partial f'}{\partial v'_j}\right\rbrace
\end{eqnarray}
or, equivalently, as
\begin{eqnarray}
\label{qles2}
\frac{\partial f}{\partial t}=\frac{\partial}{\partial
v_i}\left \lbrack D_{ij}\frac{\partial f}{\partial v_j}- f \left
(1-\kappa
\frac{f}{\eta_0}\right ) F_i^{\rm pol} \right \rbrack,
\end{eqnarray}
where $D_{ij}$ and ${\bf F}_{\rm pol}$ are defined by Eqs. (\ref{q4}) and
(\ref{q5}) with $m_b=m$ and $f_b=f$. If the
DF is
isotropic, i.e.,  $f=f(v,t)$, the quantum Landau equation reduces to the form
\begin{eqnarray}
\label{qles3}
\frac{\partial f}{\partial t}=\frac{1}{v^2}\frac{\partial}{\partial
v}\left \lbrack v^2 \left (D_{\|}\frac{\partial f}{\partial v}- f \left
(1-\kappa
\frac{f}{\eta_0}\right ) F^{\rm pol} \right )\right \rbrack,
\end{eqnarray}
where $D_{\|}$ and $F^{\rm pol}$ are given by Eqs. (\ref{q15}) and
(\ref{q17}). In that case, the quantum Landau equation can be written
explicitly as
\begin{eqnarray}
\label{qles4}
\frac{\partial f}{\partial t}=16\pi^2
G^2m\ln\Lambda\frac{1}{v^2}\frac{\partial}{\partial
v}\left \lbrack A\frac{\partial f}{\partial v}+B f \left
(1-\kappa
\frac{f}{\eta_0}\right )\right \rbrack
\end{eqnarray}
with
\begin{eqnarray}
\label{qles5}
A(v,t)= \frac{1}{3v}\int_0^v
v_1^4 f(v_1,t) \left (1-\kappa \frac{f(v_1,t)}{\eta_0}\right )\,
dv_1+\frac{v^2}{3}\int_v^{+\infty} v_1 f(v_1,t) \left (1-\kappa
\frac{f(v_1,t)}{\eta_0}\right )\, dv_1
\end{eqnarray}
and
\begin{eqnarray}
\label{qles6}
B(v,t)=\int_0^v f(v_1,t) v_1^2\,
dv_1.
\end{eqnarray}
This is a self-consistent (integrodifferential) equation for $f({v},t)$.
In the case of bosons ($\kappa=-1$), this equation
should display the process of Bose-Einstein condensation when $E<E_c$, where
$E_c=(3/2)M\sigma_c^2$ is the critical energy corresponding to $T=T_c$ (see Appendix \ref{sec_bec}).
In that case, the DF can be written as
$f(v,t)=f_{\rm gas}(v,t)+M_c(t)\delta(v)/(4\pi v^2)$ where $f_{\rm
gas}({v},t)$ is the DF of the uncondensed bosons (gas) and $M_c(t)=M-M_{\rm
gas}(t)=M-\int_0^{+\infty} f_{\rm gas}(v,t)4\pi v^2\, dv$ is the mass of the
condensed bosons. The mass of the Dirac peak (condensate) increases with time
until a statistical equilibrium state comprising both uncondensed and condensed
bosons is reached [see Eq. (\ref{bec4})]. Since  the bosonic Landau equation
conserves the energy, it corresponds to a microcanonical description of
Bose-Einstein
condensation. This can be contrasted from the canonical situation studied by
Sopik {\it et al.} \cite{sopik} (see the Remark at the end of Sec.
\ref{sec_qtb}).

It is
convenient to rewrite the quantum Landau  equation (\ref{qles4}) in terms
of the individual energy $\epsilon=(1/2)mv^2$ instead of the velocity $v$ by using the
relation $F(\epsilon,t)\,
d\epsilon=4\pi v^2
f(v,t)\, dv$  yielding
\begin{eqnarray}
\label{qles7}
F(\epsilon,t)=\frac{4\pi}{m} v f(v,t).
\end{eqnarray}
With this change of variables, we obtain
\begin{eqnarray}
\label{qles8}
\frac{\partial F}{\partial t}=16\pi^2 G^2m^3\ln\Lambda\frac{\partial}{\partial
\epsilon}\left \lbrack A\frac{\partial F}{\partial
\epsilon}-\frac{AF}{2\epsilon}+\frac{BF}{\sqrt{2m\epsilon}} \left
(1-\kappa
\frac{m^{3/2}F}{4\pi\eta_0 \sqrt{2\epsilon}}\right )\right \rbrack
\end{eqnarray}
with
\begin{equation}
\label{qles9}
A(\epsilon,t)= \frac{1}{6\pi\sqrt{2m\epsilon}}\int_0^\epsilon
\epsilon_1 F(\epsilon_1,t) \left (1-\kappa
\frac{m^{3/2}F(\epsilon_1,t)}{4\pi\eta_0
\sqrt{2\epsilon_1}}\right )\,
d\epsilon_1+\frac{\epsilon}{6\pi}\int_\epsilon^{+\infty}
\frac{1}{\sqrt{2m\epsilon_1}} F(\epsilon_1,t) \left (1-\kappa
\frac{m^{3/2}F(\epsilon_1,t)}{4\pi\eta_0
\sqrt{2\epsilon_1}}\right )\, d\epsilon_1
\end{equation}
and
\begin{eqnarray}
\label{qles10}
B(\epsilon,t)=\frac{1}{4\pi}\int_0^\epsilon  F(\epsilon_1,t) \,
d\epsilon_1.
\end{eqnarray}

If we consider the case of bosons ($\kappa=-1$) with
$E\ll E_c$, we can make the approximation
\begin{eqnarray}
\label{qles11}
f \left (1+\frac{f}{\eta_0}\right )\simeq  \frac{f^2}{\eta_0}.
\end{eqnarray}
In that case, the bosonic Landau equation (\ref{qles8}) reduces to\footnote{We
can make a similar simplification in Eqs. (\ref{qles4})-(\ref{qles6}).}
\begin{eqnarray}
\label{qles12}
\frac{\partial F}{\partial t}=16\pi^2 G^2m^3\ln\Lambda\frac{\partial}{\partial
\epsilon}\left \lbrack A\frac{\partial F}{\partial
\epsilon}+\left
(\frac{BmF}{4\pi\eta_0}-A\right )\frac{F}{2\epsilon}\right
\rbrack
\end{eqnarray}
with 
\begin{eqnarray}
\label{qles13}
A(\epsilon,t)=\frac{m}{48\pi^2\eta_0}\left\lbrack \frac{1}{\sqrt{\epsilon}}
\int_0^\epsilon
\sqrt{\epsilon_1} F(\epsilon_1,t)^2 \,
d\epsilon_1+\epsilon\int_\epsilon^{+\infty}
\frac{1}{\epsilon_1} F(\epsilon_1,t)^2 \, d\epsilon_1 \right\rbrack
\end{eqnarray}
and
\begin{eqnarray}
\label{qles14}
B(\epsilon,t)=\frac{1}{4\pi}\int_0^\epsilon  F(\epsilon_1,t) \,
d\epsilon_1.
\end{eqnarray}
When $E<E_c$, the mass of condensed bosons is $M_{\rm BEC}(t)=M-M_{\rm gas}(t)$
with $M_{\rm gas}(t)=\int_0^{+\infty} F(\epsilon,t)\, d\epsilon$. Eqs.
(\ref{qles12})-(\ref{qles14})
are equivalent to those derived by Levkov {\it et al.} \cite{levkov2} from the
Schr\"odinger-Poisson equations. They used them to study the process of
Bose-Einstein condensation in the context of DM, resulting in the formation of a
Bose star. Using scaling arguments, they showed that when $E\ll E_c$ the
relaxation (condensation) time is given by Eq. (\ref{qle11}).

\begin{figure}[!h]
\begin{center}
\includegraphics[clip,scale=0.3]{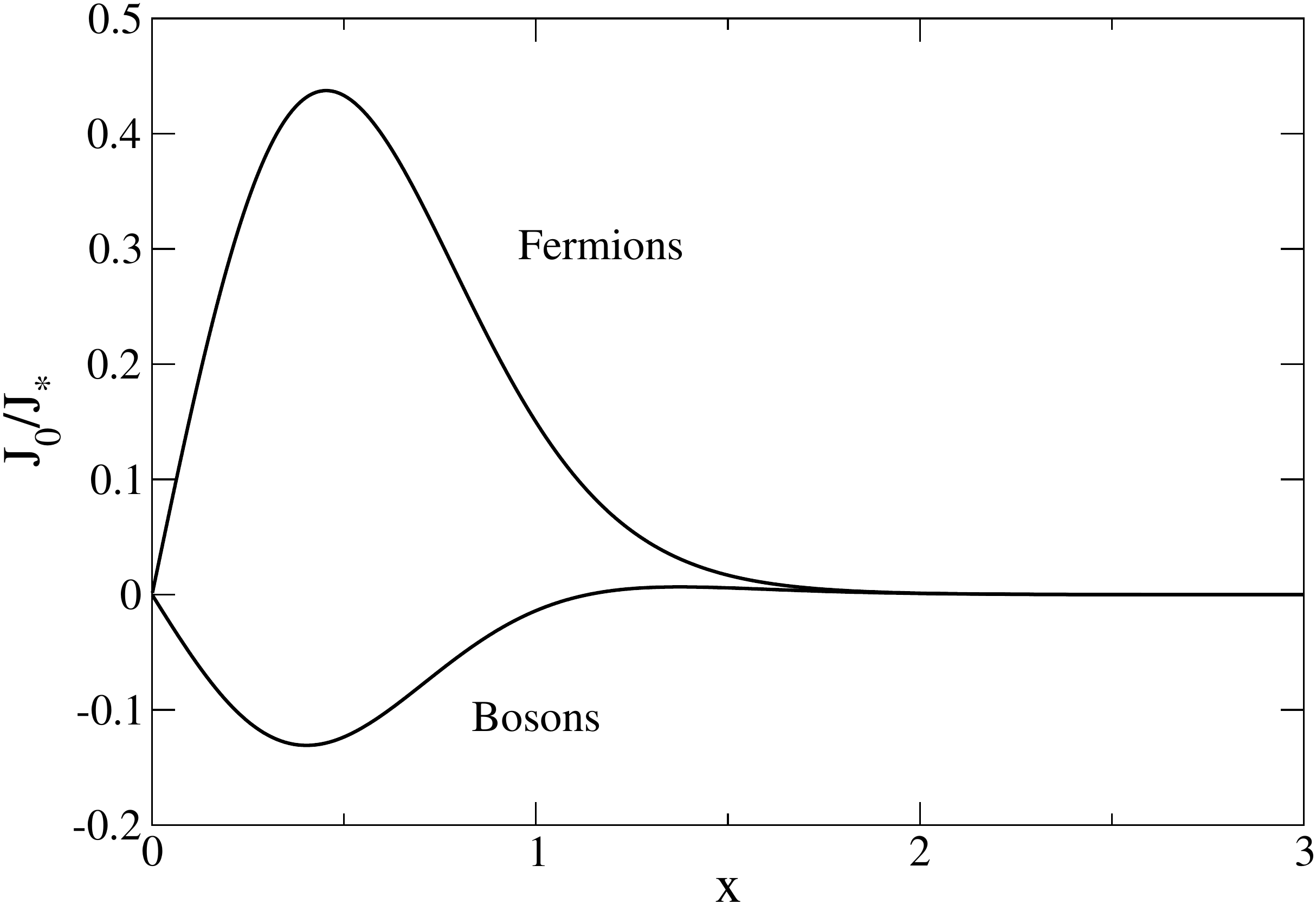}
\caption{Normalized initial
current of collisions as a function of the normalized velocity
$x=v/\sqrt{2}\sigma$. The collisions reduce the DF at small $v$ in the case of
fermions ($J_0>0$) and increase the DF at small $v$ in the case of
bosons ($J_0<0$).}
\label{F}
\end{center}
\end{figure}

{\it Remark:} It is interesting to calculate the initial current of collisions
\begin{eqnarray}
\label{qles15}
(J_i)_0=-\left \lbrack D_{ij}\frac{\partial f}{\partial v_j}- f \left
(1-\kappa
\frac{f}{\eta_0}\right ) F_i^{\rm pol} \right \rbrack_{t=0}
\end{eqnarray}
in the case of fermions and bosons. Assuming that the initial DF is 
Maxwellian (interpreted as an out-of-equilibrium DF)
\begin{eqnarray}
\label{qles16}
f_0({\bf v})=\frac{\rho}{(2\pi \sigma^2)^{3/2}}
e^{-\frac{v^2}{2\sigma^2}},
\end{eqnarray}
and noting that the Maxwellian DF is the stationary solution of the classical
Landau equation ($\kappa=0$), we conclude that the initial current of quantum
particles involves only terms proportional to $\kappa$. Therefore, it takes
the form
\begin{eqnarray}
\label{qles17}
(J_i)_0=-\left \lbrack D_{ij}^{(\kappa)}\frac{\partial f}{\partial v_j}+\kappa
\frac{f^2}{\eta_0} F_i^{\rm pol} \right \rbrack_{t=0},
\end{eqnarray}
where $F_i^{\rm pol}$ is given by Eq. (\ref{q5}) without the subscript $b$ and
\begin{eqnarray}
\label{qles18}
D_{ij}^{(\kappa)}=- 2\pi \kappa G^2 m \ln\Lambda \int d{\bf v}'
K_{ij} \frac{{f'_b}^2}{\eta_b}.
\end{eqnarray}
We can therefore use the results of Sec. \ref{sec_gbmt} both for bosons and
fermions. We get
\begin{eqnarray}
\label{qles19}
(J_i)_0=-\left \lbrack - D_{\|}^{(\kappa)}f\frac{v_i}{\sigma^2}+\kappa
\frac{f^2}{\eta_0} F_i^{\rm pol} \right \rbrack_{t=0},
\end{eqnarray}
where $D_{\|}^{(\kappa)}$ and $F_i^{\rm pol}$ are given by Eqs. (\ref{m31}) and
(\ref{m35}) without 
the subscript $b$, and we have used Eq. (\ref{qles16}) to calculate the
derivative. Introducing the variable $x$ defined by Eq. (\ref{m36}) and writing
${\bf J}_0=J_0 {\bf v}/v$, the normalized initial current of collisions is
given by
\begin{eqnarray}
\label{qles20}
\frac{J_0}{J_*}=e^{-x^2}\left\lbrack G(\sqrt{2}x)+\kappa\,
2\sqrt{2}e^{-x^2}G(x)\right\rbrack
\end{eqnarray}
with
\begin{eqnarray}
\label{qles21}
J_*=\frac{\sqrt{2}\pi G^2\rho^3\hbar^3\ln\Lambda}{g m^3\sigma^8}.
\end{eqnarray}
Using Eq. (\ref{c35}), it can be written explicitly as
\begin{eqnarray}
\label{qles22}
\frac{J_0}{J_*}=\frac{1}{4x^2}\left\lbrack e^{-x^2}{\rm erf}(\sqrt{2}x)
+4\sqrt{2}\kappa e^{-2x^2}{\rm erf}(x)-2\sqrt{\frac{2}{\pi}}(1+4\kappa) x
e^{-3x^2}\right\rbrack.
\end{eqnarray}
It is represented in Fig. \ref{F}. In the case of fermions, we see that the
collisions tend to reduce the value of the DF for small $v$ (where $f$ is
large) since $J_0/J_*\sim 2(2/\pi)^{1/2}x >0$. This is a dynamical consequence
of Pauli's blocking. Inversely, in the
case of bosons,  collisions tend to increase the value of the DF for small
$v$ since $J_0/J_*\sim -(2/3)(2/\pi)^{1/2}x<0$.
This is a dynamical consequence of 
Bose stimulation. Finally, we note that, in this particular
situation, the typical timescale of evolution is given by Eq. (\ref{qle11}) [see
Eq.
(\ref{qles21})] both
for fermions and bosons (without having to assume $f/\eta_0\gg 1$). This
timescale should not be interpreted, however, as a time of
relaxation towards thermal equilibrium.

\subsection{Classical Landau equation in energy space}
\label{sec_cles}

The classical Landau equation (\ref{cle2}) is recovered from Eq. (\ref{qles1})
in
the dilute limit $f\ll \eta_0$ where we can make the approximation
\begin{eqnarray}
\label{cles1}
f \left (1-\kappa\frac{f}{\eta_0}\right )\simeq  f.
\end{eqnarray}
If the DF is isotropic, i.e., $f=f(v,t)$, the Landau equation can be written
explicitly as\footnote{This
equation is implicit in the papers of Chandrasekhar \cite{chandra1} and Rosenbluth
{\it et al.} \cite{rosen}. It was first explicitly written by MacDonald {\it
et al.} \cite{macdo} and King
\cite{kingL}.}
\begin{eqnarray}
\label{cles2}
\frac{\partial f}{\partial t}=16\pi^2
G^2m\ln\Lambda\frac{1}{v^2}\frac{\partial}{\partial
v}\left ( A\frac{\partial f}{\partial v}+B f\right )
\end{eqnarray}
with
\begin{eqnarray}
\label{cles3}
A(v,t)= \frac{1}{3v}\int_0^v
v_1^4 f(v_1,t) \,
dv_1+\frac{v^2}{3}\int_v^{+\infty} v_1 f(v_1,t) \, dv_1
\end{eqnarray}
and
\begin{eqnarray}
\label{cles4}
B(v,t)=\int_0^v f(v_1,t) v_1^2\,
dv_1.
\end{eqnarray}
In  terms
of the energy, i.e., for $F=F(\epsilon,t)$, it takes the form
\begin{eqnarray}
\label{cles5}
\frac{\partial F}{\partial t}=16\pi^2 G^2m^3\ln\Lambda\frac{\partial}{\partial
\epsilon}\left ( A\frac{\partial F}{\partial
\epsilon}-\frac{AF}{2\epsilon}+\frac{BF}{\sqrt{2m\epsilon}} \right )
\end{eqnarray}
with
\begin{eqnarray}
\label{cles6}
A(\epsilon,t)= \frac{1}{6\pi\sqrt{2m\epsilon}}\int_0^\epsilon
\epsilon_1 F(\epsilon_1,t) \,
d\epsilon_1+\frac{\epsilon}{6\pi}\int_\epsilon^{+\infty}
\frac{1}{\sqrt{2m\epsilon_1}} F(\epsilon_1,t)\, d\epsilon_1
\end{eqnarray}
and
\begin{eqnarray}
\label{cles7}
B(\epsilon,t)=\frac{1}{4\pi}\int_0^\epsilon  F(\epsilon_1,t) \,
d\epsilon_1.
\end{eqnarray}

\subsection{Kramers-Chandrasekhar equation in energy space}
\label{sec_tbes}

We now assume that the field particles are at statistical equilibrium and make a
thermal bath approximation. This transforms a self-consistent
integrodifferential  equation (Landau) into 
a differential equation (Fokker-Planck). Using the
fact that the functions $A$ and $B$ are defined by
\begin{eqnarray}
\label{tbes1}
D_{\|}=16\pi^2G^2m\ln\Lambda\frac{1}{v^2}A,\qquad
F_{\rm pol}=-16\pi^2G^2m\ln\Lambda\frac{1}{v^2}B,
\end{eqnarray}
the Einstein relation from Eq. (\ref{q25}) can be rewritten as
\begin{eqnarray}
\label{tbes2}
B=A\beta m  v.
\end{eqnarray}

In the thermal bath approximation, using Eq. (\ref{tbes2}), the quantum Landau
equation (\ref{qles4}) is transformed into the quantum Kramers-Chandrasekhar
equation
\begin{eqnarray}
\label{tbes3}
\frac{\partial f}{\partial t}=\frac{1}{v^2}\frac{\partial}{\partial
v}\left \lbrace v^2 D_{\|} \left \lbrack\frac{\partial f}{\partial v}+\beta m  f
\left
(1-\kappa
\frac{f}{\eta_0}\right ) v \right \rbrack\right \rbrace,
\end{eqnarray}
where $D_{\|}$ is given by Eq. (\ref{q36}). For classical particles
($\kappa=0$), Eq. (\ref{tbes3}) reduces to the Kramers-Chandrasekhar
\begin{eqnarray}
\label{tbes4}
\frac{\partial f}{\partial t}=\frac{1}{v^2}\frac{\partial}{\partial
v}\left \lbrack v^2 D_{\|} \left (\frac{\partial f}{\partial v}+\beta m  f
v \right )\right \rbrack,
\end{eqnarray}
where $D_{\|}$ is given by Eq. (\ref{c32}).  For bosons
($\kappa=-1$) with $E\ll E_c$, making the approximation from Eq. (\ref{qles11}), we obtain
\begin{eqnarray}
\label{tbes3b}
\frac{\partial f}{\partial t}=\frac{1}{v^2}\frac{\partial}{\partial
v}\left \lbrack v^2 D_{\|} \left (\frac{\partial f}{\partial v}+\beta \frac{m}{\eta_0} f^2 v \right )\right \rbrack,
\end{eqnarray}
where $D_{\|}$ is given by Eq. (\ref{q34h}).

On the other hand, the quantum Landau equation in energy space (\ref{qles8}) is
transformed into the quantum Kramers-Chandrasekhar equation in energy
space
\begin{eqnarray}
\label{tbes5}
\frac{\partial F}{\partial t}=2m\frac{\partial}{\partial
\epsilon}\left \lbrace \epsilon D_{\|}\left\lbrack \frac{\partial F}{\partial
\epsilon}-\frac{F}{2\epsilon}+\beta F \left
(1-\kappa
\frac{m^{3/2}F}{4\pi\eta_0 \sqrt{2\epsilon}}\right )\right \rbrack\right
\rbrace,
\end{eqnarray}
where $D_{\|}$ is given by Eq. (\ref{q36}) with the change of variables
$\epsilon=mv^2/2$. For classical particles ($\kappa=0$), Eq. (\ref{tbes5})
reduces to
\begin{eqnarray}
\label{tbes6}
\frac{\partial F}{\partial t}=2m\frac{\partial}{\partial
\epsilon}\left \lbrack \epsilon D_{\|}\left ( \frac{\partial F}{\partial
\epsilon}-\frac{F}{2\epsilon}+\beta F \right )\right \rbrack,
\end{eqnarray}
where $D_{\|}$ is given by Eq. (\ref{c32}) with the change of variables
$\epsilon=mv^2/2$.  For bosons
($\kappa=-1$) with $E\ll E_c$, making the approximation from Eq. (\ref{qles11}), we obtain
\begin{eqnarray}
\label{tbes5b}
\frac{\partial F}{\partial t}=2m\frac{\partial}{\partial
\epsilon}\left \lbrace \epsilon D_{\|}\left\lbrack \frac{\partial F}{\partial
\epsilon}-\frac{F}{2\epsilon}+
\frac{m^{3/2}\beta F^2}{4\pi\eta_0 \sqrt{2\epsilon}}\right \rbrack\right
\rbrace,
\end{eqnarray}
where $D_{\|}$ is given by Eq. (\ref{q34h}) with the change of variables
$\epsilon=mv^2/2$.

{\it Remark:} If we make the change of variables
$\epsilon=mv^2/2$ but work in terms of $f(\epsilon,t)$ rather than in terms of
$F(\epsilon,t)$, we get
\begin{eqnarray}
\label{tbes7}
\frac{\partial f}{\partial
t}=2m\frac{1}{\sqrt{\epsilon}}\frac{\partial}{\partial
\epsilon}\left \lbrace \epsilon^{3/2} D_{\|} \left \lbrack\frac{\partial
f}{\partial \epsilon}+\beta  f
\left
(1-\kappa
\frac{f}{\eta_0}\right )  \right \rbrack\right \rbrace,
\end{eqnarray}
where $D_{\|}$ is given by Eq. (\ref{q36}) with the change of variables
$\epsilon=mv^2/2$.  For classical particles ($\kappa=0$), Eq. (\ref{tbes7})
reduces to
\begin{eqnarray}
\label{tbes8}
\frac{\partial f}{\partial
t}=2m\frac{1}{\sqrt{\epsilon}}\frac{\partial}{\partial
\epsilon}\left \lbrack \epsilon^{3/2} D_{\|} \left (\frac{\partial
f}{\partial \epsilon}+\beta  f \right )\right \rbrack,
\end{eqnarray}
where $D_{\|}$ is given by Eq. (\ref{c32}) with the change of variables
$\epsilon=mv^2/2$. Since $D_{\|}\propto v^{-3}$ when $v\rightarrow +\infty$ (see Sec. \ref{sec_qkm}), we see that $\epsilon^{3/2} D_{\|}\rightarrow {\rm
cst}$ when
$\epsilon\rightarrow +\infty$. With this approximation, the stationary solution
of Eq. (\ref{tbes7}), corresponding to a constant current $J$, returns
the quantum King model of Sec. \ref{sec_qkm}.\footnote{For classical particles,
the stationary solution of Eq. (\ref{tbes8}) or, equivalently, Eq.
(\ref{tbes4}) leads to the King model \cite{kingmodel} without
being required to make the approximation $\epsilon^{3/2} D_{\|}\rightarrow {\rm
cst}$ (see \cite{king} and  Appendix D of \cite{lcstar}).} We note that Eq.
(\ref{tbes7}) is an exact equation for the time-dependent evolution of
$f(\epsilon,t)$. However, it is restricted to spatially homogeneous
self-gravitating systems.

\section{Conclusion}
\label{sec_conclusion}

In this paper, we have developed the kinetic
theory of self-gravitating fermions and bosons. Although we have worked at a
general level, our study can have applications in the context of fermionic or
bosonic DM. For simplicity, we have considered an infinite homogeneous system
and we have neglected collective effects. We have studied the self-consistent
evolution of the system ``as a whole'' and the relaxation of test particles
experiencing collisions with field particles that are either at statistical
equilibrium (thermal bath) or in an out-of-equilibrium quasistationary state. We
have shown that the quantum  nature of the particles affects the relaxation time
with respect to the case of classical particles (assuming for comparison  that
they have the same mass and the same DF). The diffusion of the test particles is
different depending on whether the field particles are classical or quantum. In
this respect, quantum  particles of mass $m_b$ are  equivalent to classical
quasiparticles of mass $m_{\rm eff}$.\footnote{In the case of bosons, these
quasiparticles \cite{hui} correspond to incoherent granular density
fluctuations produced by wave interference \cite{ch2,ch3}. In the case of
fermions, they are more difficult to interpret physically.} The heating time due
to collisions with quantum particles is larger than the heating time due to
collisions with classical particle if the quantum particles are fermions
($m_{\rm eff}<m_b$) and smaller if they are bosons ($m_{\rm eff}>m_b$). The
friction by polarisation experienced by a test particle has the same expression
for classical and quantum field particles (except for a small change in the
Coulomb logarithm). As a result, the cooling time is not affected by the quantum
nature of the particles. This confirms the results obtained by
Bar-Or {\it et al.} \cite{bft} in the context of FDM. For
FDM, the regime of most interest is when $T\ll T_c$. In that case, $m_{\rm
eff}\gg m_b$ and the heating time is strongly reduced with respect to CDM,
becoming smaller than the age of the universe.  Therefore, collisional effects
affect the evolution of test particles (stars, globular clusters, black
holes...) traveling through  FDM halos \cite{bft}. They  also induce the secular
evolution of the halo and the slow growth of the mass of the soliton which is
fed by the halo \cite{levkov2}. A detailed description of the interaction
between the soliton and the halo should be considered in future works.

We have shown that the results of Levkov {\it et al.} \cite{levkov2}
and Bar-Or {\it et al.} \cite{bft}  could be understood in a unified manner from
the quantum Landau equation (\ref{qle5}) introduced in \cite{kingen}. In the
case of bosons, this equation is more general than the one considered
by Levkov {\it et al.} \cite{levkov2} because we do not necessarily make the
approximation $f(1+{f}/{\eta_0})\simeq  {f^2}/{\eta_0}$. As a result, the
bosonic Landau equation (\ref{qle5}) relaxes towards the Bose-Einstein DF when
$E>E_c$ and displays the phenomenon of Bose-Einstein condensation when $E<E_c$.
In the regime  $E\ll E_c$, relevant to FDM, we can make the approximation
$f(1+{f}/{\eta_0})\simeq  {f^2}/{\eta_0}$ and we recover the results of Levkov
{\it et al.} \cite{levkov2}. On the other hand, if we make a bath approximation,
we can derive from the  quantum Landau equation (\ref{qle5}) a
nonlinear Fokker-Planck
equation which generalizes the classical Kramers-Chandrasekhar equation. If the
field particles are at statistical equilibrium (thermal bath), the diffusion and
friction coefficients satisfy an Einstein relation and the Fokker-Planck
equation relaxes towards the Fermi-Dirac or Bose-Einstein DF. If the field
particles are not at statistical equilibrium, the Fokker-Planck equation relaxes
towards a different DF. In the case of FDM, if we assume that the bosons have an
out-of-equilibrium Maxwellian DF (resulting from a process of collisionless
violent relaxation) and make the approximation $f(1+{f}/{\eta_0})\simeq
{f^2}/{\eta_0}$, we recover the results of Bar-Or {\it et al.} \cite{bft} that
were obtained in a different manner. Although we
have focused in this paper on the
kinetic theory of quantum particles because of its possible applications to
fermionic and bosonic DM, it is shown in  \cite{kingen} that the results
of kinetic theory can be extended to other situations where the system is
described by a generalized
entropy (e.g., the Tsallis entropy). On the other hand, it
would be interesting to generalize the quantum kinetic  theory to inhomogenous
systems and take collective effects into account in order to extend the
classical inhomogeneous Lenard-Balescu equation \cite{heyvaerts,angleaction2} to
the quantum realm.

\vskip0.5cm

{\it Note added:} While this paper was in course of finalization,
I came across the very interesting paper of Bar-Or {\it et al.} \cite{bft2} who,
independently, derived similar results. In particular, they derived the bosonic
Landau equation (\ref{qle5}) from a heuristic approach (equivalent to the one
presented in \cite{kingen}) and also directly from the Schr\"odinger-Poisson or
Wigner-Poisson equations by using a quasilinear theory. They used this equation
to directly obtain
the coefficients of diffusion and friction as done in the present paper (and
previously in Ref. \cite{kingen} at a general level). The scope of the two
papers is relatively different so they complete each other. A draft of the
present paper was transmitted to J.B. Fouvry on 21 October
2020 and I thank him for his kind correspondence.

\appendix

\section{Application of the kinetic theory to FDM: validity of the approximations}
\label{sec_fdm}

The kinetic theory developed in this paper can be applied in the context of
fermionic and bosonic DM. In this Appendix, we focus on the case of bosonic DM.
It has been proposed that DM may be made of bosons like ultralight axions with a
mass of the order of $m_b\sim 10^{-22}\, {\rm eV/c^2}$ (see, e.g., \cite{hui}).
For such a small mass, the wave nature of the bosons manifests itself at the
scale of DM halos. This has been called FDM \cite{hu}. As recalled in the
Introduction, two situations have been considered in the literature.

In the first situation, studied by Levkov {\it et al.} \cite{levkov2},
it is assumed that the gas of bosons is spatially homogeneous  and dynamically
stable (virialized). In that case, it can only evolve under the effect of
gravitational interactions (``collisions''). This is a  slow process taking
place  on a secular timescale. The evolution of the DF $f({\bf v},t)$ is
governed by the self-consistent bosonic Landau equation (\ref{qles8}). In most
applications of astrophysical interest, we are far below the condensation
temperature so we can make the approximation $f(1+f/\eta_0)\simeq f^2/\eta_0$
leading to the simplified self-consistent bosonic Landau equation
(\ref{qles12}). This equation describes the condensation of the boson gas. As
shown by Levkov {\it et al.} \cite{levkov2}, this condensation process leads to
the formation of a Bose star associated  with the emergence of a Dirac peak in
the velocity DF  (i.e., in $k$-space) surrounded by a homogeneous (or weakly
inhomogeneous) halo of
uncondensed bosons that feeds it.

In the second situation, studied
by \cite{ch2,ch3,schwabe,mocz,moczSV,veltmaat,moczprl,moczmnras,veltmaat2}, the
system is
initially
unsteady or dynamically unstable so it rapidly becomes spatially inhomogeneous.
In that case, it experiences two successive kinds of relaxation. In a first
regime, it undergoes a process of gravitational cooling
\cite{seidel94,gul0,gul} and violent collisionless relaxation \cite{lb} on a few
dynamical times, arising from the strong fluctuations of the gravitational
potential accompanying the formation of the DM halo. This leads to a DM halo
with a core-halo structure made of a quantum core (soliton) and an approximately
isothermal halo (see the Introduction for more details). The halo has a granular
structure arising from wave interferences \cite{ch2,ch3}. It is made of
quasiparticles of large mass $m_{\rm eff}\gg m_b$ due to Bose stimulation
\cite{hui}. These quasiparticles induce a slow collisional relaxation of the DM
halo on a secular timescale like in the case of globular clusters. The initial
DF of the halo can be approximated by the Maxwell-Boltzmann distribution with a
velocity dispersion $\sigma_b$ corresponding to an effective temperature $T_{\rm
eff}\ll T_c$. This is an out-of-equilibrium DF arising from the process of
violent relaxation.\footnote{This is a state of dynamical
equilibrium with respect to a collisionless evolution (virialized state or
quasistationary state) but not a state of thermodynamical equilibrium with
respect to a collisional relaxation.} In a sense, this isothermal DF can be
justified by Lynden-Bell's statistical theory. Treating the halo as
approximately homogeneous (i.e., making a local approximation), its secular
evolution is governed by the self-consistent bosonic Landau equation
(\ref{qles8}), or its simplified form (\ref{qles12}), as discussed above. The DF
of the halo is expected to slowly change with time due to gravitational
interactions. Since $T_{\rm eff}\ll T_c$, it also undergoes a process of
condensation feeding the solitonic core. This corresponds to the problem studied
by Levkov {\it et al.} \cite{levkov2} except that the system (core $+$ halo) is fundamentally
inhomogeneous in the present case. On the other hand, if we consider the
evolution of classical test particles in the halo, like stars, globular clusters
of black holes, their evolution is governed by a classical Fokker-Planck
equation that can be obtained from the Landau equation (\ref{m1}) by making a
bath approximation, i.e.,  by replacing the DF of the bosonic field particles by
their (out-of-equilibrium) Maxwell-Boltzmann DF. Since the amplification factor
is large we can also make the approximation $f(1+f/\eta_0)\simeq f^2/\eta_0$.
This is the situation considered by Bar-Or {\it et al.} \cite{bft} who studied
different applications of the kinetic theory to DM halos.

Let us now make orders of magnitude estimates in order to justify the approximations that are relevant for FDM. The DF of the halo [see Eq. (\ref{m22})] can be estimated by
\begin{eqnarray}
\label{fdm1}
f_b\sim \frac{\rho_b}{\sigma_b^3},
\end{eqnarray}
where $\rho_b$ is the mass density of the halo and $\sigma_b$ is the velocity dispersion of the bosons that compose it. Recalling that
\begin{eqnarray}
\label{fdm2}
\eta_b=\frac{m_b^4}{h^3},
\end{eqnarray}
the amplification factor due to Bose stimulation is given by
\begin{eqnarray}
\label{fdm3}
\chi\equiv \frac{f_b}{\eta_b}=\frac{\rho_b}{\eta_b\sigma_b^3}=\frac{\rho_b
h^3}{\sigma_b^3
m_b^4}.
\end{eqnarray}
Assuming that the halo is approximately isothermal with an effective temperature $T_{\rm eff}$, we have
\begin{eqnarray}
\label{fdm4}
\sigma_b^2\sim \frac{k_B T_{\rm eff}}{m_b}.
\end{eqnarray}
On the other hand, recalling the expression of the condensation temperature $T_c$ from Eq. (\ref{bec3}), we can write
\begin{eqnarray}
\label{fdm5}
\sigma_c^2\sim \frac{k_B T_c}{m_b}\sim \left (\frac{\rho_b}{\eta_b}\right
)^{2/3}.
\end{eqnarray}
Therefore, the amplification factor can be written in terms of the effective temperature as
\begin{eqnarray}
\label{fdm6}
\chi\sim \left (\frac{T_c}{T_{\rm eff}}\right )^{3/2}\sim  \left
(\frac{\sigma_c}{\sigma_b}\right )^{3}.
\end{eqnarray}
The condition that the amplification due to Bose stimulation is important can be expressed as
\begin{eqnarray}
\label{fdm7}
\chi\gg 1   \qquad \Leftrightarrow \qquad f_b\gg\eta_b \qquad \Leftrightarrow
\qquad T_{\rm eff}\ll T_c.
\end{eqnarray}
We will see that this condition is fulfilled in FDM halos. We can also obtain a more explicit expression of $\chi$ as follows. The velocity dispersion of the bosons in the halo can be estimated from the virial theorem yielding
\begin{eqnarray}
\label{fdm8}
\sigma_b^2\sim \frac{GM_h}{r_h},
\end{eqnarray}
where $M_h$ is the halo mass and $r_h$ is the halo radius. They are connected by the relation
\begin{eqnarray}
\label{fdm9}
M_h\sim \Sigma_0 r_h^2,
\end{eqnarray}
where $\Sigma_0=141\, M_{\odot}/{\rm pc}^2$ is the universal surface density of
the DM halos (see,
e.g., \cite{modeldm} for a more detailed discussion). Therefore,
$\sigma_b^2\sim G\sqrt{\Sigma_0 M_h}$. Substituting this relation into Eq.
(\ref{fdm3}) and using $\rho_b\sim
{M_h}/{r_h^3}\sim \Sigma_0^{3/2}/M_h^{1/2}$, we get
\begin{eqnarray}
\label{fdm10}
\chi\sim \frac{\Sigma_0^{3/4}h^3}{G^{3/2}m_b^4M_h^{5/4}}.
\end{eqnarray}
This equation determines the amplification factor $\chi$ in  a halo of mass
$M_h$.\footnote{If we apply the condition $\chi\gg 1$ to the ``minimum halo''
of mass $(M_h)_{\rm min}\sim (\hbar^4\Sigma_0/G^2m^4)^{1/3}\sim 10^8\,
M_{\odot}$ \cite{mcmh}, we get the condition $m_{b}\ll
(\hbar^4\Sigma_0/G^2)^{1/7}\sim 10^{10}\, {\rm GeV}/c^2$. Using $\Sigma_0\sim
c\sqrt{\Lambda}/G$ \cite{mcmh}, this condition can be rewritten as $m_b\ll
(\hbar^4c\sqrt{\Lambda}/G^3)^{1/7}$, where $\Lambda=1.00\times 10^{-35}\, {\rm
s^{-2}}$ is the cosmological constant.} When $\chi\gg 1$, the
halo acts as if it
were composed of quasiparticles with effective mass (see Sec. \ref{sec_gbm})
\begin{eqnarray}
\label{fdm11}
m_{\rm eff}\sim \chi m_b\sim \frac{\rho_b h^3}{\sigma_b^3m_b^3}\sim \rho_b \lambda_{\rm dB}^3,
\end{eqnarray}
where  $\lambda_{\rm dB}=h/(m_b\sigma_b)$ is the de
Broglie length. We see that $\chi$ is equal to the
occupation number ${\cal N}= n_b \lambda_{\rm dB}^3$ (number of bosons in
the de Broglie sphere) so that $m_{\rm eff}\sim {\cal N}m_b$. For $\chi\gg1$,
the effective mass $m_{\rm eff}$ of the quasiparticles is much larger than the
mass $m_b$ of the bosons. This is a striking effect of Bose stimulation.
The number of quasiparticles in the DM halo is $N_{\rm eff}=N/m_{\rm eff}$ and the relaxation time induced by these quasiparticles can be written as (see Sec. \ref{sec_qle})
\begin{eqnarray}
\label{fdm12}
t_R\sim \frac{N_{\rm eff}}{\ln N_{\rm eff}}t_D,
\end{eqnarray}
where $t_D$ is the dynamical time
\begin{eqnarray}
\label{fdm13}
t_D\sim \frac{R}{\sigma}\sim \frac{1}{(G\rho)^{1/2}}.
\end{eqnarray}
Because of  Bose stimulation, the effective number of particles and the collisional relaxation time of FDM halos are strongly reduced as compared to classical CDM halos.

{\it Numerical application:} For a FDM halo of mass $M_h\sim 10^{11}\,
M_{\odot}$, radius $r_h\sim 30\, {\rm kpc}$, density $\rho_b\sim 4\times
10^{-3}\, M_{\odot}/{\rm pc}^3$ and velocity dispersion $\sigma_b\sim 100\, {\rm
km/s}$ (medium spiral), and
for a boson mass
$m_b\sim 10^{-22}\, {\rm eV/c^2}$, we get, $T_{\rm eff}\sim 10^{-25}\, {\rm K}$ 
 and $T_c\sim 10^{38}\, {\rm K}$.\footnote{The very small temperature $T_{\rm
eff}\sim 10^{-25}\, {\rm K}$ of the bosons is clearly unphysical confirming that
$T_{\rm eff}$ is an effective temperature. On the other hand, since $T_{\rm
eff}\ll T_c$, the isothermal halo is an out-of-equilibrium structure otherwise
it would be condensed (or have a non-Maxwellian DF corresponding to the
Bose-Einstein DF below $T_c$). As we have previously indicated, the isothermal
DF of the halo is established by a process of violent collisionless relaxation
\cite{lb}, not by a collisional process. FDM is a striking example where the
out-of-equilibrium isothermal Lynden-Bell DF is very different from the thermal
equilibrium state of the system. For classical particles, the Lynden-Bell DF and
the Maxwell-Boltzmann DF  are similar (in the nondegenerate limit) even though
they have a very different interpretation and a very different domain of
validity.} This yields $\chi\sim (T_c/T_{\rm eff})^{3/2}\sim 
10^{94}\gg 1$. Therefore, the amplification factor  due to Bose stimulation is
gigantic. The description of FDM halos in terms of quasiparticles is fully
justified, as well as the approximation $f(1+f/\eta_b)\simeq f^2/\eta_b$. The de
Broglie length is $\lambda_{\rm
dB}\sim 1\, {\rm kpc}$, corresponding to the typical size of the
soliton.\footnote{The mass-radius relation of the soliton is $M_cR_c\sim
h^2/Gm_b^2$ (see, e.g., \cite{prd1}). Using the velocity dispersion tracing
relation $\sigma_b^2\sim GM_c/R_c\sim GM_h/R_h$ \cite{mcmh}, we get $R_c\sim
h/m_b\sigma_b=\lambda_{\rm dB}$.} 
The occupation number is ${\cal N}=n_b\lambda_{\rm dB}^3=10^{94} \gg 1$
justifying to treat the wavefunction as a classical field. The effective mass
of the
quasiparticles is $m_{\rm eff}\sim {\cal N} m_b\sim 10^6\, M_{\odot}$. Because
of Bose
stimulation, the quasiparticles are much heavier than the bosons. The number of
bosons in a FDM halo is $N\sim  10^{99}$ but the number of
quasiparticles is only $N_{\rm eff}\sim 10^5$, which is comparable to the
number of stars in a globular cluster.  The dynamical time is $t_D\sim 10^8\,
{\rm yrs}$ and  the relaxation time is
$t_R\sim 10^{12}\, {\rm yrs}$, which is comparable to the age of the
Universe (the relaxation time due to particles of mass $m_b\sim 10^{-22}\, {\rm
eV/c^2}$ would be $t_R\sim 10^{105}\, {\rm yrs}$). Therefore,
``collisions'' of quasiparticles  in FDM halos are relevant on a secular
timescale of the order of the age of the universe, similarly to ``collisions''
of stars in globular clusters (recall that our numerical application is
indicative since the effective mass $m_{\rm eff}$ depends on the position and is
larger in regions of high density implying a shorter relaxation time in the
core of the galaxy). Finally, as a comparison, we note that $\chi\sim 0.1$  for
fermions of mass $m\sim 100\, {\rm eV/c^2}$, so that quantum effects --
Pauli's exclusion principle -- are important in the core of the galaxy (where
the density is higher than the average estimate made above).

\section{Conservation laws and  $H$-theorem}
\label{sec_cl}

In this Appendix, we establish general properties of the multi-species quantum
Landau equation (\ref{qle6}).

\subsection{Conservation of the mass of each species}

Since $\partial f_a/\partial t$ is the divergence of a current, the mass of
each species
\begin{eqnarray}
\label{cl1}
M_{a}=\int f_a\, d{\bf r}d{\bf v}
\end{eqnarray}
is conserved.

\subsection{Conservation of the total impulse}

The quantum Landau equation (\ref{qle6}) conserves the total impulse
\begin{eqnarray}
\label{cl2}
{\bf P}=\sum_a\int f_a {\bf v}\, d{\bf r}d{\bf v}.
\end{eqnarray}
For each component we have
\begin{eqnarray}
\dot P_k=\sum_a\int d{\bf r}d{\bf v}\, v_k \frac{\partial
f_a}{\partial t},
\end{eqnarray}
hence
\begin{eqnarray}
\label{cl3}
\dot P_k
&=&2\pi G^2\ln\Lambda\sum_{a,b}\int
d{\bf r}d{\bf v}\,  v_k
 \frac{\partial}{\partial
v_i}\int d{\bf v}' K_{ij}\left\lbrace m_b f'_b \left (1-\kappa_b
\frac{f'_b}{\eta_b}\right )\frac{\partial f_a}{\partial v_j}- m_a f_a \left
(1-\kappa_a
\frac{f_a}{\eta_a}\right )\frac{\partial f'_b}{\partial
v'_j}\right\rbrace\nonumber\\
&=&-2\pi G^2\ln\Lambda\sum_{a,b}\int
d{\bf r}d{\bf v}d{\bf v}'\, K_{kj}\left\lbrace m_b f'_b \left
(1-\kappa_b
\frac{f'_b}{\eta_b}\right )\frac{\partial f_a}{\partial v_j}- m_a f_a \left
(1-\kappa_a
\frac{f_a}{\eta_a}\right )\frac{\partial f'_b}{\partial
v'_j}\right\rbrace\nonumber\\
&=&2\pi G^2\ln\Lambda\sum_{a,b}\int
d{\bf r}d{\bf v}d{\bf v}'\,  K_{kj}\left\lbrace m_b f'_b \left
(1-\kappa_b
\frac{f'_b}{\eta_b}\right )\frac{\partial f_a}{\partial v_j}- m_a f_a \left
(1-\kappa_a
\frac{f_a}{\eta_a}\right )\frac{\partial f'_b}{\partial
v'_j}\right\rbrace=0.
\end{eqnarray}
The third line is obtained by integrating by parts and the conservation law
results from the antisymmetry of the collision current with respect to the
interchange $(a,{\bf v})\leftrightarrow  (b,{\bf v}')$.

\subsection{Conservation of the total energy}

The quantum Landau equation (\ref{qle6}) conserves the total
energy\footnote{There is no 
potential (gravitational) energy since we have assumed that the system is
infinite and homogeneous.}
\begin{eqnarray}
\label{cl4}
E=\sum_a\int f_a\frac{v^2}{2}\, d{\bf r}d{\bf v}.
\end{eqnarray}
We have
\begin{eqnarray}
\dot E&=&\sum_a\int d{\bf r}d{\bf v}\, \frac{v^2}{2} \frac{\partial
f_a}{\partial t},
\end{eqnarray}
hence
\begin{eqnarray}
\label{cl5}
\dot E &=&2\pi G^2\ln\Lambda\sum_{a,b}\int
d{\bf r}d{\bf v}\, \frac{v^2}{2}
 \frac{\partial}{\partial
v_i}\int d{\bf v}' K_{ij}\left\lbrace m_b f'_b \left (1-\kappa_b
\frac{f'_b}{\eta_b}\right )\frac{\partial f_a}{\partial v_j}- m_a f_a \left
(1-\kappa_a
\frac{f_a}{\eta_a}\right )\frac{\partial f'_b}{\partial
v'_j}\right\rbrace\nonumber\\
&=&-2\pi G^2\ln\Lambda\sum_{a,b}\int
d{\bf r}d{\bf v}d{\bf v}'\, v_i K_{ij}\left\lbrace m_b f'_b \left
(1-\kappa_b
\frac{f'_b}{\eta_b}\right )\frac{\partial f_a}{\partial v_j}- m_a f_a \left
(1-\kappa_a
\frac{f_a}{\eta_a}\right )\frac{\partial f'_b}{\partial
v'_j}\right\rbrace\nonumber\\
&=&2\pi G^2\ln\Lambda\sum_{a,b}\int
d{\bf r}d{\bf v}d{\bf v}'\, v'_i K_{ij}\left\lbrace m_b f'_b \left
(1-\kappa_b
\frac{f'_b}{\eta_b}\right )\frac{\partial f_a}{\partial v_j}- m_a f_a \left
(1-\kappa_a
\frac{f_a}{\eta_a}\right )\frac{\partial f'_b}{\partial
v'_j}\right\rbrace\nonumber\\
&=&\pi G^2\ln\Lambda\sum_{a,b}\int
d{\bf r}d{\bf v}d{\bf v}'\, u_i K_{ij}\left\lbrace m_b f'_b \left
(1-\kappa_b
\frac{f'_b}{\eta_b}\right )\frac{\partial f_a}{\partial v_j}- m_a f_a \left
(1-\kappa_a
\frac{f_a}{\eta_a}\right )\frac{\partial f'_b}{\partial
v'_j}\right\rbrace=0.
\end{eqnarray}
The third line is obtained by integrating by parts and the conservation
law results from the interchange $(a,{\bf v})\leftrightarrow  (b,{\bf v}')$
and the identity $K_{ij}u_j=0$.

\subsection{The Fermi-Dirac or Bose-Einstein distribution is a stationary
solution}

The Fermi-Dirac or Bose-Einstein distribution
\begin{eqnarray}
\label{cl6}
f_a^{\rm eq}=\frac{\eta_a}{\lambda_a e^{\beta m_a ({\bf v}-{\bf
U})^2/2}+\kappa_a},
\end{eqnarray}
where ${\bf U}$ and $\beta$ are the same for all the species,
is a steady state of the quantum Landau equation (\ref{qle6}). When the DF is
given by Eq. (\ref{cl6}) we can easily establish that
\begin{eqnarray}
\label{cl7}
\frac{\partial f_a}{\partial {\bf v}}=-f_a  \left
(1-\kappa_a \frac{f_a}{\eta_a}\right ) \beta m_a ({\bf v}-{\bf U}).
\end{eqnarray}
As a result
\begin{eqnarray}
\label{cl8}
\left\lbrace m_b f'_b \left (1-\kappa_b
\frac{f'_b}{\eta_b}\right )\frac{\partial f_a}{\partial v_j}- m_a f_a \left
(1-\kappa_a
\frac{f_a}{\eta_a}\right )\frac{\partial f'_b}{\partial v'_j}\right\rbrace=-
\beta m_a f_a  \left
(1-\kappa_a \frac{f_a}{\eta_a}\right ) m_b f'_b \left (1-\kappa_b
\frac{f'_b}{\eta_b}\right )(v_j-v'_j).
\end{eqnarray}
Substituting this identity into Eq.  (\ref{qle6}) and using $K_{ij}u_j=0$, we
find that the DF is stationary: $\partial_t
f_a=0$.

\subsection{$H$-theorem}

The quantum Landau equation  (\ref{qle6}) satisfies an $H$-theorem for the
Fermi-Dirac or
Bose-Einstein entropy\footnote{In the classical limit $f_a/\eta_a\ll 1$, we recover the Boltzmann entropy
\begin{eqnarray}
\label{cl9class}
S=-k_B \sum_a \int \left (\frac{f_a}{m_a}\ln
\frac{f_a}{\eta_a}-\frac{f_a}{m_a}\right ) \, d{\bf r}d{\bf v}.
\end{eqnarray}
}
\begin{eqnarray}
\label{cl9}
S=-k_B \sum_a \frac{gm_a^3}{h^3}\int \left\lbrace \frac{f_a}{\eta_a}\ln
\frac{f_a}{\eta_a}+\kappa_a \left (1-\kappa_a \frac{f_a}{\eta_a}\right )\ln
\left (1-\kappa_a \frac{f_a}{\eta_a}\right )\right\rbrace \, d{\bf r}d{\bf v}.
\end{eqnarray}
We have
\begin{eqnarray}
\dot S=-k_B \sum_a \int d{\bf r}d{\bf v}\, \ln\left
(\frac{f_a/\eta_a}{1-\kappa_a f_a/\eta_a}\right )\frac{1}{m_a} \frac{\partial
f_a}{\partial t},
\end{eqnarray}
hence
\begin{eqnarray}
\label{cl10}
\dot S
&=&-2\pi k_B G^2\ln\Lambda  \sum_{a,b}\int
d{\bf r}d{\bf v}\, \ln\left
(\frac{f_a/\eta_a}{1-\kappa_a f_a/\eta_a}\right
)\frac{1}{m_a}\frac{\partial}{\partial
v_i}\int d{\bf v}' K_{ij}\left\lbrace m_b f'_b \left (1-\kappa_b
\frac{f'_b}{\eta_b}\right )\frac{\partial f_a}{\partial v_j}- m_a f_a \left
(1-\kappa_a
\frac{f_a}{\eta_a}\right )\frac{\partial f'_b}{\partial
v'_j}\right\rbrace\nonumber\\
&=&2\pi k_B G^2\ln\Lambda  \sum_{a,b}\int
d{\bf r}d{\bf v}d{\bf v}'\, \frac{1}{m_a f_a \left (1-\kappa_a
\frac{f_a}{\eta_a}\right )} \frac{\partial f_a}{\partial v_i} K_{ij}\left\lbrace
m_b f'_b \left (1-\kappa_b
\frac{f'_b}{\eta_b}\right )\frac{\partial f_a}{\partial v_j}- m_a f_a \left
(1-\kappa_a
\frac{f_a}{\eta_a}\right )\frac{\partial f'_b}{\partial
v'_j}\right\rbrace\nonumber\\
&=&-2\pi k_B G^2\ln\Lambda  \sum_{a,b}\int
d{\bf r}d{\bf v}d{\bf v}'\, \frac{1}{m_b f'_b \left (1-\kappa_b
\frac{f'_b}{\eta_b}\right )} \frac{\partial f'_b}{\partial v'_i}
K_{ij}\left\lbrace
m_b f'_b \left (1-\kappa_b
\frac{f'_b}{\eta_b}\right )\frac{\partial f_a}{\partial v_j}- m_a f_a \left
(1-\kappa_a
\frac{f_a}{\eta_a}\right )\frac{\partial f'_b}{\partial
v'_j}\right\rbrace\nonumber\\
&=&\pi k_B G^2\ln\Lambda  \sum_{a,b}\int
d{\bf r}d{\bf v}d{\bf v}'\, \frac{1}{m_a  f_a \left (1-\kappa_a
\frac{f_a}{\eta_a} \right ) m_b  f'_b \left (1-\kappa_b
\frac{f'_b}{\eta_b}\right )} X_i
K_{ij} X_j,
\end{eqnarray}
where
\begin{eqnarray}
\label{cl11}
X_i\equiv m_b f'_b \left (1-\kappa_b
\frac{f'_b}{\eta_b}\right )\frac{\partial f_a}{\partial v_i}- m_a f_a \left
(1-\kappa_a
\frac{f_a}{\eta_a}\right )\frac{\partial f'_b}{\partial
v'_i}.
\end{eqnarray}
The third line is obtained by integrating by parts and the last line results
from the interchange $(a,{\bf v})\leftrightarrow  (b,{\bf v}')$.
Since
\begin{eqnarray}
\label{cl12}
X_iK_{ij}X_j=\frac{u^2X^2-({\bf X}\cdot {\bf u})^2}{u^3}\ge 0,
\end{eqnarray}
we conclude that $\dot S\ge 0$ with equality if, and only if, ${\bf X}$ is
parallel to ${\bf u}={\bf v}'-{\bf v}$. Therefore, the entropy increases
($H$-theorem). On the other hand, we show in the next
subsection that $S$ is bounded from above so that $S\le S_{\rm max}$,
where $S_{\rm max}$ corresponds to the value of the entropy calculated with the
Fermi-Dirac or Bose-Einstein distribution $f_a^{\rm eq}$ (maximum entropy
state). Finally, we show below that $\dot S=0$ if, and only if, $f_a$ is the
Fermi-Dirac or
Bose-Einstein distribution $f_a^{\rm eq}$. As a result, we can conclude that
$f_a({\bf v},t)$ tends to the Fermi-Dirac or Bose-Einstein distribution
$f_a^{\rm eq}$ when $t\rightarrow +\infty$.\footnote{In the case of bosons, this is true only for $E\ge E_c$. For $E<E_c$, the total DF involves a Dirac-$\delta$ which accounts for the process of Bose-Einstein condensation (see Appendix \ref{sec_bec}).}

{\it Proof:} The condition $\dot S=0$ is
equivalent to the condition that ${\bf X}$ is parallel to ${\bf v}'-{\bf v}$.
Using
\begin{eqnarray}
\frac{\partial}{\partial {\bf v}}\ln \left
(\frac{f/\eta}{1-\kappa f/\eta}\right
)=\frac{1}{f(1-\kappa f/\eta)}\frac{\partial f}{\partial {\bf v}},
\end{eqnarray}
this condition can be rewritten as
\begin{eqnarray}
\label{cl13}
\frac{1}{m_a}\frac{\partial}{\partial {\bf v}}\ln \left
(\frac{f_a/\eta_a}{1-\kappa_a f_a/\eta_a}\right
)-\frac{1}{m_b}\frac{\partial}{\partial {\bf v}'}\ln \left
(\frac{f'_b/\eta_b}{1-\kappa_b f'_b/\eta_b}\right )=-A_{ab}({\bf v},{\bf
v}')({\bf v}-{\bf v}').
\end{eqnarray}
From the symmetry of the left hand side of Eq. (\ref{cl13}), it can be shown
\cite{lenard} that $A_{ab}({\bf
v},{\bf
v}')$ must be a constant $A$. This implies that
\begin{eqnarray}
\label{cl14}
\frac{1}{m_a}\frac{\partial}{\partial {\bf v}}\ln \left
(\frac{f_a/\eta_a}{1-\kappa_a f_a/\eta_a}\right
)=-A {\bf v}+{\bf B},
\end{eqnarray}
where ${\bf B}$ is another constant. Eq. (\ref{cl14}) may be rewritten as
\begin{eqnarray}
\label{cl15}
\frac{\partial}{\partial {\bf v}}\ln \left
(\frac{f_a/\eta_a}{1-\kappa_a f_a/\eta_a}\right
)=-\beta m_a ({\bf v}-{\bf U})
\end{eqnarray}
with $\beta=A$ and ${\bf U}={\bf B}/A$.
These constants are the same for all the species. After
integration, Eq. (\ref{cl15}) leads  to the Fermi-Dirac or Bose-Einstein
distribution (\ref{cl6}).

\subsection{Maximum entropy state}

From very general considerations, we expect that the statistical equilibrium
state in the microcanonical ensemble is the ``most probable'' state, i.e., the
state that maximizes the entropy at fixed total energy, total impulse and fixed mass
of each species. We thus have to solve the maximization problem
\begin{eqnarray}
\label{cl16}
\max\ \lbrace {S}\, |\,  E, {\bf P}, M_a \,\, {\rm fixed} \rbrace.
\end{eqnarray}
As extremum of entropy at fixed total energy, total impulse and fixed mass
of each species is determined by the variational problem (first
variations)
\begin{eqnarray}
\label{cl17}
\frac{\delta S}{k_B}-\beta\delta E+\beta {\bf U}\cdot \delta {\bf
P}+\sum_a
\frac{\alpha_a}{m_a}\delta M_a=0,
\end{eqnarray}
where $\beta$, ${\bf U}$ and $\alpha_a=\mu_a/k_B T$ are Lagrange multipliers
taking
into account the constraints of fixed $E$, ${\bf P}$ and $M_a$. They  have the interpretation of an inverse
temperature, a velocity of translation and a chemical potential divided by $k_B
T$. Using
\begin{eqnarray}
\label{cl18}
\delta S=-k_B \sum_a \int d{\bf r}d{\bf v}\, \ln\left
(\frac{f_a/\eta_a}{1-\kappa_a f_a/\eta_a}\right )\frac{1}{m_a}
\delta f_a,
\end{eqnarray}
\begin{eqnarray}
\label{cl19}
\delta  E=\sum_a\int d{\bf r}d{\bf v}\, \frac{v^2}{2} \delta f_a,
\end{eqnarray}
\begin{eqnarray}
\label{cl20}
\delta{\bf P}=\sum_a\int d{\bf r}d{\bf v}\, {\bf v} \delta f_a,
\end{eqnarray}
\begin{eqnarray}
\label{cl21}
\delta  M_a=\int d{\bf r}d{\bf v}\, \delta f_a,
\end{eqnarray}
we obtain
\begin{eqnarray}
\label{cl22}
\ln\left
(\frac{f_a/\eta_a}{1-\kappa_a f_a/\eta_a}\right )=-\beta m_a
\frac{v^2}{2}+\beta m_a {\bf U}\cdot {\bf v}+\alpha_a,
\end{eqnarray}
yielding the Fermi-Dirac or Bose-Einstein distribution
\begin{eqnarray}
\label{cl23}
f_a^{\rm eq}({\bf v})=\frac{\eta_a}{\lambda_a e^{\beta m_a ({\bf v}-{\bf
U})^2/2}+\kappa_a},
\end{eqnarray}
where $\lambda_a=e^{-\alpha_a-\beta m_a {\bf U}^2/2}$ in the inverse fugacity.
The second variations of
entropy write
\begin{eqnarray}
\label{cl24}
\delta^2 S=-\frac{k_B}{2}\sum_a \frac{g m_a^3}{h^3}\int \frac{\left (\frac{\delta f_a}{\eta_a}\right )^2}{\frac{f_a}{\eta_a}\left (1-\kappa_a \frac{f_a}{\eta_a}\right )}\, d{\bf r}d{\bf v},
\end{eqnarray}
and they are clearly negative. Therefore, the Fermi-Dirac or Bose-Einstein
distribution is the unique maximum of $S$ at fixed  $E$,
${\bf P}$ and  $M_a$. Hence it is a global maximum.

{\it Remark:} In the case of bosons, as in the previous section (see footnote
54), we have implicitly assumed that $E>E_c$ otherwise a Dirac peak occurs in
the DF in relation to the process of Bose-Einstein condensation.

\subsection{Equation of state}
\label{sec_aeos}

Let us consider a single species system of fermions or bosons at statistical
equilibrium and take ${\bf U}={\bf 0}$ for simplicity. In that case, the
Fermi-Dirac or
Bose-Einstein distribution writes
\begin{eqnarray}
\label{eos1}
f({\bf v})=\frac{\eta_0}{\lambda e^{\beta m
\frac{v^2}{2}}+\kappa}\qquad {\rm with}\qquad \eta_0=g\frac{m^4}{h^3}.
\end{eqnarray}
We recall that $\lambda=e^{-\alpha}$ with $\alpha=\beta\mu=\mu/k_B T$. The density $\rho=\int f\, d{\bf v}$  and the pressure $P=\frac{1}{3}\int f v^2\, d{\bf v}$  are given by
\begin{eqnarray}
\label{eos2}
\rho=\frac{4\pi\sqrt{2}\eta_0}{(\beta m)^{3/2}}I_{1/2}(\lambda),\qquad P=\frac{8\pi\sqrt{2}\eta_0}{3(\beta m)^{5/2}}I_{3/2}(\lambda),
\end{eqnarray}
where $I_n(t)$ are the Fermi or Bose integrals 
\begin{eqnarray}
\label{eos4}
I_n(t)=\int_0^{+\infty} \frac{x^n}{te^x+\kappa}\, dx,
\end{eqnarray}
with $\kappa=+1$ for fermions and $\kappa=-1$ for bosons. For fermions, they are defined for $t\ge 0$. For bosons, they are defined for $t\ge 1$ when $n>0$ and for $t>1$ when $-1<n\le 0$. We have the identity
\begin{eqnarray}
\label{iprime}
I_n'(t)=-\frac{n}{t}I_{n-1}(t).
\end{eqnarray}
The inverse fugacity
$\lambda$ is determined by the mass density
$\rho$ through the first relation of Eq. (\ref{eos2}). On the other hand, the
two relations from Eq. (\ref{eos2}) determine the quantum equation of state
$P=P(\rho)$. Finally, the velocity 
dispersion in one direction is
\begin{eqnarray}
\label{eos5}
\sigma^2=\frac{\langle
v^2\rangle}{3}=\frac{1}{3\rho}\int f v^2\,
d{\bf v}=\frac{P}{\rho}=\frac{2}{3}\frac{k_B
T}{m}\frac{I_{3/2}(\lambda)}{I_{1/2}(\lambda)}
\end{eqnarray}
and the squared speed of sound is
\begin{eqnarray}
\label{eos5b}
c_s^2=P'(\rho)=\frac{2}{\beta m}\frac{I_{1/2}(\lambda)}{I_{-1/2}(\lambda)},
\end{eqnarray}
where we have used Eq. (\ref{iprime}). The Jeans wavenumber for an infinite
homogeneous self-gravitating gas of fermions or bosons (neglecting the quantum
potential)\footnote{See \cite{prd1} for more general results taking into
account the quantum
potential.} is \cite{nyquistgrav}
\begin{eqnarray}
\label{eos5c}
k_J=\frac{\sqrt{4\pi G\rho}}{c_s}.
\end{eqnarray}
The Jeans length $\lambda_J=2\pi/k_J\sim c_s t_D$
corresponds to the typical distance traveled by a sound wave at velocity $c_s$
during the dynamical time $t_D\sim (G\rho)^{-1/2}$.

\subsection{Classical limit}
\label{sec_class}

In the classical limit, the statistical equilibrium state is determined by the Boltzmann distribution
\begin{eqnarray}
\label{eos5qq}
f({\bf v})=\frac{\eta_0}{\lambda} e^{-\beta m
\frac{v^2}{2}}.
\end{eqnarray}
The density and the pressure are given by
\begin{eqnarray}
\label{eos6}
\rho= \left (\frac{2\pi}{\beta m}\right )^{3/2}  \frac{\eta_0}{\lambda},\qquad
P=\frac{1}{2\pi} \left (\frac{2\pi}{\beta m}\right )^{5/2}
\frac{\eta_0}{\lambda},
\end{eqnarray}
leading to the classical (linear) equation of state
\begin{eqnarray}
\label{eos7}
P=\rho \frac{k_B T}{m}
\end{eqnarray}
and to the velocity dispersion
\begin{eqnarray}
\label{fermi5}
\sigma^2=\frac{\langle v^2\rangle}{3}=\frac{P}{\rho}=\frac{k_B T}{m}.
\end{eqnarray}
The squared speed of sound is $c_s^2=k_B T/m$. The classical
limit corresponds to $T\rightarrow +\infty$ (implying $\lambda\rightarrow
+\infty$ according to Eq. (\ref{eos6}), $\alpha\rightarrow -\infty$ and
$\mu\rightarrow -\infty$). The transition between the classical regime and the
quantum regime corresponds to $\lambda\sim 1$,  $f\sim\eta_0$ (i.e. $\chi\sim
1$) and $T\sim T_Q$, where the quantum temperature $T_Q$ is approximately
determined by
\begin{eqnarray}
\label{eos8}
\frac{k_B T_Q}{m}\sim \left (\frac{\rho}{\eta_0}\right )^{2/3}\qquad {\rm i.e. }\qquad T_Q\sim \frac{\hbar^2n^{2/3}}{m k_B}.
\end{eqnarray}
Physically, the classical regime  is valid
when $T\gg T_Q$, $\lambda\gg 1$ and $f\ll\eta_0$  (i.e. $\chi\ll 1$).

{\it Remark:} The
classical results can be recovered from
the general
formulae of Appendix \ref{sec_aeos} by using the
identity \cite{kingen}
\begin{eqnarray}
\label{eos9}
I_n(t)\sim \frac{1}{t}\Gamma(n+1), \qquad (t\rightarrow +\infty).
\end{eqnarray}
More precisely, for $t\gg 1$, we have the expansion
\begin{eqnarray}
\label{eos9b}
I_n(t)=\Gamma(n+1)\sum_{k=1}^{+\infty}(-\kappa)^{k-1}\left (\frac{1}{t}\right )^k \frac{1}{k^{n+1}}.
\end{eqnarray}

\subsection{Bose-Einstein condensation}
\label{sec_bec}

The Bose-Einstein DF is given by
\begin{eqnarray}
\label{bec1}
f({\bf v})=\frac{\eta_0}{\lambda e^{\beta m
\frac{v^2}{2}}-1}\qquad (T\ge T_c)
\end{eqnarray}
with $\eta_0={m^4}/{h^3}$ and
$\lambda=e^{-\alpha}=e^{-\beta
\mu}$. The value of the DF for ${\bf v}={\bf 0}$ is $f({\bf
0})=\eta_0/(\lambda-1)$ so we need to impose $\lambda\ge 1$ (i.e., $\alpha\le 0$
and $\mu\le 0$) to have $f$ positive for all $v$.  The condensation temperature $T_c$
corresponds to $\lambda=1$ (i.e., $\alpha=\mu=0$). Using Eq. (\ref{eos2}), we
obtain
\begin{eqnarray}
\label{bec3}
\frac{k_B T_c}{m}=\frac{1}{2\pi \zeta(3/2)^{2/3}}\left
(\frac{\rho}{\eta_0}\right )^{2/3}\qquad { \rm i.e.}\qquad
T_c=\frac{2\pi\hbar^2n^{2/3}}{m k_B \zeta(3/2)^{2/3}},
\end{eqnarray}
where we have used $I_{1/2}(1)=\frac{\sqrt{\pi}}{2}\zeta(3/2)$ with
$\zeta(3/2)=2.612...$. At the condensation point,
\begin{eqnarray}
\label{bec2}
f({\bf v})=\frac{\eta_0}{e^{\beta m \frac{v^2}{2}}-1}\qquad (T=T_c).
\end{eqnarray}
For $v\rightarrow 0$, we get
\begin{eqnarray}
\label{bec2b}
f({\bf v})\sim \frac{2\eta_0}{\beta m}v^{-2}\qquad (T=T_c),
\end{eqnarray}
which is similar to the Rayleigh-Jeans spectrum of radiation (see Appendix
\ref{sec_logo}). On the other hand,
the pressure and the velocity dispersion at the condensation point are
\begin{eqnarray}
\label{bec4b}
P_c=\frac{(2\pi)^{3/2}\eta_0}{(\beta_c m)^{5/2}}\zeta(5/2),
\qquad \sigma_c^2=\frac{P_c}{\rho}=\frac{\zeta(5/2)}{\zeta(3/2)}\frac{k_B
T_c}{m},
\end{eqnarray}
where we have used $I_{3/2}(1)=\frac{3\sqrt{\pi}}{4}\zeta(5/2)$ with
$\zeta(5/2)=1.341...$. The energy of condensation is $E_c=(3/2)M\sigma_c^2$. We note that $\sigma_c^2$ is of the order of $k_B T_c/m$.
This is because, at the condensation point, $P_{c}\sim P_{\rm
class}$. For $T\rightarrow T_c^+$,  using Eqs. (\ref{we1}) and
(\ref{we2}), we have $T/T_c\simeq
1+[4\sqrt{\pi}/3\zeta(3/2)]\sqrt{\ln \lambda}$ (see Eq. (\ref{nm20})) and
$c_s^2\sim [\zeta(3/2)/\sqrt{\pi}\beta_c m]\sqrt{\ln \lambda}$  (see Eq.
(\ref{eos5b})), leading to
\begin{eqnarray}
c_s^2\sim \frac{3\zeta\left (\frac{3}{2}\right )^2}{4\pi}\frac{k_B
T_c}{m}\left (\frac{T}{T_c}-1\right ).
\end{eqnarray}
The squared speed of sound vanishes at the critical temperature
(this is also the case for the Jeans length from Eq.
(\ref{eos5c})). When $T>T_c$,
the bosons are uncondensed and their DF is given by Eq.
(\ref{bec1}). When $T<T_c$, a fraction of  the bosons is
condensed in the state ${\bf v}={\bf 0}$. The total DF is
\begin{eqnarray}
\label{bec4}
f({\bf v})=\frac{\eta_0}{e^{\beta m \frac{v^2}{2}}-1}+\rho\left\lbrack 1-\left
(\frac{T}{T_c}\right )^{3/2}\right\rbrack \delta({\bf v})
\qquad (T\le T_c).
\end{eqnarray}
The first term corresponds to uncondensed bosons with $\lambda=1$ and the
second term corresponds to condensed bosons (the proportion of condensed bosons is obtained from the
normalization condition $\rho=\int f\, d{\bf v}$). At $T=0$, all the bosons are condensed:
$f=\rho\delta({\bf v})$. We note that the velocity dispersion (or pressure) vanishes at $T=0$.

The completely condensed limit corresponds to $T=0$ and $\lambda=1$ (implying
$\alpha=\mu=0$). Physically, the system is completely condensed  when $T\ll T_c$, $\lambda=1$
and $f\gg \eta_0$ (i.e., $\chi\gg 1$). The classical limit corresponds to $T\gg
T_c$, $\lambda\gg 1$ and $f\ll \eta_0$  (i.e., $\chi\ll 1$). The transition
between the completely condensed  limit and the classical limit
corresponds to $T\sim T_c$, $\lambda\sim 1$ and $f\sim\eta_0$
($\chi\sim 1$).

{\it Remark:}  The de Broglie (thermal) wavelength
of a particle is $\lambda_{\rm dB}=\sqrt{2\pi\hbar^2/mk_B T}$. Particles become
correlated with each other when their wavelengths overlap, that is, when the
thermal wavelength is greater than the mean inter-particle distance
$l=n^{-1/3}$. The condition $\lambda_{\rm dB}>l$ can be written $n\lambda_{\rm
dB}^3>1$ which yields the inequality $T<T_c$ with $T_c\sim
2\pi\hbar^2n^{2/3}/mk_B$ corresponding, up to a factor of order unity, to the condensation temperature from Eq. (\ref{bec3}). Close to the condensation temperature, i.e., for $t\rightarrow 1^+$, we have the Robinson expansion
\begin{eqnarray}
\label{we1gen}
I_{n}(t)=\Gamma(n+1)\left\lbrack \Gamma(-n)(\ln t)^n+\sum_{k=0}^{+\infty}\frac{(-1)^k}{k!}\zeta(n+1-k)(\ln t)^k\right\rbrack,
\end{eqnarray}
which is valid for $n<0$ and for non-integer $n>0$. In
particular,
\begin{eqnarray}
\label{we1}
I_{-1/2}(t)\sim \frac{\Gamma(1/2)^2}{\sqrt{\ln
t}},
\end{eqnarray}
\begin{eqnarray}
\label{we2}
I_{1/2}(t)=\Gamma(3/2)\lbrack \zeta(3/2)+\Gamma(-1/2)\sqrt{\ln t}+...\rbrack,
\end{eqnarray}
\begin{eqnarray}
\label{we3}
I_{3/2}(t)=\Gamma(5/2)\lbrack \zeta(5/2)+\Gamma(-3/2)(\ln t)^{3/2}+...\rbrack,
\end{eqnarray}
with $\Gamma(1/2)=\sqrt{\pi}$,
$\Gamma(3/2)=\sqrt{\pi}/2$, $\Gamma(-1/2)=-2\sqrt{\pi}$, $\zeta(3/2)=2.612...$,
$\Gamma(5/2)=3\sqrt{\pi}/4$, $\Gamma(-3/2)=4\sqrt{\pi}/3$ and
$\zeta(5/2)=1.341...$

\subsection{Completely degenerate fermions}
\label{sec_compdeg}

The Fermi-Dirac DF is given by
\begin{eqnarray}
\label{fermi1}
f({\bf v})=\frac{\eta_0}{\lambda e^{\beta m
\frac{v^2}{2}}+1}
\end{eqnarray}
with  $\eta_0=g {m^4}/{h^3}$ and $\lambda=e^{-\alpha}=e^{-\beta \mu}$.
In the completely degenerate limit, the Fermi-Dirac DF reduces to the step function
\begin{eqnarray}
\label{fermi2}
f({\bf v})=\eta_0 H(v-v_F)\qquad (T=0),
\end{eqnarray}
where $v_F=(2\mu/m)^{1/2}$ is the Fermi velocity (the Fermi energy $\epsilon_F=m v_F^2/2=\mu$ is equal to the chemical potential) and $H$ is the
Heaviside function defined such that $H(x)=1$ if $x<0$ and $H(x)=0$ if $x>0$. The density and the pressure are given by
\begin{eqnarray}
\label{fermi3}
\rho=\frac{4\pi}{3} \eta_0 v_F^3,\qquad P=\frac{4\pi}{15} \eta_0 v_F^5.
\end{eqnarray}
This leads to the polytropic equation of state
\begin{eqnarray}
\label{fermi4}
P=\frac{1}{5}\left (\frac{3}{4\pi\eta_0}\right )^{2/3}\rho^{5/3},
\end{eqnarray}
to the velocity dispersion
\begin{eqnarray}
\label{fermi5g}
\sigma^2=\frac{\langle v^2\rangle}{3}=\frac{P}{\rho}=\frac{1}{5}\left
(\frac{3}{4\pi\eta_0}\right
)^{2/3}\rho^{2/3}=\frac{v_F^2}{5},
\end{eqnarray}
and to the squared speed of sound
\begin{eqnarray}
\label{fermi5b}
c_s^2=P'(\rho)=\frac{1}{3}\left
(\frac{3}{4\pi\eta_0}\right
)^{2/3}\rho^{2/3}=\frac{v_F^2}{3}.
\end{eqnarray}
We note that the pressure, the velocity dispersion and the speed of sound do not
vanish at $T=0$ 
as a consequence of the Pauli exclusion principle (this is also
the case for the Jeans length from Eq.
(\ref{eos5c})).

If we define the Fermi temperature $T_F$ by the relation $k_B
T_F=m v_F^2/2=\mu$,  we obtain
\begin{eqnarray}
\label{fermi6}
\frac{k_B T_F}{m}=\frac{1}{2}\left (\frac{3\rho}{4\pi\eta_0}\right
)^{2/3}\qquad {\rm i.e.}\qquad T_F=\frac{1}{2}\left (\frac{6\pi^2}{g}\right
)^{2/3}\frac{\hbar^2n^{2/3}}{m k_B}.
\end{eqnarray}
Physically, the Fermi temperature determines the transition
between the completely degenerate limit and the nondegenerate limit:
\begin{eqnarray}
\label{fermi7}
P_{\rm deg}\sim P_{\rm class} \quad \Rightarrow \qquad \frac{1}{5}\left (\frac{3}{4\pi\eta_0}\right )^{2/3}\rho^{5/3}\sim \rho \frac{k_B T}{m}\qquad \Rightarrow \qquad T\sim T_F.
\end{eqnarray}

The completely degenerate limit corresponds to $T=0$ and $\mu>0$ finite
(implying $\alpha\rightarrow +\infty$ and
$\lambda\rightarrow 0$). Physically, it is valid when $T\ll
T_F$, $\lambda\ll 1$ and $f\sim \eta_0$ (i.e., $\chi\sim 1$). The nondegenerate
(classical) limit corresponds to $T\gg T_F$, $\lambda\gg 1$ and $f\ll \eta_0$
(i.e., $\chi\ll 1$). The transition between
the  completely degenerate limit and the nondegenerate limit corresponds to
$T\sim T_F$, $\lambda\sim 1$ and $f\sim \eta_0$ (i.e. $\chi\sim 1$).

{\it Remark:} The completely degenerate limit can be recovered from the general
formulae of Appendix \ref{sec_aeos} by using the
identity \cite{kingen}
\begin{eqnarray}
\label{fermi8}
I_n(t)\sim \frac{(-\ln t)^{n+1}}{n+1}\qquad (t\rightarrow 0).
\end{eqnarray}
More precisely, for $t\ll 1$, we have the Sommerfeld expansion
\begin{eqnarray}
I_n(t)=\frac{(-\ln t)^{n+1}}{n+1}\left\lbrack 1+\sum_{k=1}^{+\infty}2C_{2k-1}^n (n+1)(-\ln t)^{-2k}\Gamma(2k)\zeta(2k)\left (1-\frac{1}{2^{2k-1}}\right )\right\rbrack.
\end{eqnarray}

\section{Kinetic theory of bosons in the approximation $f/\eta\gg 1$}
\label{sec_logo}

In this Appendix, we consider the case of bosons in the limit of large
occupation numbers $f/\eta\gg 1$ 
(see Appendix \ref{sec_fdm}) where we can make the approximation
\begin{eqnarray}
\label{logo1}
f\left (1+\frac{f}{\eta}\right )\simeq \frac{f^2}{\eta}.
\end{eqnarray}

\subsection{Log-entropy and Rayleigh-Jeans distribution}

When $f/\eta\gg 1$, the Bose-Einstein entropy from Eq. (\ref{cl9}) reduces to
\begin{eqnarray}
\label{tlogo1}
S=k_B\sum_a\frac{gm_a^3}{h^3}\int \ln\left (\frac{f_a}{\eta_a}\right )\, d{\bf r}d{\bf v}.
\end{eqnarray}
In the context of generalized thermodynamics, it corresponds to
a generalized entropy of the form $S\sim \int \ln f\ d{\bf r}d{\bf v}$ that we called the log-entropy \cite{logo}. Extremizing this entropy at
fixed energy and fixed mass of each species, we obtain the DF:
\begin{eqnarray}
\label{tlogo2}
f_{a}({\bf v})=\frac{\eta_a}{\beta m_a \frac{v^2}{2}-\alpha_a}.
\end{eqnarray}
This DF is positive provided that $\alpha_a\le 0$. In that case, it reduces
to the Cauchy or Lorentz distribution \cite{logo}. In the present context, we
will call it the Rayleigh-Jeans distribution
because of its similarity with the Rayleigh-Jeans spectrum of
radiation for a vanishing chemical potential ($\alpha=0$).\footnote{Note that
the Rayleigh-Jeans spectrum of radiation corresponds to ultrarelativistic bosons
(photons) while we consider here nonrelativistic bosons.}  The DF from Eq.
(\ref{tlogo2}) is the limit form of the Bose-Einstein DF from Eq. (\ref{bec1})
when  $v\rightarrow 0$ and $\alpha_a\rightarrow 0^-$. Therefore, it becomes
exact for small velocities close to the condensation temperature $T_c$. In that
case, we have $f_a/\eta_a\gg 1$.\footnote{If we consider the core  of the
Bose-Einstein  DF, we have $f({\bf 0})=\eta_0/(\lambda-1)$.
Therefore, the limit $f({\bf 0})/\eta=1/(\lambda-1)\gg 1$ is valid for
$\lambda\rightarrow 1$ hence for $T\rightarrow T_c$.}  It has to be noted that
the Rayleigh-Jeans DF decreases as $v^{-2}$ for $v\rightarrow +\infty$ so it is
not normalizable. Therefore, the density $\rho_b$ is not defined, except if we
introduce a 
cut-off at $V_{\rm max}$ \cite{sopik}.  In that case, for a
given density $\rho_a$, the chemical potential is related to the temperature by
the formula
\begin{eqnarray}
\label{tlogo2a}
\rho_a=4\pi\eta_a |\alpha_a|^{1/2}\left (\frac{2}{\beta m_a}\right )^{3/2}
\left\lbrack \sqrt{\frac{\beta m_a}{2|\alpha_a|}}V_{\rm max}-\tan^{-1}\left
(\sqrt{\frac{\beta m_a}{2|\alpha_a|}}V_{\rm max}\right )\right\rbrack.
\end{eqnarray}
The condensation temperature
$T_c$, corresponding to $\alpha_a=0$,  is given by
\begin{eqnarray}
\label{tlogo2b}
k_B T_c=\frac{m_a\rho_a}{8\pi\eta_a V_{\rm max}}.
\end{eqnarray}
At that point, the DF from Eq. (\ref{tlogo2}) reduces to
\begin{eqnarray}
\label{tlogo2c}
f_{a}({\bf v})=\frac{2\eta_a}{\beta m_a v^{2}}.
\end{eqnarray}
On the other hand, Eq. (\ref{tlogo2a}) can be rewritten as
\begin{eqnarray}
\label{tlogo2s}
\frac{T_c}{T}=1-\frac{\tan^{-1}(X)}{X},\quad {\rm with}\quad X=\sqrt{\frac{\beta
m_a}{2|\alpha_a|}}V_{\rm max}.
\end{eqnarray}

\subsection{Bosonic Landau equation}

With the approximation from Eq. (\ref{logo1}), the bosonic Landau
equation (\ref{q1}) becomes
\begin{eqnarray}
\label{logo2}
\frac{\partial f}{\partial t}=2\pi G^2\ln\Lambda\frac{\partial}{\partial
v_i}\int d{\bf v}' K_{ij}\left ( \frac{m_b}{\eta_b} {f'_b}^2 \frac{\partial f}{\partial v_j}- \frac{m}{\eta_0} f^2\frac{\partial f'_b}{\partial v'_j}\right ).
\end{eqnarray}
In the single-species case, we obtain the self-consistent kinetic equation
\begin{eqnarray}
\label{logo2b}
\frac{\partial f}{\partial t}=2\pi G^2 \frac{m}{\eta_0}\ln\Lambda\frac{\partial}{\partial
v_i}\int d{\bf v}' K_{ij}\left ( {f'}^2 \frac{\partial f}{\partial v_j}- f^2\frac{\partial f'}{\partial v'_j}\right ),
\end{eqnarray}
which can be used as an approximation of Eq. (\ref{qles1}) to study the process of Bose-Einstein condensation when $E<E_c$ (see Sec. \ref{sec_qles}).

\subsection{Bosonic Fokker-Planck equation}

The bosonic Landau equation (\ref{logo2}) can be written under the form of a bosonic
Fokker-Planck equation
\begin{eqnarray}
\label{logo3}
\frac{\partial f}{\partial t}=\frac{\partial}{\partial
v_i}\left ( D_{ij}\frac{\partial f}{\partial v_j}- \frac{f^2}{\eta_0} F_i^{\rm pol} \right )
\end{eqnarray}
with a diffusion tensor
\begin{eqnarray}
\label{logo4}
D_{ij}=2\pi G^2 m_b \ln\Lambda \int d{\bf v}'
K_{ij} \frac{{f'_b}^2}{\eta_b}
\end{eqnarray}
and a friction by polarisation
\begin{eqnarray}
\label{logo5}
F_i^{\rm pol}=2\pi G^2 m \ln\Lambda\int d{\bf v}'
K_{ij}\frac{\partial f'_b}{\partial v'_j}.
\end{eqnarray}
The relation between the friction by polarisation and the
true friction is still given by Eq. (\ref{q8}). The results from Secs. \ref{sec_qrp} and \ref{sec_qib} remain valid with the approximation from Eq. (\ref{logo1}).

\subsection{Thermal bath}

We now assume that the field particles are in a statistical equilibrium state
described by the Rayleigh-Jeans DF
\begin{eqnarray}
\label{logo6}
f_{b}({\bf v})=\frac{\eta_b}{\beta m_b \frac{v^2}{2}+|\alpha_b|}.
\end{eqnarray}
We can check that Eq. (\ref{q24}) remains valid so the bosonic
Kramers-Chandrasekhar equation takes the form
\begin{eqnarray}
\label{logo7}
\frac{\partial f}{\partial t}=\frac{\partial}{\partial
v_i}\left \lbrack D_{ij} \left (\frac{\partial f}{\partial v_j}+\beta \frac{m}{\eta_0} f^2 v_j\right ) \right \rbrack.
\end{eqnarray}
This equation relaxes towards the Rayleigh-Jeans  DF
\begin{eqnarray}
\label{logo8}
f_{\rm eq}({\bf v})=\frac{\eta_0}{\beta m \frac{v^2}{2}+|\alpha|}.
\end{eqnarray}
At statistical equilibrium, the test particles and the field
particles have the same temperature.

\subsection{Diffusion and friction terms in the thermal bath
approximation}

For a thermal bath [see Eq. (\ref{logo6})], the diffusion coefficients of the test
particles obtained from Eqs. (\ref{q29})-(\ref{q33}) with the approximation from Eq. (\ref{logo1})
are given by
\begin{eqnarray}
\label{q29h}
D_{\|}=\frac{16\pi^2}{3}G^2m_b\ln\Lambda\frac{1}{v^3}( L_4+v^3
M_1)
\end{eqnarray}
and
\begin{eqnarray}
\label{q30h}
D_{\perp}=\frac{16\pi^2}{3}G^2m_b\ln\Lambda\frac{1}{v}\left
(3L_2-\frac{1}{v^2}L_4+2v M_1 \right )
\end{eqnarray}
with
\begin{eqnarray}
M_1=\frac{1}{\eta_b}\int_v^{+\infty} v_1 f_b(v_1)^2 \,
dv_1=\frac{\eta_b}{\beta
m_b}\frac{1}{|\alpha_b|+\beta m_b \frac{v^2}{2}},
\end{eqnarray}
\begin{eqnarray}
\label{q32h}
L_2=\frac{1}{\eta_b}\int_0^{v} v_1^2 f_b(v_1)^2\,
dv_1=-\frac{\eta_b}{\beta
m_b}\frac{v}{|\alpha_b|+\beta m_b
\frac{v^2}{2}}+\frac{\eta_b}{\beta
m_b}\left (\frac{2}{\beta m_b |\alpha_b|}\right )^{1/2}\tan^{-1}\left\lbrack
\left (\frac{\beta m_b}{2|\alpha_b|}\right
)^{1/2}v\right\rbrack,
\end{eqnarray}
\begin{eqnarray}
L_4=\frac{1}{\eta_b}\int_0^{v} v_1^4 f_b(v_1)^2\,
dv_1=-\frac{\eta_b}{\beta
m_b}\frac{v^3}{|\alpha_b|+\beta m_b
\frac{v^2}{2}}+\frac{6\eta_b}{(\beta
m_b)^2}\left\lbrace v-\left (\frac{2|\alpha_b|}{\beta m_b}\right
)^{1/2}\tan^{-1}\left\lbrack \left (\frac{\beta m_b}{2|\alpha_b|}\right
)^{1/2}v\right\rbrack\right\rbrace.
\end{eqnarray}
This leads to the explicit expressions 
\begin{eqnarray}
\label{q34h}
D_{\|}=32\pi^2 G^2m_b\ln\Lambda\frac{1}{v^3}\frac{\eta_b}{(\beta
m_b)^2}\left\lbrace v-\left (\frac{2|\alpha_b|}{\beta m_b}\right
)^{1/2}\tan^{-1}\left\lbrack \left (\frac{\beta m_b}{2|\alpha_b|}\right
)^{1/2}v\right\rbrack\right\rbrace
\end{eqnarray}
and
\begin{eqnarray}
\label{q35h}
D_{\perp}=16\pi^2 G^2\ln\Lambda\frac{\eta_b}{\beta v}\Biggl
\lbrack\left (\frac{2}{\beta m_b |\alpha_b|}\right
)^{1/2}\tan^{-1}\left\lbrack
\left (\frac{\beta m_b}{2|\alpha_b|}\right
)^{1/2}v\right\rbrack\nonumber\\
-\frac{2}{\beta
m_b v^2}\left\lbrace v-\left (\frac{2|\alpha_b|}{\beta m_b}\right
)^{1/2}\tan^{-1}\left\lbrack \left (\frac{\beta m_b}{2|\alpha_b|}\right
)^{1/2}v\right\rbrack\right\rbrace\Biggr\rbrack.
\end{eqnarray}
Despite the fact that the density $\rho_b$ is not defined (the Rayleigh-Jeans 
DF is not normalisable), the diffusion coefficients are perfectly well-defined.
We note that $D_{\|}$ decreases as $v^{-2}$, instead of $v^{-3}$ in Eq.
(\ref{q36}), when $v\rightarrow +\infty$. Using Eqs. (\ref{q17})-(\ref{q19}) and
(\ref{q25}), we also obtain
\begin{eqnarray}
{\bf F}_{\rm pol}=-32\pi^2 G^2m\ln\Lambda\frac{1}{v^3}\frac{\eta_b}{\beta
m_b}\left\lbrace v-\left (\frac{2|\alpha_b|}{\beta m_b}\right
)^{1/2}\tan^{-1}\left\lbrack \left (\frac{\beta m_b}{2|\alpha_b|}\right
)^{1/2}v\right\rbrack\right\rbrace {\bf v},
\end{eqnarray}
\begin{eqnarray}
\frac{\partial D_{ij}}{\partial v_j}=-16\pi^2 G^2m_b\ln\Lambda\frac{v_i}{v^3}\frac{\eta_b}{\beta
m_b}\left\lbrace -\frac{v}{|\alpha_b|+\beta m_b
\frac{v^2}{2}}+\left (\frac{2}{\beta m_b |\alpha_b|}\right )^{1/2}\tan^{-1}\left\lbrack
\left (\frac{\beta m_b}{2|\alpha_b|}\right
)^{1/2}v\right\rbrack\right\rbrace,
\end{eqnarray}
\begin{eqnarray}
{\bf F}_{\rm friction}={\bf F}_{\rm pol}+\frac{\partial D_{ij}}{\partial v_j}.
\end{eqnarray}

At the critical temperature $T_c$, corresponding to $\alpha_b\rightarrow 0$, the
diffusion coefficient in the direction parallel to the velocity of the test 
particle converges and achieves the expression
\begin{eqnarray}
\label{q34htc}
D_{\|}=32\pi^2 G^2m_b\ln\Lambda\frac{1}{v^2}\frac{\eta_b}{(\beta_c
m_b)^2},
\end{eqnarray}
while the diffusion coefficient in the direction perpendicular to the velocity of the test particle diverges like
\begin{eqnarray}
\label{q35htc}
D_{\perp}\sim 8\pi^3 G^2\ln\Lambda\frac{\eta_b}{\beta_c v}\left
(\frac{2}{\beta_c m_b |\alpha_b|}\right
)^{1/2}.
\end{eqnarray}
Similarly, the friction by polarisation converges at $T_c$ while the total
friction diverges (this 
is also the case when the bosons are described by the exact Bose-Einstein DF).
One can also define an effective mass of the particles. From Eqs. (\ref{m16})
and (\ref{logo6}), we obtain\footnote{Note that the integral in the numerator
of Eq. (\ref{m16}) converges while the integral in the denominator diverges in
the absence of a cut-off $V_{\rm max}$.}
\begin{eqnarray}
\label{meffrj}
m_{\rm eff}=\frac{2\pi^2\eta_b}{\beta^{3/2}\rho_b}\left (\frac{2}{ m_b |\alpha_b|}\right
)^{1/2}.
\end{eqnarray}
We can then express
$D_{\perp}$ close to $T_c$ under the form
\begin{eqnarray}
\label{q35hmeff}
D_{\perp}\sim 4\pi G^2\ln\Lambda\frac{\rho_b}{v}m_{\rm eff}.
\end{eqnarray}
Using Eq. (\ref{tlogo2s}), we see that the effective mass diverges like
\begin{eqnarray}
\label{meffrjdiv}
m_{\rm eff}\sim \frac{\pi^2}{4}\frac{1}{\beta_c V_{\rm max}^2}\left
(\frac{T}{T_c}-1\right )^{-1}
\end{eqnarray}
when $T\rightarrow T_c$.

{\it Remark:} The  Rayleigh-Jeans (Lorentz or Cauchy) DF is a rare case, with
the Maxwellian, where the diffusion and friction coefficients can be calculated
analytically (for a thermal bath).

\section{Rate of change of the energy of classical particles}
\label{sec_dre}

In this Appendix we compute the rate of change of the energy of a
classical system of particles of mass $m$ (test particles) due to collisions
with
classical  particles of mass $m_b$ (field particles).\footnote{We note that the
collisions of the test particles between themselves do not change their energy.
We also note that the total energy (test $+$ field particles) is conserved if the two populations evolve self-consistently (see
Appendix \ref{sec_cl}).} We use
this
result to estimate the relaxation time of the system towards statistical
equilibrium (thermalization).  We also study the evolution of the velocity dispersion
 of the test particles due to heating and cooling (see Appendix D of
\cite{hfcp} for more general results).

The (kinetic) energy
of the test particles is
\begin{eqnarray}
\label{dre1}
E=\int f\frac{v^2}{2}\, d{\bf v}.
\end{eqnarray}
Using the Fokker-Planck equation (\ref{c2}) and performing integrations by
parts, we
find that its rate of change is
\begin{eqnarray}
\label{dre2}
\dot E&=&\int \frac{\partial f}{\partial t}\frac{v^2}{2}\, d{\bf v}\nonumber\\
&=&\int \frac{\partial}{\partial
v_i}\left ( D_{ij}\frac{\partial f}{\partial v_j}- f  F_i^{\rm pol}
\right )
\frac{v^2}{2}\, d{\bf v}\nonumber\\
&=&-\int \left ( D_{ij}\frac{\partial f}{\partial v_j}-
f  F_i^{\rm pol}
\right ) v_i\, d{\bf v}\nonumber\\
&=&-\int  D_{ij}v_i\frac{\partial f}{\partial v_j}\, d{\bf v}+\int
f  ({\bf F}_{\rm pol}\cdot {\bf v})\, d{\bf v}.
\end{eqnarray}
The mean change in energy arises from the competition between diffusion (heating) and dynamical friction (cooling). We
assume that, initially, the test particles have a Maxwellian DF
\begin{eqnarray}
\label{dre3}
f_0({\bf v})=\rho \left (\frac{\beta_0 m}{2\pi}\right )^{3/2}e^{-\beta_0 m
\frac{v^2}{2}}
\end{eqnarray}
with a
temperature $T_0$.  Using
\begin{eqnarray}
\label{dre4}
\frac{\partial f_0}{\partial {\bf v}}=-\beta_0 m f_0 {\bf v},
\end{eqnarray}
we find that the initial rate of change of the energy of the test particles is given by
\begin{eqnarray}
\label{dre5}
(\dot E)_0=\beta_0 m \int  D_{ij}v_i v_j f_0\, d{\bf
v}+\int
f_0  ({\bf F}_{\rm pol}\cdot {\bf v})\, d{\bf v}.
\end{eqnarray}
If the DF of the field particles is isotropic, using the identity
$D_{ij}v_iv_j=D_{\|} v^2$ [see Eq. (\ref{c17})], we obtain
\begin{eqnarray}
\label{dre7}
(\dot E)_0=\beta_0 m \int D_{\|} v^2  f_0\, d{\bf
v}+\int
f_0  ({\bf F}_{\rm pol}\cdot {\bf v})\, d{\bf v}.
\end{eqnarray}
If the field particles are at statistical equilibrium  with a temperature $T$ (thermal
bath) then, using Eq. (\ref{c28}), we get
\begin{eqnarray}
\label{dre8}
{\bf F}_{\rm pol}\cdot {\bf v}= - \beta m D_{\|} v^2.
\end{eqnarray}
In that case, the initial rate of change of the energy of the test particles can be
written as\footnote{We can obtain this equation directly from the Kramers-Chandrasekhar equation (\ref{c30}).}
\begin{eqnarray}
\label{dre10}
(\dot E)_0=(\beta_0-\beta) m \int D_{\|} v^2  f_0\, d{\bf
v}.
\end{eqnarray}
From this formula, we see that $(\dot E)_0<0$ if $T_0>T$
and $(\dot
E)_0>0$ if $T_0<T$. In the first case, the test particles have to lose
energy in
order to acquire the bath temperature $T$. In the second case they have to
gain energy. Using Eqs. (\ref{c32}) and (\ref{dre3}), we have
\begin{eqnarray}
\label{dre11}
\int f_0 D_{\|} v^2  \, d{\bf v}=\rho \left (\frac{\beta_0 m}{2\pi}\right
)^{3/2} 4\pi G^2 m_b\ln\Lambda \rho_b 4\pi \left (\frac{2}{\beta
m_b}\right )^2
I\left (\frac{mT}{m_b T_0}\right ),
\end{eqnarray}
where $I(a)$ is the integral
\begin{eqnarray}
\label{dre12}
I(a)=\int_0^{+\infty} e^{-ax^2}G(x)x^3\, dx.
\end{eqnarray}
Recalling the definition of $G(x)$ from Eq. (\ref{c35}) and integrating by part,
we get
\begin{eqnarray}
\label{dre13}
I(a)=\frac{1}{a\sqrt{\pi}}\int_0^{+\infty} e^{-(1+a)x^2}x^2\, dx.
\end{eqnarray}
With the change of variables $y=\sqrt{1+a}\, x$, this integral can be
rewritten as
\begin{eqnarray}
\label{dre14}
I(a)=\frac{1}{a\sqrt{\pi}(1+a)^{3/2}}\int_0^{+\infty} e^{-y^2}y^2\, dy.
\end{eqnarray}
Using the identity
\begin{eqnarray}
\label{dre15}
\int_0^{+\infty} e^{-y^2}y^2\, dy=\frac{\sqrt{\pi}}{4},
\end{eqnarray}
we finally obtain
\begin{eqnarray}
\label{dre16}
I(a)=\frac{1}{4a(1+a)^{3/2}}.
\end{eqnarray}
Collecting the foregoing results, we find that
\begin{eqnarray}
\label{dre11b}
\int f_0 D_{\|} v^2  \, d{\bf v}=4\sqrt{2\pi} \rho_b G^2 \ln\Lambda \rho
k_B T \frac{k_B T_0}{m}\frac{1}{\left (\frac{k_B T_0}{m}+\frac{k_B
T}{m_b}\right
)^{3/2}}.
\end{eqnarray}
Therefore, the initial rate of change of the energy of the test particles is
\begin{eqnarray}
\label{dre17}
(\dot E)_0=4\sqrt{2\pi} \rho_b G^2 \ln\Lambda \rho
k_B \frac{T-T_0}{\left (\frac{k_B T_0}{m}+\frac{k_B T}{m_b}\right
)^{3/2}}.
\end{eqnarray}
This equation is exact at $t=0$ under the preceding assumptions.
Now, assuming that the DF of the test particles can be approximated at any time
by a
Maxwellian distribution with temperature $T_t(t)$,\footnote{The DF which is the
solution of the Fokker-Planck equation (\ref{c30}) is not exactly Maxwellian
(see Appendix \ref{sec_ssss}) but we
may expect that a Maxwellian ansatz  provides a reasonable approximation of the exact
DF to evaluate $\dot E$.} and using the relation
\begin{eqnarray}
\label{dre18}
E(t)=\frac{3}{2}\rho \frac{k_B T_t(t)}{m},
\end{eqnarray}
we can extend Eq. (\ref{dre17}) to any time and write
\begin{eqnarray}
\label{dre19}
\frac{dT_t}{dt}=\frac{8}{3}\sqrt{2\pi} \rho_b G^2 m \ln\Lambda
\frac{T-T_t}{\left (\frac{k_B T_t}{m}+\frac{k_B T}{m_b}\right
)^{3/2}}.
\end{eqnarray}
This
corresponds to the
result of Spitzer \cite{spitzer1940,spitzerplasma,spitzerastro} (see also Landau \cite{landau} and Eq. (2.377) of Chandrasekhar \cite{chandra}) obtained in a
different manner. This equation shows that the temperature of the test particles
relaxes towards
the temperature of the bath ($T_t=T$). For short times, we have
\begin{eqnarray}
\label{dre19a}
T_t(t)=T_0+\frac{8}{3}\sqrt{2\pi} \rho_b G^2 m \ln\Lambda
\frac{T-T_0}{\left (\frac{k_B T_0}{m}+\frac{k_B T}{m_b}\right
)^{3/2}}t+...
\end{eqnarray}
For late times, we have $T_t\simeq T$ (thermalization), and we can write Eq.
(\ref{dre19}) as
\begin{eqnarray}
\label{dre20}
\frac{dT_t}{dt}=\frac{T-T_t}{t_{\rm relax}},
\end{eqnarray}
where, following Spitzer
\cite{spitzerplasma,spitzer1969}, we have introduced the relaxation (or equipartition)
time
\begin{eqnarray}
\label{dre21}
t_{\rm relax}=\frac{3\left (\frac{k_B T}{m_b}\right
)^{3/2}}{8\sqrt{2\pi} \rho_b G^2 m \ln\Lambda}\left (1+\frac{m_b}{m}\right
)^{3/2}.
\end{eqnarray}
The solution of Eq. (\ref{dre20}) is
\begin{eqnarray}
\label{dre22}
T_t(t)=T+(T_0-T)e^{-t/t_{\rm relax}}.
\end{eqnarray}
Close to equilibrium, the temperature of the test particles relaxes
exponentially rapidly towards the temperature of the bath on a time scale
$t_{\rm relax}$. If we introduce the velocity dispersion (in one direction) of the test and field
particles defined by
\begin{eqnarray}
\label{dre23}
\sigma_t^2=\frac{k_B T_t}{m}=\frac{1}{\beta_t m}=\frac{1}{3\rho}\int f
v^2\, d{\bf v} \quad {\rm
and} \quad \sigma_b^2=\frac{k_B
T}{m_b}=\frac{1}{\beta m_b}=\frac{1}{3\rho_b}\int f_b
v^2\, d{\bf v},
\end{eqnarray}
we can rewrite Eq. (\ref{dre19}) as
\begin{eqnarray}
\label{dre24}
\frac{d\sigma_t^2}{dt}=\frac{8}{3}\sqrt{2\pi} \rho_b G^2
\ln\Lambda
\frac{m_b\sigma_b^2-m\sigma_t^2}{\left (\sigma_t^2+\sigma_b^2\right
)^{3/2}}.
\end{eqnarray}
At equilibrium, we have equipartition of energy:
\begin{eqnarray}
\label{dre25b}
T_t=T \qquad {\rm and}  \qquad m\sigma_t^2=m_b\sigma_b^2.
\end{eqnarray}
On the other hand, the
relaxation time from Eq. (\ref{dre21}) can be written as
\begin{eqnarray}
\label{dre25}
t_{\rm relax}=\frac{3\sigma_b^3}{8\sqrt{2\pi} \rho_b G^2 m \ln\Lambda}\left
(1+\frac{m_b}{m}\right
)^{3/2}.
\end{eqnarray}
For a single species system ($m_b=m$) it
reduces to\footnote{This result may look paradoxical at first sight since it is
based on the rate of change of energy [see Eq. (\ref{dre17})] while the energy of a single species
system of particles is conserved. Actually, it gives the relaxation time of a
test particle (or an ensemble of noninteracting test particles) of mass $m$ in
collision with field particles of the same mass $m$. The test particles are initially out-of-equilibrium while the field particles are at statistical equilibrium (this is how they can be distinguished).
It also provides the typical relaxation time of the system ``as a whole''
evolving according to the self-consistent Landau equation (\ref{cle2}). In that case, it can
be obtained qualitatively from scaling arguments (see Sec. \ref{sec_cle}).}
\begin{eqnarray}
\label{dre26}
t_{\rm relax}=\frac{3\sigma^3}{4\sqrt{\pi} \rho G^2 m \ln\Lambda}.
\end{eqnarray}
We stress that Eqs. (\ref{dre20}), (\ref{dre21}) and (\ref{dre25}) are valid only close to the equilibrium state so they may not accurately characterize the whole relaxation process. To obtain a better description of the relaxation process we have to come back to Eq. (\ref{dre19}) or, equivalently Eq. (\ref{dre24}), as discussed in the following sections.

\subsection{General case}

In the general case, Eq. (\ref{dre24}) can be written as
\begin{eqnarray}
\label{dre27}
\frac{d\sigma_t^2}{dt}=\left (\frac{\sigma_b^2}{t_{\rm
heat}}-\frac{\sigma_t^2}{t_{\rm cool}}\right )\frac{1}{\left
(1+\frac{\sigma_t^2}{\sigma_b^2}\right )^{3/2}},
\end{eqnarray}
where, following Spitzer  \cite{spitzer1940},\footnote{These characteristic times are implicit in the work of  Spitzer  \cite{spitzer1940} (we note that this paper, which includes the effects of diffusion and dissipation was written before the seminal paper of Chandrasekhar \cite{chandra1} on dynamical friction). They have also been
introduced by Bar-Or {\it et al.} \cite{bft} in the case of FDM
 (see Sec. \ref{sec_gbmt} and Appendix \ref{sec_mapp}).} we have introduced
the heating time
\begin{eqnarray}
\label{dre28}
t_{\rm heat}=\frac{3\sigma_b^3}{8\sqrt{2\pi} \rho_b G^2 m_b
\ln\Lambda}
\end{eqnarray}
and the cooling time
\begin{eqnarray}
\label{dre29}
t_{\rm cool}=\frac{3\sigma_b^3}{8\sqrt{2\pi} \rho_b G^2 m
\ln\Lambda}.
\end{eqnarray}
They satisfy the relations
\begin{eqnarray}
\label{dre30}
\frac{t_{\rm cool}}{t_{\rm heat}}=\frac{m_b}{m}\qquad {\rm and}\qquad
\frac{t_{\rm relax}}{t_{\rm cool}}=\left (1+\frac{m_b}{m}\right
)^{3/2}.
\end{eqnarray}
We can also write Eq. (\ref{dre27}) as
\begin{eqnarray}
\label{dre31}
\frac{d\sigma_t^2}{dt}=\frac{1}{t_{\rm cool}}\left
(\frac{m_b}{m}\sigma_b^2-\sigma_t^2\right
)\frac{1}{\left
(1+\frac{\sigma_t^2}{\sigma_b^2}\right )^{3/2}}.
\end{eqnarray}
Setting $x=\sigma_t^2/\sigma_b^2$ and $a=m_b/m$, the solution of this equation can be written as
\cite{spitzer1940} 
\begin{eqnarray}
\label{dre32}
\int \frac{(1+x)^{3/2}}{x-a}\, dx=-\frac{t}{t_{\rm
cool}}
\end{eqnarray}
where the integral is explicitly given by
\begin{eqnarray}
\label{dre32exp}
\int \frac{(1+x)^{3/2}}{x-a}\,
dx=\frac{2}{3}\sqrt{1+x}(4+3a+x)-2(1+a)^{3/2}\tanh^{
-1 } \left
(\sqrt{\frac{1+x}{1+a}}\right )+C.
\end{eqnarray}
Let us consider particular limits of this equation.

(i) When $\sigma_t\gg \sigma_b$ ($x\gg 1$),  Eq. (\ref{dre32}) reduces to
\begin{eqnarray}
\label{dre33}
\int \frac{x^{3/2}}{x-a}\, dx=-\frac{t}{t_{\rm cool}}
\end{eqnarray}
and the integral is explicitly given by
\begin{eqnarray}
\label{dre34}
\int \frac{x^{3/2}}{x-a}\,
dx=2a\sqrt{x}+\frac{2}{3}x^{3/2}-2a^{3/2}\tanh^{-1}\left
(\sqrt{\frac{x}{a}}\right )+C.
\end{eqnarray}
We recall that, at equilibrium, $(\sigma_t^2/\sigma_b^2)_{\rm eq}=m_b/m$.
Therefore, if we want the condition $\sigma_t\gg \sigma_b$ to be valid at any
time, we have to assume that $m_b\gg m$. In that case, $t_{\rm
cool}\simeq (m/m_b)^{3/2}
t_{\rm relax}\gg t_{\rm heat}$.

(ii) When $\sigma_t\ll \sigma_b$ ($x\ll 1$), Eq. (\ref{dre31}) reduces to
\begin{eqnarray}
\label{dre35}
\frac{d\sigma_t^2}{dt}=\frac{1}{t_{\rm cool}}\left
(\frac{m_b}{m}\sigma_b^2-\sigma_t^2\right
)
\end{eqnarray}
and its solution is \cite{spitzer1940}
\begin{eqnarray}
\label{dre36}
\sigma_t^2(t)=\frac{m_b}{m}\sigma_b^2+\left\lbrack
\sigma_t^2(0)-\frac{m_b}{m}\sigma_b^2\right\rbrack e^{-t/t_{\rm cool}}.
\end{eqnarray}
At equilibrium, $(\sigma_t^2/\sigma_b^2)_{\rm eq}=m_b/m$.
Therefore, if we want the condition $\sigma_t\ll \sigma_b$ to be
valid at any time, we have to assume that $m_b\ll m$. In that case, $t_{\rm
cool}\simeq t_{\rm
relax}\ll t_{\rm heat}$.

(iii) For a single species system ($m_b=m$), Eq. (\ref{dre31}) reduces to (see
footnote 62)
\begin{eqnarray}
\label{dre37}
\frac{d\sigma_t^2}{dt}=\frac{1}{t_{\rm
cool}}\frac{\sigma_b^2-\sigma_t^2}{\left
(1+\frac{\sigma_t^2}{\sigma_b^2}\right )^{3/2}}
\end{eqnarray}
with $t_{\rm cool}=t_{\rm heat}=t_{\rm relax}/2^{3/2}$. At equilibrium
$\sigma_t=\sigma_b$. Setting $x=\sigma_t^2/\sigma_b^2$,
the solution of this
equation can be
written as
\begin{eqnarray}
\label{dre38}
\int \frac{(1+x)^{3/2}}{x-1}\, dx=-\frac{t}{t_{\rm cool}},
\end{eqnarray}
where the integral is explicitly given by
\begin{eqnarray}
\label{dre39}
\int \frac{(1+x)^{3/2}}{x-1}\,
dx=\frac{2}{3}\sqrt{1+x}\, (7+x)-4\sqrt{2}\, \tanh^{-1}\left
(\sqrt{\frac{1+x}{2}}\right )+C.
\end{eqnarray}
When $\sigma_t\gg \sigma_b$
we get 
\begin{eqnarray}
\label{dre40}
\frac{\sigma_t^2}{\sigma_b^2}=\left
\lbrack \frac{\sigma_t(0)^3}{\sigma_b^3}-\frac{3}{2}\frac {t}{t_{\rm
cool}}\right
\rbrack^{2/3}.
\end{eqnarray}
When $\sigma_t\ll \sigma_b$ we obtain 
\begin{eqnarray}
\label{dre41}
\frac{\sigma_t^2}{\sigma_b^2}=1+\left
\lbrack \frac{\sigma_t(0)^2}{\sigma_b^2}-1\right\rbrack\, e^{-t/t_{\rm cool}}.
\end{eqnarray}
These two approximate expressions are only valid for sufficiently short times
since $\sigma_t=\sigma_b$ at equilibrium.

\subsection{$m_b\gg m$: diffusion only (heating)}

When $m_b\gg m$, we have $t_{\rm cool}\gg t_{\rm heat}$. This is the situation
studied
by Spitzer and Schwarzschild \cite{ss} (see also Appendix \ref{sec_ssss}). In
that case, the
friction can be neglected and the evolution of the test particles is purely
diffusive (on the timescale $t_{\rm heat}$). Eqs. (\ref{dre24}) and
(\ref{dre27}) reduce to
\begin{eqnarray}
\label{dre42}
\frac{d\sigma_t^2}{dt}=\frac{8}{3}\sqrt{2\pi} \rho_b G^2 m_b
\ln\Lambda
\frac{\sigma_b^2}{\left (\sigma_t^2+\sigma_b^2\right
)^{3/2}}
\end{eqnarray}
and
\begin{eqnarray}
\label{dre43}
\frac{d\sigma_t^2}{dt}=\frac{\sigma_b^2}{t_{\rm heat}}\frac{1}{\left
(1+\frac{\sigma_t^2}{\sigma_b^2}\right )^{3/2}}.
\end{eqnarray}
The solution of Eq. (\ref{dre43}) is 
\cite{spitzer1940,ss,spitzer1958}
\begin{eqnarray}
\label{dre44}
\frac{\sigma_t^2(t)}{\sigma_b^2}=\left\lbrack \left
(1+\frac{\sigma_t^2(0)}{\sigma_b^2}\right )^{5/2}+\frac{5}{2}\frac{t}{t_{\rm
heat}}\right\rbrack^{2/5}-1.
\end{eqnarray}
For $t\rightarrow +\infty$ (formally), we get
${\sigma_t^2(t)}/{\sigma_b^2}\sim (5t/2t_{\rm heat})^{2/5}$. If we take
$\sigma_t(0)=\sigma_b$ we find that $\sigma_t/\sigma_b=1.15$ at $t=t_{\rm
heat}$.
On the other hand, $\sigma_t/\sigma_b=2$ at $t=20.1\, t_{\rm heat}$.

\subsection{$m\gg m_b$: friction only (cooling)}

When $m\gg m_b$, we have $t_{\rm heat}\gg t_{\rm cool}$. In that case, the
diffusion can be neglected and the test particles just feel the effect
of the friction (on the timescale $t_{\rm cool}$). Eqs. (\ref{dre24}) and
(\ref{dre27}) reduce to
\begin{eqnarray}
\label{dre45}
\frac{d\sigma_t^2}{dt}=-\frac{8}{3}\sqrt{2\pi} \rho_b G^2 m \ln\Lambda
\frac{\sigma_t^2}{\left (\sigma_t^2+\sigma_b^2\right
)^{3/2}}
\end{eqnarray}
and
\begin{eqnarray}
\label{dre46}
\frac{d\sigma_t^2}{dt}=-\frac{\sigma_t^2}{t_{\rm
cool}}\frac{1}{\left
(1+\frac{\sigma_t^2}{\sigma_b^2}\right )^{3/2}}.
\end{eqnarray}
Setting
$x=\sigma_t^2/\sigma_b^2$, the solution of Eq. (\ref{dre46}) can be written as
\begin{eqnarray}
\label{dre47}
\int (1+x)^{3/2}\, \frac{dx}{x}=-\frac{t}{t_{\rm cool}},
\end{eqnarray}
where the integral is explicitly given by
\begin{eqnarray}
\label{dre48}
\int (1+x)^{3/2}\, \frac{dx}{x}=\frac{2}{3}\sqrt{1+x}\,
(4+x)+\ln\left
(\frac{1-\sqrt{1+x}}{1+\sqrt{1+x}}\right )+C.
\end{eqnarray}
For $t\rightarrow +\infty$ (formally), we get $\sigma_t^2(t)\propto e^{-t/t_{\rm
coll}}$.

\section{Rate of change of the energy of classical particles in FDM halos}
\label{sec_mapp}

We consider classical particles of mass $m$ in collision with bosons of mass
$m_b$ -- or  quasiparticles of effective mass $m_{\rm eff}$ -- in FDM halos (see
Sec. \ref{sec_gbmt}). 
The Fokker-Planck equation (\ref{m2}) has the
same form as Eq. (\ref{c2}) so the results (\ref{dre1})-(\ref{dre7}) of Appendix \ref{sec_dre}
remain valid. The initial rate of change of the energy of the test particles is
\begin{eqnarray}
\label{am1}
(\dot E)_0=\beta_0 m \int D_{\|} v^2  f_0\, d{\bf
v}+\int
f_0  ({\bf F}_{\rm pol}\cdot {\bf v})\, d{\bf v}.
\end{eqnarray}
It is the sum of two terms, heating and cooling. To evaluate these terms, we
make the same
approximations as in Sec. \ref{sec_gbmt}. We assume that the DF of the field
particles (bosons) is approximately Maxwellian with a velocity dispersion
$\sigma_b^2\ll \sigma_c^2$ [see
Eq. (\ref{m22})] and we make the approximation $f_b(1+f_b/\eta_b)\simeq
f_b^2/\eta_b$ [see Eq. (\ref{m19})] in the diffusion tensor from Eq. (\ref{m3}). When the corresponding expressions of $D_{\|}$ and ${\bf F}_{\rm pol}$ are substituted into Eq. (\ref{am1}), we recover the results of Bar-Or {\it et al.} \cite{bft} obtained in a slightly different manner.

\subsection{Heating}

We first consider the rate of energy due to heating:
\begin{eqnarray}
\label{am2}
(\dot E)_0^{\rm heat}=\beta_0 m \int D_{\|} v^2  f_0\, d{\bf
v}.
\end{eqnarray}
We have to compute the diffusion coefficient $D_{\|}$ of the test
particles due to the collisions with the bosons. As explained in Sec.
\ref{sec_gbmt}, the diffusion coefficient is the same as
the one created by classical particles of effective mass $m_{\rm eff}$ and
Maxwellian DF with a velocity dispersion $\sigma_b^2/2$ \cite{bft}. Replacing
$m_b$ by
$m_{\rm eff}$ and $\sigma_b^2$ by $\sigma_b^2/2$ in Eq. (\ref{dre42}), we obtain
\begin{eqnarray}
\label{am3}
\frac{d\sigma_t^2}{dt}=\frac{4}{3}\sqrt{2\pi}G^2 m_{\rm
eff}\ln\Lambda\rho_b\sigma_b^2 \frac{1}{\left
(\sigma_t^2+\frac{1}{2}\sigma_b^2\right )^{3/2}}.
\end{eqnarray}
This equation can be rewritten as 
\begin{eqnarray}
\label{am4}
\frac{d\sigma_t^2}{dt}=\frac{\sigma_b^2}{t_{\rm
heat}}\frac{1}{\left
(1+\frac{2\sigma_t^2}{\sigma_b^2}\right )^{3/2}},
\end{eqnarray}
where we have introduced  the heating time \cite{bft}
\begin{eqnarray}
\label{am5}
t_{\rm heat}=\frac{3\sigma_b^3}{16\sqrt{\pi}G^2\rho_b m_{\rm eff}\ln\Lambda}.
\end{eqnarray}
This expression differs from the classical heating time (\ref{dre28}) by the
fact that $m_b$ is replaced by $m_{\rm eff}$.
Since $m_{\rm eff}\gg m_b$, the heating time due to collisions with bosons of mass $m_b$  -- or quasiparticles of effective mass $m_{\rm eff}$ -- is much shorter than the heating time due to collisions with classical
particles of mass $m_b$ as in the CDM model. This is due to Bose enhancement.
Using Eq. (\ref{m29}), we can rewrite Eq. (\ref{am5}) as
\begin{eqnarray}
\label{am6}
t_{\rm heat}=\frac{3 m_b^3\sigma_b^6}{16\pi^2G^2\rho_b^2 \hbar^3 \ln\Lambda}.
\end{eqnarray}
The solution of Eq. (\ref{am4}) is
\begin{eqnarray}
\label{dre44eff}
\frac{\sigma_t^2(t)}{\sigma_b^2}=\frac{1}{2}\left\lbrack \left
(1+2\frac{\sigma_t^2(0)}{\sigma_b^2}\right )^{5/2}+5\frac{t}{t_{\rm
heat}}\right\rbrack^{2/5}-\frac{1}{2},
\end{eqnarray}
which may be compared with Eq. (\ref{dre44}). For $t\rightarrow
+\infty$ (formally), we get
${\sigma_t^2(t)}/{\sigma_b^2}\sim \frac{1}{2}(5t/t_{\rm heat})^{2/5}$. If we
take
$\sigma_t(0)=\sigma_b$ we find that $\sigma_t/\sigma_b=1.08$ at $t=t_{\rm
heat}$. On
the other hand, $\sigma_t/\sigma_b=2$ at $t=45.5\, t_{\rm heat}$.

\subsection{Cooling}

We now consider the rate of energy due to cooling:
\begin{eqnarray}
\label{m7g}
(\dot E)_0^{\rm cool}=\int
f_0  ({\bf F}_{\rm pol}\cdot {\bf v})\, d{\bf v}.
\end{eqnarray}
We have to compute the friction by polarisation  ${\bf F}_{\rm pol}$ experienced
by the test particles due to the collisions with the bosons. As explained in Sec.
\ref{sec_gbmt}, the friction by  polarisation  is the same as
the one created by classical particles of mass $m_b$ and
Maxwellian DF with a velocity dispersion $\sigma_b^2$. Therefore,  Eq.
(\ref{dre45}) remains unchanged and we obtain
\begin{eqnarray}
\label{am8}
\frac{d\sigma_t^2}{dt}=-\frac{8}{3}\sqrt{2\pi}G^2
m\ln\Lambda\rho_b\sigma_t^2 \frac{1}{\left
(\sigma_t^2+\sigma_b^2\right )^{3/2}}.
\end{eqnarray}
This equation can be rewritten as 
\begin{eqnarray}
\label{am9}
\frac{d\sigma_t^2}{dt}=-\frac{\sigma_t^2}{t_{\rm cool}}\frac{1}{\left
(1+\frac{\sigma_t^2}{\sigma_b^2}\right )^{3/2}},
\end{eqnarray}
where we have introduced the cooling time \cite{bft}
\begin{eqnarray}
\label{am10}
t_{\rm cool}=\frac{3\sigma_b^3}{8\sqrt{2\pi}G^2\rho_b m\ln\Lambda}.
\end{eqnarray}
It corresponds to the classical cooling time (\ref{dre29}). Therefore, the cooling time due to collisions with bosons of
mass $m_b$
is the same as the cooling time due to collisions with classical
particles of mass $m_b$ as in the CDM model (it is independent of the mass of
the field particles)
except for a change in the Coulomb logarithm \cite{bft}.  The solution of Eq.
(\ref{am9}) is given by Eqs. (\ref{dre47}) and (\ref{dre48}).

\subsection{General case}

If we account simultaneously for the processes of heating and cooling, and
combine Eqs. (\ref{am4}) and (\ref{am9}), we obtain the following equation 
\begin{eqnarray}
\label{m11}
\frac{d\sigma_t^2}{dt}=\frac{\sigma_b^2}{t_{\rm
heat}}\frac{1}{\left
(1+\frac{2\sigma_t^2}{\sigma_b^2}\right )^{3/2}}-\frac{\sigma_t^2}{t_{\rm
cool}}\frac{1}{\left
(1+\frac{\sigma_t^2}{\sigma_b^2}\right )^{3/2}}.
\end{eqnarray}
According to Eqs. (\ref{am5}) and (\ref{am10}), we have
\begin{eqnarray}
\label{m11a}
\frac{t_{\rm cool}}{t_{\rm
heat}}=\sqrt{2}\, \frac{m_{\rm eff}}{m}.
\end{eqnarray}
The diffusion (heating) dominates the friction (cooling) when
$m_{\rm eff}\gg m$. The friction (cooling) dominates the diffusion
(heating) when $m\gg m_{\rm eff}$.

Setting $x=\sigma_t^2/\sigma_b^2$ and $a=m_{\rm eff}/m$, the formal solution of Eq. (\ref{m11}) is
\begin{eqnarray}
\label{m11b}
\int \frac{dx}{\frac{x}{(1+x)^{3/2}}-\frac{\sqrt{2}a}{(1+2x)^{3/2}}}=-\frac{t}{t_{\rm cool}}.
\end{eqnarray}
When $x\gg 1$ the solution is given by Eqs. (\ref{dre33}) and (\ref{dre34})
with the substitution $m_b\rightarrow m_{\rm eff}/2$.  When $x\ll 1$ the
solution is given by Eq. (\ref{dre36}) with the substitution
$m_b\rightarrow \sqrt{2} m_{\rm eff}$.

According to Eq. (\ref{m11}), the normalized velocity dispersion of the test particles relaxes towards the
equilibrium value $x_e=(\sigma_t^2)_{\rm eq}/\sigma_b^2$ determined  by the equation
\begin{eqnarray}
\label{m11c}
\frac{x_e(1+2x_e)^{3/2}}{(1+x_e)^{3/2}}=\sqrt{2}\, \frac{m_{\rm eff}}{m}.
\end{eqnarray}
When $m\ll m_{\rm eff}$, we get $x_e\sim m_{\rm eff}/2m$. When $m\gg m_{\rm
eff}$, we get $x_e\sim \sqrt{2}\, m_{\rm eff}/m$. When $m_{\rm eff}=m$, we
obtain $x_e=0.812$. We recall that these results are approximate since they
assume that the DF of the test particles is Maxwellian which is usually not the
case [see Eq. (\ref{m41})]. This is, however, the case when $m\ll m_{\rm eff}$
and when $m\gg m_{\rm
eff}$, so we recover the results of footnote 43.

\begin{figure}[!h]
\begin{center}
\includegraphics[clip,scale=0.3]{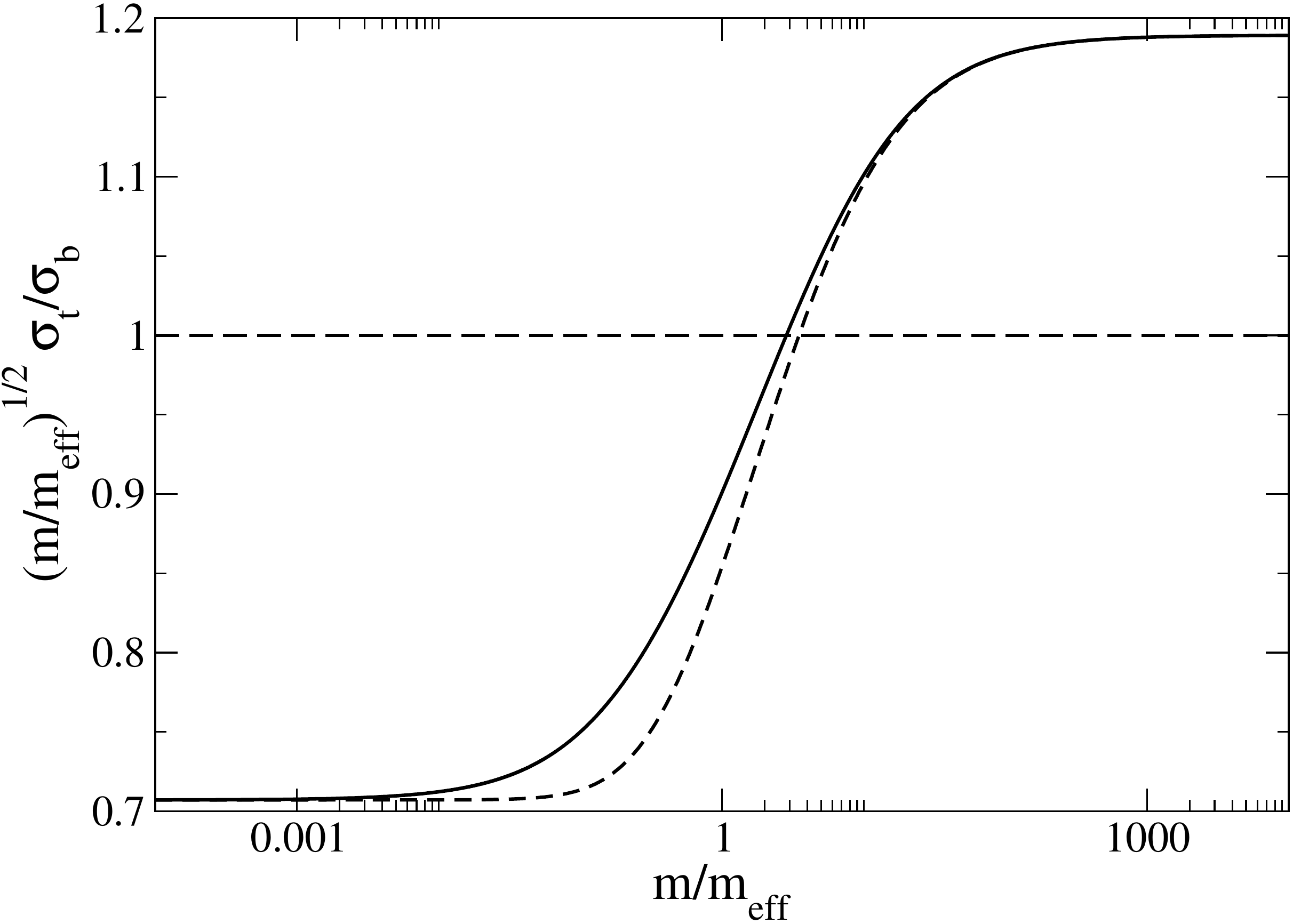}
\caption{Normalized velocity dispersion of the test particles at equilibrium as
a function of the mass ratio $m/m_{\rm eff}$ obtained from Eq. (\ref{m11c}).
The short dashed line corresponds to the normalized velocity dispersion
computed from the exact DF (\ref{m41}).  It
can be compared with Fig. 2 of \cite{bft}.}
\label{disp}
\end{center}
\end{figure}

\section{Self-similar solution of the Spitzer-Schwarzschild equation}
\label{sec_ssss}

For $x\rightarrow +\infty$, using Eq. (\ref{diff12}), the Spitzer-Schwarzschild
equation (\ref{ss1})
reduces to 
\begin{eqnarray}
\label{ssss1}
\frac{\partial f}{\partial
t}=\frac{\pi^{3/2}}{t_H}\frac{1}{x^2}\frac{\partial}{\partial
x}\left ( \frac{1}{x}\frac{\partial f}{\partial x}\right ).
\end{eqnarray}
Measuring the time in units of $t_H/\pi^{3/2}$ and measuring the DF in units
of $\rho/(2^{3/2}\sigma_b^3)$, we obtain
\begin{eqnarray}
\label{ssss2}
\frac{\partial
f}{\partial t}=\frac{1}{x^2}\frac{\partial}{\partial
x}\left ( \frac{1}{x}\frac{\partial f}{\partial x}\right )
\end{eqnarray}
with the normalization condition $\int f\, d{\bf x}=1$. On the other hand, the
velocity dispersion of the test particles [see Eq. (\ref{dre23})] is given by
\begin{eqnarray}
\label{ssss3}
\frac{\sigma_t^2}{\sigma_b^2}=\frac{2}{3}\int f x^2\, d{\bf x}.
\end{eqnarray}
The diffusion equation (\ref{ssss2}), which was not explicitly written in
\cite{ss},
admits a self-similar solution of the form\footnote{Coming back to the
Spitzer-Schwarzschild
equation (\ref{ss1}), the
self-similar solution (\ref{ssss4})
is valid for $t\rightarrow +\infty$, when the typical value of $x$
is large (see the scaling below), so that the approximation from  Eq.
(\ref{ssss1})
is justified.}
\begin{eqnarray}
\label{ssss4}
f(x,t)=t^{-\alpha}F\left (\frac{x}{t^{\beta}}\right ).
\end{eqnarray}
The scaling $x\sim t^{1/5}$ deduced from Eq. (\ref{ssss2}) yields
$\beta=1/5$. On the other hand, the
normalization condition $\int f\, d{\bf x}= 1$ implies
$t^{-\alpha} t^{3\beta}\sim 1$, hence $\alpha=3\beta=3/5$. Therefore, we can
rewrite Eq. (\ref{ssss4}) as
\begin{eqnarray}
\label{ssss5}
f(x,t)=t^{-3/5}F\left (\frac{x}{t^{1/5}}\right ),
\end{eqnarray}
with $\int_0^{+\infty} F(X) 4\pi X^2\, d{X}=1$.
Substituting Eq. (\ref{ssss5}) into Eq. (\ref{ssss2}), we find that the
invariant profile $F(X)$ is
determined by the differential equation
\begin{eqnarray}
\label{ssss6}
\frac{1}{X^2}\frac{d}{dX}\left (\frac{1}{X}\frac{dF}{dX}\right
)+\frac{1}{5}X\frac{dF}{dX}+\frac{3}{5}F=0
\end{eqnarray}
or, equivalently,
\begin{eqnarray}
\label{ssss7}
\frac{d^2F}{dX^2}+\left (\frac{1}{5}X^4-\frac{1}{X}\right
)\frac{dF}{dX}+\frac{3}{5}X^3F=0.
\end{eqnarray}
If we make the change of variables
\begin{eqnarray}
\label{ssss8}
F(X)=e^{-X^5/25} V(X),
\end{eqnarray}
we find that $V(X)$ satisfies
\begin{eqnarray}
\label{ssss9}
\frac{d^2V}{dX^2}-\left (\frac{1}{5}X^4+\frac{1}{X}\right
)\frac{dV}{dX}=0.
\end{eqnarray}
This is a first order differential equation for $V'(X)$ whose solution is
\begin{eqnarray}
\label{ssss10}
V'(X)=A X e^{X^5/25}.
\end{eqnarray}
Therefore, the general solution of Eq. (\ref{ssss7}) is
\begin{eqnarray}
\label{ssss11}
F(X)=A e^{-X^5/25} \int_{0}^X  w e^{w^5/25} \, dw+B e^{-X^5/25},
\end{eqnarray}
where $A$ and $B$ are integration constants. Defining the function
\begin{eqnarray}
\label{ssss12}
\Phi(s,x)=\int_0^x  t^{s-1}  e^{t} \, dt,
\end{eqnarray}
we can rewrite Eq. (\ref{ssss11}) as
\begin{eqnarray}
\label{ssss13}
F(X)=A e^{-X^5/25} \Phi\left (\frac{2}{5},\frac{X^5}{25}\right )+B e^{-X^5/25}.
\end{eqnarray}
The physical solution corresponds to $A=0$ leading to
\begin{eqnarray}
\label{ssss14}
F(X)=B e^{-X^5/25}.
\end{eqnarray}
The constant $B$ is determined by the normalization condition $\int_0^{+\infty}
F(X)
4\pi X^2\, d{X}=1$ yielding
\begin{eqnarray}
\label{ssss15}
B=\frac{1}{4\pi 5^{1/5}\Gamma(3/5)}=0.0387298...
\end{eqnarray}
According to Eqs. (\ref{ssss3}) and (\ref{ssss5}), the velocity dispersion of  the test
particles is given by
\begin{eqnarray}
\label{ssss16}
\frac{\sigma_t^2}{\sigma_b^2}\sim\frac{2}{3}t^{2/5}\int_0^{+\infty} F(X) 4\pi
X^4\,
d{X}.
\end{eqnarray}
With the results from Eqs. (\ref{ssss14}) and (\ref{ssss15}), we
obtain
\begin{eqnarray}
\label{ssss17}
\frac{\sigma_t^2}{\sigma_b^2}\sim\frac{2\times
5^{4/5}}{3\Gamma(3/5)}t^{2/5}=1.62231...\, t^{2/5}.
\end{eqnarray}
This asymptotic result, valid for $t\rightarrow +\infty$,  is in very good agreement with formula
(18) of Spitzer
and Schwarzschild \cite{ss} obtained by solving the diffusion
equation (\ref{ss1}) numerically. We also
note
that the velocity distribution (\ref{ssss14}) is non-Maxwellian.

If we
come back to the
original variables, we get
\begin{eqnarray}
\label{ssss18}
\frac{\sigma_t^2}{\sigma_b^2}\sim 3.22421...\left (\frac{t}{t_H}\right )^{2/5}.
\end{eqnarray}
Now, comparing Eqs. (\ref{dke4}) and (\ref{dre28}), we find ${t_H}/{t_{\rm
heat}}={16\pi}/{3}$, yielding
\begin{eqnarray}
\label{ssss19}
\frac{\sigma_t^2}{\sigma_b^2}\sim 1.04415...\left (\frac{t}{t_{\rm heat}}\right
)^{2/5}.
\end{eqnarray}
This exact result may be compared with the approximate formula [see
Eq. (\ref{dre44})]:
\begin{eqnarray}
\left (\frac{\sigma_t^2}{\sigma_b^2}\right )_{\rm app}\sim \left
(\frac{5}{2}\, \frac{t}{t_{\rm heat}}\right
)^{2/5}\sim 1.4427...\left (\frac{t}{t_{\rm heat}}\right
)^{2/5},
\end{eqnarray}
based on the assumption that the velocity distribution of the test
particles is always Maxwellian (which is not true for large times as we have
seen above). This
Maxwellian assumption provides the correct exponent $2/5$ but the prefactor is
slightly in error.

{\it Remark:} These results are also valid for the diffusion of
light particles with mass  $m\ll m_{\rm eff}$ in FDM halos (see Sec.
\ref{sec_gbmt}) provided that we replace $t_H$ in Eq. (\ref{ssss1}) by
$2\sqrt{2}t_H$, where $t_H$ is now given by Eq. (\ref{m43}).

\end{document}